\documentclass[letterpaper]{article}

\usepackage{arxiv}

\usepackage[utf8]{inputenc}
\usepackage{float}
\usepackage{enumitem}
\usepackage[english]{babel}
\usepackage{amssymb}
\usepackage{amsmath,mathtools}
\usepackage{xcolor}
\usepackage{graphicx}
\usepackage{multirow}
\usepackage{overpic}
\usepackage{psfrag}
\usepackage{bm}
\usepackage{url}
\usepackage{epsfig}
\usepackage{marvosym}
\usepackage{verbatim}
\usepackage{subcaption}
\usepackage{caption}
\usepackage{hyperref}
\hypersetup{colorlinks=true,linkcolor=red,citecolor=blue}
\usepackage{ulem}
\usepackage{lmodern}
\usepackage{booktabs} 
\usepackage{moresize} 
\usepackage{siunitx}
\usepackage{lineno}
\usepackage{tikz}
\usepackage{pgfplots}
\usepgfplotslibrary{colormaps} 
\usetikzlibrary{pgfplots.colormaps} 
\pgfplotsset{compat=1.14}
\usepgfplotslibrary{fillbetween}
\usetikzlibrary{shapes, arrows, arrows.meta, calc, positioning, decorations,
decorations.pathmorphing, decorations.text, decorations.markings, fit, plotmarks}
\pgfplotsset{translate gnuplot=true}
\usepackage{etex} 								
\usepgfplotslibrary{external} 						
\tikzsetexternalprefix{ext/}							

\definecolor{myblue1}	{RGB}{0,177,234}				
\definecolor{myblue2}	{RGB}{76,200,239}				
\definecolor{myblue3}	{RGB}{127,215,244}				
\definecolor{myblue4}	{RGB}{178,231,248}				
\definecolor{myblue5}	{RGB}{198,251,255}				
\definecolor{mybluegray1}{RGB}{0,127,167}				
\definecolor{mybluegray2}{RGB}{76,165,193}				
\definecolor{mybluegray3}{RGB}{127,191,211}				
\definecolor{mybluegray4}{RGB}{178,216,228}				
\definecolor{mygray1}	{RGB}{76,84,93}				
\definecolor{mygray2}	{RGB}{129,135,141}				
\definecolor{mygray3}	{RGB}{165,169,174}				
\definecolor{mygray4}	{RGB}{201,203,206}				
\definecolor{myorange1}	{RGB}{255,126,46}				
\definecolor{myorange2}	{RGB}{255,164,108}				
\definecolor{myorange3}	{RGB}{255,190,150}				
\definecolor{myorange4}	{RGB}{255,216,192}				
\definecolor{mypurple1}{RGB}{89,89,171}
\definecolor{mypurple4}{RGB}{189,189,231}

\newcommand\red[1]{\textcolor{black}{#1}}

\pgfplotsset{
    colormap={custom_map}{[5pt]
            rgb255(0pt)=(255,126,46);
            rgb255(500pt)=(255,190,150);
            rgb255(1000pt)=(0,177,234);
            rgb255(1500pt)=(127,215,244);
    },
}

\newcommand{\V}[1]{\textbf{#1}}
\newcommand{\GV}[1]{\boldsymbol{#1}}
\newcommand\ie{\textit{i.e.}\,\,}

\renewcommand{\emph}[1]{\textit{#1}}
\setenumerate{label=\textcolor{mybluegray1}{$\bm{\diamond}$},itemsep=0.0pt,topsep=5pt}

\pgfplotsset{
	colormap/rdbur/.style={
		colormap={rdbur}{
			rgb255(0cm)=(20,46,97); 
			rgb255(1cm)=(52,100,171); 
			rgb255(2cm)=(83,146,194); 
			rgb255(3cm)=(153,197,222); 
			rgb255(4cm)=(211,229,240);
			rgb255(5cm)=(247,247,247);
			rgb255(6cm)=(250,219,200);
			rgb255(7cm)=(238,165,132);
			rgb255(8cm)=(207,94,80);
			rgb255(9cm)=(171,10,46);
			rgb255(10cm)=(99,0,32);
			}
		}
	}

\newcommand{\colorbar}[4][rdbur]{
\begin{tikzpicture}
	\pgfmathsetmacro{\nsamp}{100};
	\pgfmathsetmacro{\ntick}{3};
	\pgfmathsetmacro{\step}{#2+(#3-#2)/(\ntick-1)};
	\begin{axis}[	xmin=0, xmax=1, ymin=0, ymax=1,
             			hide axis, scale only axis,
				height=0pt, width=0pt,
				colormap/#1, colorbar sampled, colorbar,
				point meta min=#2, point meta max=#3,
				colorbar style={
					scaled y ticks = false,
					tick label style={/pgf/number format/fixed},
					samples=\nsamp,
					height=#4cm,
					ytick={#2,\step,...,#3},
					/pgf/number format/precision=3
				}]
	\end{axis}
\end{tikzpicture}
}

\title{A twin-decoder structure for incompressible laminar flow reconstruction with uncertainty estimation around 2D obstacles}

\def\size{7.2cm}
\author{
	\parbox{\size}{\centering J. Chen}\\
	MINES Paristech, CEMEF\\
	PSL - Research University
\And
	\parbox{\size}{\centering J. Viquerat\thanks{Corresponding author}}\\
	MINES Paristech, CEMEF\\
	PSL - Research University\\
	\texttt{jonathan.viquerat@mines-paristech.fr}\\
\And
	\parbox{\size}{\centering F. Heymes}\\
	MINES Al\`es\\
	Institute for Risk Science
\And
	\parbox{\size}{\centering E. Hachem}\\
	MINES Paristech, CEMEF\\
	PSL - Research University
}

\begin{document}
\newgeometry{left=3cm,right=3cm,top=3cm,bottom=3cm}
\maketitle

\begin{abstract} 
Over the past few years, deep learning methods have proved to be of great interest for the computational fluid dynamics community, especially when used as surrogate models, either for flow reconstruction, turbulence modeling, or for the prediction of aerodynamic coefficients. Overall, exceptional levels of accuracy have been obtained, but the robustness and reliability of the proposed approaches remain to be explored, particularly outside the confidence region defined by the training dataset. In this contribution, we present an autoencoder architecture with twin decoder for incompressible laminar flow reconstruction with uncertainty estimation around 2D obstacles. The proposed architecture is trained over a dataset composed of numerically-computed laminar flows around 12,000 random shapes, and naturally enforces a quasi-linear relation between a geometric reconstruction branch and the flow prediction decoder. Based on this feature, two uncertainty estimation processes are proposed, allowing either a binary decision (accept or reject prediction), or proposing a confidence interval along with the flow quantities prediction ($u,v,p$). Results over dataset samples as well as unseen shapes show a strong positive correlation of this reconstruction score to the mean-squared error of the flow prediction. Such approaches offer the possibility to warn the user of trained models when the provided input shows too large deviation from the training data, making the produced surrogate model conservative for fast and reliable flow prediction.
\end{abstract}


\section{Introduction}

During the last few years, the computational fluid dynamics (CFD) community has largely benefited from the fast-paced development of the machine learning (ML) field, and more specifically from that of the neural networks (NN) domain. In many cases, a part of the usual numerical resolution process is replaced with a trained NN, in order to reduce its computational cost. Examples for these applications are the prediction of closure terms in RANS \cite{Ling2016, Tracey2015} or LES \cite{Beck2018} computations. In other situations, a supervised network is trained to directly predict a flow profile: in \cite{Guo2016}, an autoencoder is used to obtain steady state flow predictions around elementary and real-life shapes; in \cite{Jin2018}, a fusion convolutional neural network (CNN) is trained to predict velocity snapshots around a cylinder in weakly turbulent flows, using the time history of pressure around the cylinder as an input; in \cite{Lee2019}, a neural network is trained to predict unsteady flow around a circular cylinder, by minimizing a physical loss function composed of regression error and conservation laws. CNNs were also directly applied to predict lift and drag coefficients of 2D airfoils \cite{Zhang2018} or arbitrary shapes \cite{Viquerat2020}.

Still, in most of the proposed works, the question of the reliability of the predictions produced by the trained models is left out. Indeed, the topological complexity of the input space can make it hard to determine whether or not a given element, provided by an external user, lies within the boundaries of the dataset used during training. While very few approaches were proposed to tackle such issues in the context of NN-assisted CFD, several outlier detection techniques have been proposed in other domains. Among them, unsupervised methods have attracted much attention, as they do not require labeled data. In particular, several autoencoder (AE) based techniques were developed for medical and industrial applications: in \cite{Hawkins2002}, a fully connected AE with three hidden layers is applied to breast cancer detection; in \cite{Chow2020}, a convolutional AE (CAE) is used to detect cracking and spalling defects on concrete structures; in \cite{Ke2017}, CAE is used to detect miss-printed logo images on mobile phones, and in \cite{Baur2018}, the authors compared several variants of AE on anomaly segmentation in brain magnetic resonance images.

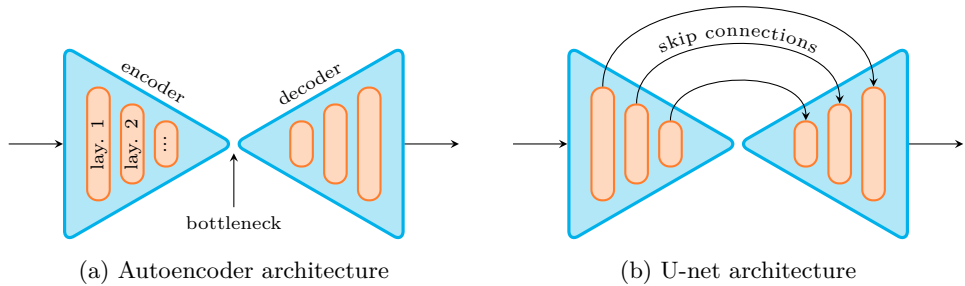
\begin{figure}
\centering

\tikzset{layer/.style={	rectangle, 	thick, rounded corners,
				fill=myorange4, draw=myorange1,
				minimum width=\w}}
\tikzset{down/.style={regular polygon, regular polygon sides=3, 	rotate=-90,
				very thick, rounded corners,
				fill=myblue4, draw=myblue1,
				minimum width=\t}}
\tikzset{up/.style={	down, rotate=180}}
\tikzset{inout/.style={	inner sep=0pt, minimum size=0pt}}
	
\def\s{0.75}
\def\t{\s*4cm}
\def\w{\s*0.4cm}
\def\lar{\s*2cm}
\def\med{\s*1.4cm}
\def\sma{\s*0.8cm}
\def\io{0.25}

\begin{subfigure}[b]{.4\textwidth}
	\centering
	
	\begin{tikzpicture}
		\node[down] 	(tri1) at (0.25*\t,0) {};
		\node[up] 		(tri2) at (\t+0.25*\t,0) {};

		\node[inout] (input) at (-\io*\t,0) {};
		\node[inout] (output) at (1.5*\t+\io*\t,0) {};
		\draw[-stealth] (input.east) -- (tri1.south);
		\draw[-stealth] (tri2.south) -- (output.west);

		\node[layer,minimum height=\lar] 	(down_1) at (0.15*\t,0) {};
		\node[layer,minimum height=\med] 	(down_2) at (0.3*\t,0) {};
		\node[layer,minimum height=\sma] 	(down_3) at (0.45*\t,0) {};
		\node[layer,minimum height=\lar] 	(up_1) at (\t+0.5*\t-0.15*\t,0) {};
		\node[layer,minimum height=\med] 	(up_2) at (\t+0.5*\t-0.3*\t,0) {};
		\node[layer,minimum height=\sma] 	(up_3) at (\t+0.5*\t-0.45*\t,0) {};
		
		\node[rotate=90,anchor=north]		at (0.075*\t,0) {\scriptsize lay. 1};
		\node[rotate=90,anchor=north]		at (0.225*\t,0) {\scriptsize lay. 2};
		\node[rotate=90,anchor=north]		at (0.4*\t,0) {\scriptsize ...};
		\node[rotate=-30,anchor=north]		at (0.43*\t,0.34*\t) {\scriptsize encoder};
		\node[rotate=30,anchor=north]		at (0.5*\t+0.55*\t,0.33*\t) {\scriptsize decoder};
		
		\node[] (bottleneck) at (0.75*\t,0) {};
		\node[] (bottleneck_txt) at (0.75*\t,-0.35*\t) {\scriptsize bottleneck};
		\draw[stealth-] (bottleneck.south) to [out=-90,in=90] (bottleneck_txt.north);
		
	\end{tikzpicture}
	\caption{Autoencoder architecture}
	\label{fig:AE}
\end{subfigure} \quad
\begin{subfigure}[b]{.4\textwidth}
	\centering
	
	\begin{tikzpicture}
		\node[down] 	(tri1) at (0.25*\t,0) {};
		\node[up] 		(tri2) at (\t+0.25*\t,0) {};

		\node[inout] (input) at (-\io*\t,0) {};
		\node[inout] (output) at (1.5*\t+\io*\t,0) {};
		\draw[-stealth] (input.east) -- (tri1.south);
		\draw[-stealth] (tri2.south) -- (output.west);

		\node[layer,minimum height=\lar] 	(down_1) at (0.15*\t,0) {};
		\node[layer,minimum height=\med] 	(down_2) at (0.3*\t,0) {};
		\node[layer,minimum height=\sma] 	(down_3) at (0.45*\t,0) {};
		\node[layer,minimum height=\lar] 	(up_1) at (\t+0.5*\t-0.15*\t,0) {};
		\node[layer,minimum height=\med] 	(up_2) at (\t+0.5*\t-0.3*\t,0) {};
		\node[layer,minimum height=\sma] 	(up_3) at (\t+0.5*\t-0.45*\t,0) {};
		
		\def\myshift#1{\raisebox{1ex}}
		\draw[-stealth] (down_1.north) to [out=90,in=90] (up_1.north);
		\draw[-stealth, postaction={	decorate,
								decoration={	text along path,
											text align=center,
											text={|\scriptsize\myshift|skip connections}}}] (down_2.north) to [out=90,in=90] (up_2.north);
		\draw[-stealth] (down_3.north) to [out=90,in=90] (up_3.north);
	\end{tikzpicture}
	\caption{U-net architecture}
	\label{fig:Unet}
\end{subfigure}

\caption{\textbf{Sketch of autoencoder architectures.} Standard autoencoders (left) are composed of an encoder and a decoder paths, and can be exploited either for end-to-end regression tasks (in a supervised way, with labels), or for the inference of latent space representations (in an unsupervised way, without labels). U-net autoencoders are a specific class of AE, in which skip connections are added from the encoder branch to the decoder one in order to mix high-level features from the latent space with low-level one from the contractive path. They usually present a superior level of performance on regression tasks.}
\label{fig:autoencoder}
\end{figure}

Autoencoder are feedforward neural networks that are widely applied to dimension reduction and feature extraction of high-dimensional data \cite{Hinton2006}. As shown in the sketch of figure \ref{fig:AE}, AEs are composed of a contractive path, named \emph{encoder}, \red{whose} role is to compress input data to a space of reduced dimension called \emph{latent space}. The latent space representation of the input variables is obtained at the \emph{bottleneck} of the structure, which is followed by a \emph{decoder} branch, mirror of the encoder one, responsible for the reconstruction of the input. Autoencoders can be trained both in unsupervised or supervised way, depending on the application scenario. In the case of unsupervised learning, AEs are usually exploited to infer the latent space structure of a given dataset. In the CFD community, \red{this functionality makes AE potential candidate tools for model reduction}. In \cite{Lee2020}, \red{AEs are used} in conjunction with convolutional layers to learn low-dimensional features of fluid systems. In \cite{Bukka2020} and \cite{Gonzalez2018}, the authors combine recurrent neural networks with CAE to learn the dynamics of the extracted low dimensional features. Adversely, in the case of supervised learning, \red{AEs are exploited to perform various full-field flow prediction tasks \cite{Guo2016, Jin2018, Lee2019}}. Among the multiple variations of AE structures, the special case of U-net \cite{UnetRonneberger} must be mentioned. As sketched in figure \ref{fig:Unet}, U-nets structures contain \emph{skip connections} from the encoder to the decoder, \red{the role of which} is to concatenate low-level features from the contractive path to the expansion path (here, concatenation means stacking tensors together along the channel axis). By allowing the mixing of low-level features with the high-level latent-space representation, U-nets usually achieve excellent performance levels on segmentation \cite{UnetRonneberger} and regression tasks: in \cite{Thuerey2020}, the authors exploit U-nets to infer the velocity and pressure fields of turbulent flow around airfoils computed in a Reynolds-averaged Navier-Stokes framework; in \cite{Fukami2018}, a U-net-like architecture is used to reconstruct turbulent flows from extremely coarse flow field images with remarkable accuracy; \red{in \cite{KAMRAVA2021119050}, a recurrent U-net architecture is trained to predict the instationary velocity and pressure fields in porous membranes.}

In the present paper, \red{we introduce an autoencoder architecture with a twin-decoder as a possible tool for outlier detection in the context of fluid flow predictions}. New contributions of this work include:

\begin{enumerate}
	\item A novel twin-decoder architecture displaying a strong correlation between the input reconstruction and the flow prediction error levels by taking advantage of proper skip connections between the two decoder branches. We find that this correlation is almost linear, at the expense of a slightly lower flow prediction accuracy than that of a U-net with similar structure;
	\item Two uncertainty estimation procedures taking advantage of the latter property: (i) a qualitative procedure based on a user-provided error threshold level, providing binary decisions regarding the prediction (accept or reject), and (ii) a quantitative procedure, providing the user with a error level interval on top of the flow prediction. Results over the considered dataset as well as unseen shapes proved these methods to be efficient to detect the applicability limits of the trained model. Both methods rely on simple concepts, and can be easily applied to other end-to-end prediction tasks.
\end{enumerate}

The paper is organized as follows: the problem settings and dataset construction are presented in section \ref{section:dataset}. Insights about the proposed twin-decoder architecture and its training procedure are given in section \ref{section:autoencoder}. The concepts of both the qualitativee and quantitativ trust-level methods are then described. In section \ref{section:results}, the performance of the method is first explored through a hyper-parameter calibration, then the best architecture is selected based on a cost-to-accuracy ratio. Finally, the correlation levels of the trained model are presented, and the two trust-level methods are put into practice. Finally, future perspectives are given. The base code used in this paper is available at \url{https://github.com/jviquerat/twin_autoencoder} (see section \ref{section:open_source} for additional details).

\section{Dataset construction}
\label{section:dataset}

This section provides insights on the random shape dataset generation used to train networks in the next sections. This dataset was initially used in \cite{Viquerat2020} (section 3.5), thus only the main lines are sketched here. For more details, the reader is referred to \cite{Viquerat2020}. First, we describe the steps to generate arbitrary shapes by means of connected Bezier curves. Then, the solving of the Navier-Stokes equations with an immersed method is presented. Finally, details about the dataset are given.

\subsection{Random shape generation}

\begin{figure}
\centering
\def\xaxis{3}
\def\yaxis{3}
\def\sc{1.0}
\def\pOneX{0.7}
\def\pOneY{0.25}
\def\pTwoX{-0.2}
\def\pTwoY{0.5}
\def\pThreeX{-0.7}
\def\pThreeY{-0.3}
\def\pFourX{0.4}
\def\pFourY{-0.6}
\def\rad{0.5}

\begin{subfigure}{.45\textwidth}
	\centering
	\begin{tikzpicture}[	scale=\sc,
					axes/.style={thick,->,-stealth},
					pt/.style={circle,inner sep=0pt,text width=2mm,fill=myblue1,draw=black}]

	\draw[axes] (-\xaxis,0) -- (\xaxis,0) node[anchor=north] {$x$};
	\draw[axes] (0,-\yaxis) -- (0,\yaxis) node[anchor=west] {$y$};

	\node[pt, label=above:$p_1$] 	(p1) at ( \pOneX*\xaxis, 	\pOneY*\yaxis) {};
	\node[pt, label=above:$p_2$] 	(p2) at (\pTwoX*\xaxis, 	\pTwoY*\yaxis) {};
	\node[pt, label=below:$p_3$] 	(p3) at (\pThreeX*\xaxis,	\pThreeY*\yaxis) {};
	\node[pt, label=below:$p_4$] 	(p4) at (\pFourX*\xaxis,	\pFourY*\yaxis) {};

	\draw[->,-stealth,dash pattern=on 1pt] (0,0) -- (0.4*\xaxis,0.7*\yaxis);
	\draw[->,-stealth,thick] (0,0) ++(0:1.0) arc (0:60:1.0) node[pos=0.45,xshift=0.15cm,yshift=0.15cm] {$\theta$};

	\end{tikzpicture}
	
	\caption{Draw $n_s$ random points, translate them around $(0,0)$ and sort them by ascending trigonometric angle}
	\label{fig:shape_generation_1}
\end{subfigure} \qquad
\begin{subfigure}{.45\textwidth}
	\centering
	\begin{tikzpicture}[	scale=\sc,
					axes/.style={thick,->,-stealth},
					pt/.style={circle,inner sep=0pt,text width=2mm,fill=myblue1,draw=black}]

	\draw[axes] (-\xaxis,0) -- (\xaxis,0) node[anchor=north] {$x$};
	\draw[axes] (0,-\yaxis) -- (0,\yaxis) node[anchor=west] {$y$};

	\node[pt, label=above:$p_1$] 	(p1) at ( \pOneX*\xaxis, 	\pOneY*\yaxis) {};
	\node[pt, label=above:$p_2$] 	(p2) at (\pTwoX*\xaxis, 	\pTwoY*\yaxis) {};
	\node[pt, label=below:$p_3$] 	(p3) at (\pThreeX*\xaxis,	\pThreeY*\yaxis) {};
	\node[pt, label=below:$p_4$] 	(p4) at (\pFourX*\xaxis,	\pFourY*\yaxis) {};
	
	\draw[->,-stealth,dash pattern=on 1pt] (p1) -- (p2);
	\draw[->,-stealth,dash pattern=on 1pt] (p2) -- (p3);
	
	\draw[dash pattern=on 0.5pt] (\pOneX*\xaxis,\pOneY*\yaxis) -- (\pTwoX*\xaxis,\pOneY*\yaxis);
	\draw[dash pattern=on 0.5pt] (\pTwoX*\xaxis,\pTwoY*\yaxis) -- (\pTwoX*\xaxis,\pOneY*\yaxis);
	\draw[->,-stealth,thick] (\pOneX*\xaxis,\pOneY*\yaxis) ++(164.5:1.25) arc (164.5:180:1.25) node[pos=0.5,xshift=-0.4cm,yshift=0.1cm] {$\theta_{12}$};
	
	\draw[dash pattern=on 0.5pt] (\pTwoX*\xaxis,\pTwoY*\yaxis) -- (\pTwoX*\xaxis,\pThreeY*\yaxis);
	\draw[dash pattern=on 0.5pt] (\pThreeX*\xaxis,\pThreeY*\yaxis) -- (\pTwoX*\xaxis,\pThreeY*\yaxis);
	\draw[->,-stealth,thick] (\pThreeX*\xaxis,\pThreeY*\yaxis) ++(0.5:0.8) arc (0.5:58:0.8) node[pos=0.5,xshift=0.3cm,yshift=0.1cm] {$\theta_{23}$};
	
	\node[pt, label=above:$p_1$] 	(p1) at ( \pOneX*\xaxis, 	\pOneY*\yaxis) {};
	\node[pt, label=above:$p_2$] 	(p2) at (\pTwoX*\xaxis, 	\pTwoY*\yaxis) {};
	\node[pt, label=below:$p_3$] 	(p3) at (\pThreeX*\xaxis,	\pThreeY*\yaxis) {};
	\node[pt, label=below:$p_4$] 	(p4) at (\pFourX*\xaxis,	\pFourY*\yaxis) {};

	\end{tikzpicture}
	
	\caption{Compute angles between random points, and compute an average angle around each point $\theta^*_i$}
	\label{fig:shape_generation_2}
\end{subfigure}%

\medskip
\medskip
\begin{subfigure}{.45\textwidth}
	\centering
	\begin{tikzpicture}[	scale=\sc,
					axes/.style={thick,->,-stealth},
					pt/.style={circle,inner sep=0pt,text width=2mm,fill=myblue1,draw=black}]

	\draw[axes] (-\xaxis,0) -- (\xaxis,0) node[anchor=north] {$x$};
	\draw[axes] (0,-\yaxis) -- (0,\yaxis) node[anchor=west] {$y$};

	\node[pt, label=above:$p_1$] 	(p1) at ( \pOneX*\xaxis, 	\pOneY*\yaxis) {};
	\node[pt, label=above:$p_2$] 	(p2) at (\pTwoX*\xaxis, 	\pTwoY*\yaxis) {};
	\node[pt, label=below:$p_3$] 	(p3) at (\pThreeX*\xaxis,	\pThreeY*\yaxis) {};
	\node[pt, label=below:$p_4$] 	(p4) at (\pFourX*\xaxis,	\pFourY*\yaxis) {};
	
	\node[pt, label=left:$p^*_2$] 	(p2s) at (\pTwoX*\xaxis-0.989*\rad*\xaxis, \pTwoY*\yaxis-0.147*\rad*\yaxis) {};
	\node[pt, label=left:$p^{**}_2$] 	(p2ss) at (\pThreeX*\xaxis-0.147*\rad*\xaxis, \pThreeY*\yaxis+0.989*\rad*\yaxis) {};
	
	\draw[->,-stealth,dash pattern=on 1pt] (p1) -- (p2);
	\draw[->,-stealth,dash pattern=on 1pt] (p2) -- (p3);
	
	\draw[dash pattern=on 0.5pt] (p2) -- (p2s);
	\draw[dash pattern=on 0.5pt] (p3) -- (p2ss);
	\node[draw=none] (p2t) at (\pTwoX*\xaxis-0.963*\rad*\xaxis, \pTwoY*\yaxis+0.267*\rad*\yaxis) {};
	\draw[dash pattern=on 1pt] (p2) -- (p2t);
	\draw[->,-stealth,thick] (\pTwoX*\xaxis,\pTwoY*\yaxis) ++(164.5:0.8) arc (164.5:188:0.8) node[pos=0.5,xshift=-0.35cm,yshift=0.03cm] {$\theta^*_2$};
	
	\draw[thick,myorange1] (p2) .. controls (p2s) and (p2ss) .. (p3);
	
	\node[pt, label=above:$p_1$] 	(p1) at ( \pOneX*\xaxis, 	\pOneY*\yaxis) {};
	\node[pt, label=above:$p_2$] 	(p2) at (\pTwoX*\xaxis, 	\pTwoY*\yaxis) {};
	\node[pt, label=below:$p_3$] 	(p3) at (\pThreeX*\xaxis,	\pThreeY*\yaxis) {};
	\node[pt, label=below:$p_4$] 	(p4) at (\pFourX*\xaxis,	\pFourY*\yaxis) {};

	\end{tikzpicture}
	
	\caption{Compute control points coordinates from averaged angles and generate cubic B\'ezier curve}
	\label{fig:shape_generation_3}
\end{subfigure} \qquad
\begin{subfigure}{.45\textwidth}
	\centering
	\begin{tikzpicture}[	scale=\sc,
					axes/.style={thick,->,-stealth},
					pt/.style={circle,inner sep=0pt,text width=2mm,fill=myblue1,draw=black},
					sample/.style={circle,inner sep=0pt,text width=1mm,fill=mygray1,draw=black}]

	\draw[axes] (-\xaxis,0) -- (\xaxis,0) node[anchor=north] {$x$};
	\draw[axes] (0,-\yaxis) -- (0,\yaxis) node[anchor=west] {$y$};

	\node[pt, label=above:$p_1$] 	(p1) at ( \pOneX*\xaxis, 	\pOneY*\yaxis) {};
	\node[pt, label=above:$p_2$] 	(p2) at (\pTwoX*\xaxis, 	\pTwoY*\yaxis) {};
	\node[pt, label=below:$p_3$] 	(p3) at (\pThreeX*\xaxis,	\pThreeY*\yaxis) {};
	\node[pt, label=below:$p_4$] 	(p4) at (\pFourX*\xaxis,	\pFourY*\yaxis) {};
	
	\node[draw=none] 	(p2s) at (\pTwoX*\xaxis-0.989*\rad*\xaxis, \pTwoY*\yaxis-0.147*\rad*\yaxis) {};
	\node[draw=none] 	(p2ss) at (\pThreeX*\xaxis-0.147*\rad*\xaxis, \pThreeY*\yaxis+0.989*\rad*\yaxis) {};
	\node[draw=none] 	(p3s) at (\pThreeX*\xaxis+0.147*\rad*\xaxis, \pThreeY*\yaxis-0.989*\rad*\yaxis) {};
	\node[draw=none] 	(p3ss) at (\pFourX*\xaxis-0.6*\rad*\xaxis, \pFourY*\yaxis-0.5*\rad*\yaxis) {};
	\node[draw=none] 	(p4s) at (\pFourX*\xaxis+0.6*\rad*\xaxis, \pFourY*\yaxis+0.5*\rad*\yaxis) {};
	\node[draw=none] 	(p4ss) at (\pOneX*\xaxis+0.4*\rad*\xaxis, \pOneY*\yaxis-0.5*\rad*\yaxis) {};
	\node[draw=none] 	(p1s) at (\pOneX*\xaxis-0.4*\rad*\xaxis, \pOneY*\yaxis+0.5*\rad*\yaxis) {};
	\node[draw=none] 	(p1ss) at (\pTwoX*\xaxis+0.989*\rad*\xaxis, \pTwoY*\yaxis+0.147*\rad*\yaxis) {};
	
	\draw[thick,myorange1] (p2) .. controls (p2s) and (p2ss) .. (p3);
	\draw[thick,myorange1] (p3) .. controls (p3s) and (p3ss) .. (p4);
	\draw[thick,myorange1] (p4) .. controls (p4s) and (p4ss) .. (p1);
	\draw[thick,myorange1] (p1) .. controls (p1s) and (p1ss) .. (p2);
	
	\draw[draw=none] (p4) .. controls (p4s) and (p4ss) .. (p1) node[sample,pos=0.1,opacity=1] {};
	\draw[draw=none] (p4) .. controls (p4s) and (p4ss) .. (p1) node[sample,pos=0.2,opacity=1] {};
	\draw[draw=none] (p4) .. controls (p4s) and (p4ss) .. (p1) node[sample,pos=0.3,opacity=1] {};
	\draw[draw=none] (p4) .. controls (p4s) and (p4ss) .. (p1) node[sample,pos=0.4,opacity=1] {};
	\draw[draw=none] (p4) .. controls (p4s) and (p4ss) .. (p1) node[sample,pos=0.5,opacity=1] {};
	\draw[draw=none] (p4) .. controls (p4s) and (p4ss) .. (p1) node[sample,pos=0.6,opacity=1] {};
	\draw[draw=none] (p4) .. controls (p4s) and (p4ss) .. (p1) node[sample,pos=0.7,opacity=1] {};
	\draw[draw=none] (p4) .. controls (p4s) and (p4ss) .. (p1) node[sample,pos=0.8,opacity=1] {};
	\draw[draw=none] (p4) .. controls (p4s) and (p4ss) .. (p1) node[sample,pos=0.9,opacity=1] {};
	
	\node[pt, label=above:$p_1$] 	(p1) at ( \pOneX*\xaxis, 	\pOneY*\yaxis) {};
	\node[pt, label=above:$p_2$] 	(p2) at (\pTwoX*\xaxis, 	\pTwoY*\yaxis) {};
	\node[pt, label=below:$p_3$] 	(p3) at (\pThreeX*\xaxis,	\pThreeY*\yaxis) {};
	\node[pt, label=below:$p_4$] 	(p4) at (\pFourX*\xaxis,	\pFourY*\yaxis) {};

	\end{tikzpicture}
	
	\caption{Sample all B\'ezier lines and export for mesh immersion}
	\label{fig:shape_generation_4}
\end{subfigure}
\caption{\textbf{Random shape generation with cubic B\'ezier curves}. }
\label{fig:shape_generation}
\end{figure}
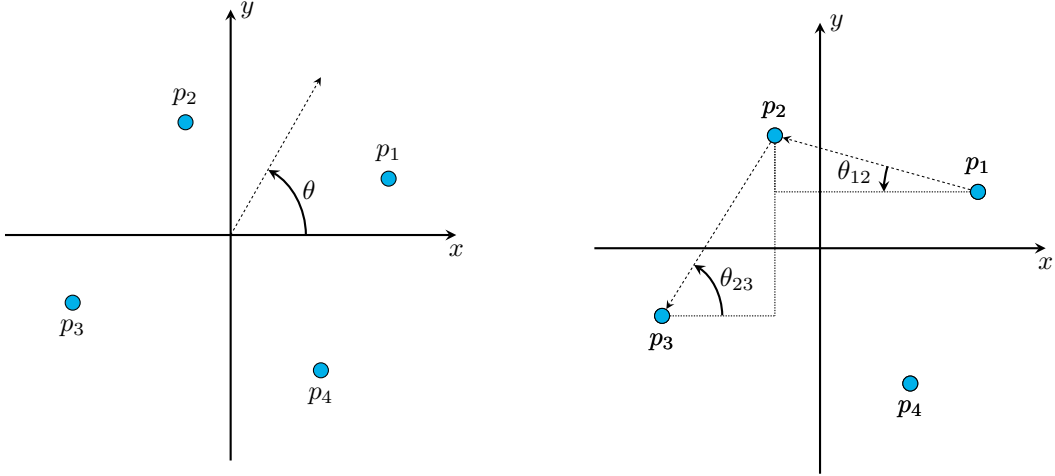
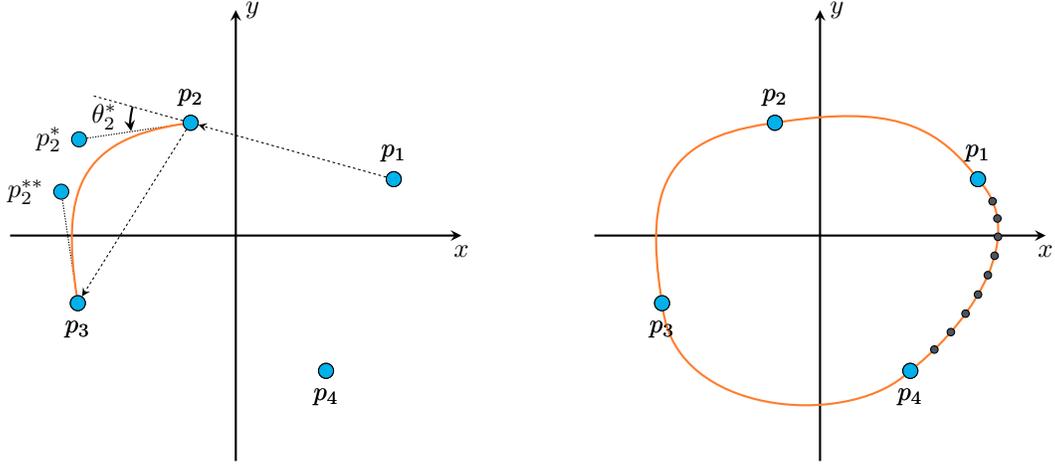

\begin{figure}
\centering
\def\w{1.5cm}
\def\arraystretch{4.5}
\setlength\tabcolsep{9pt}
\newcolumntype{C}{ >{\centering\arraybackslash} m{1.5cm} }

\makebox[\linewidth]{%
	\begin{tabular}{CCCCCCC}
		\includegraphics[width=\w]{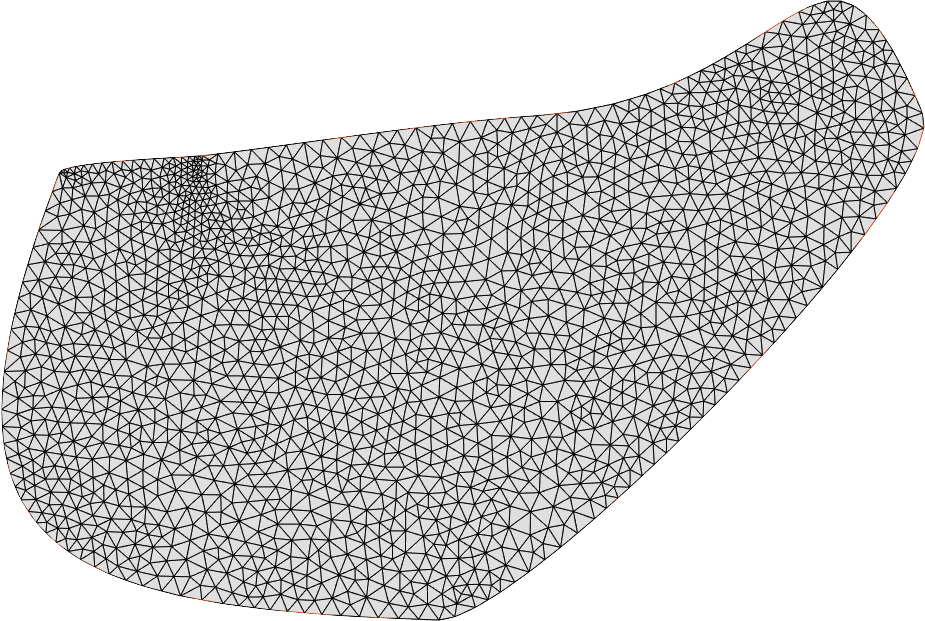} 	& 
		\includegraphics[width=\w]{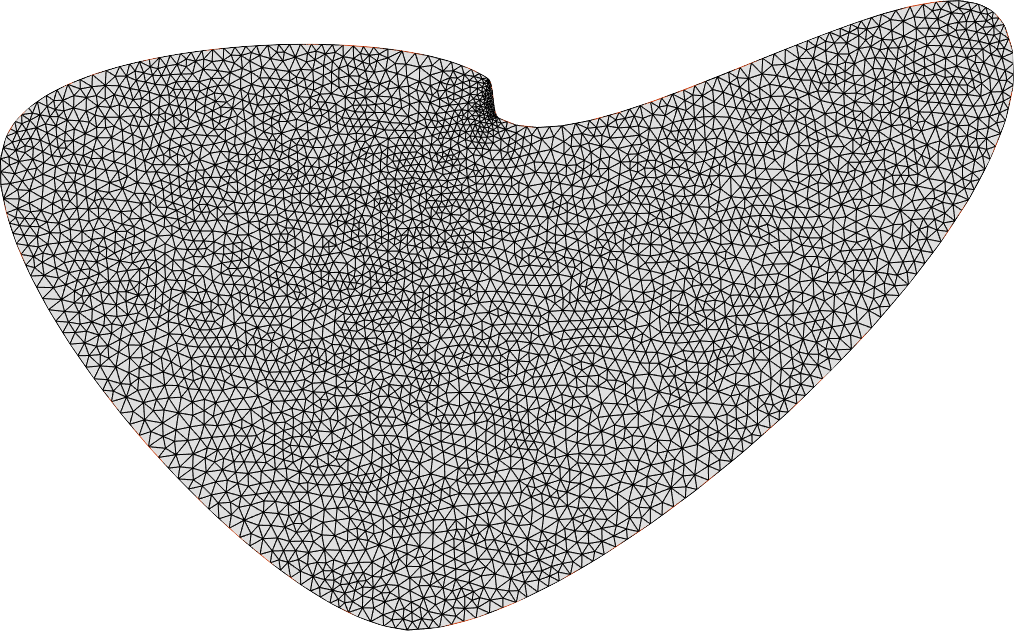}	& 
		\includegraphics[width=\w]{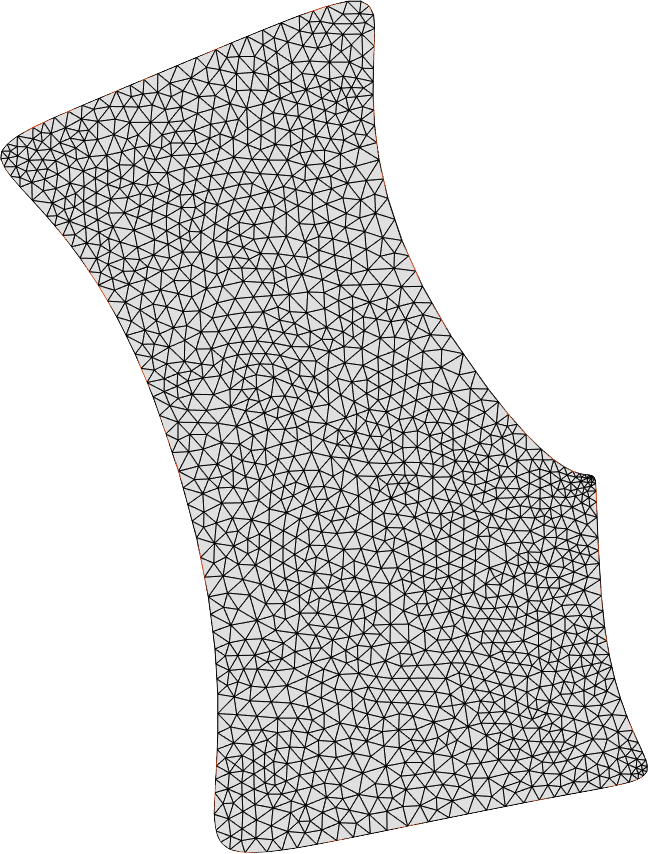}	&
		\includegraphics[width=\w]{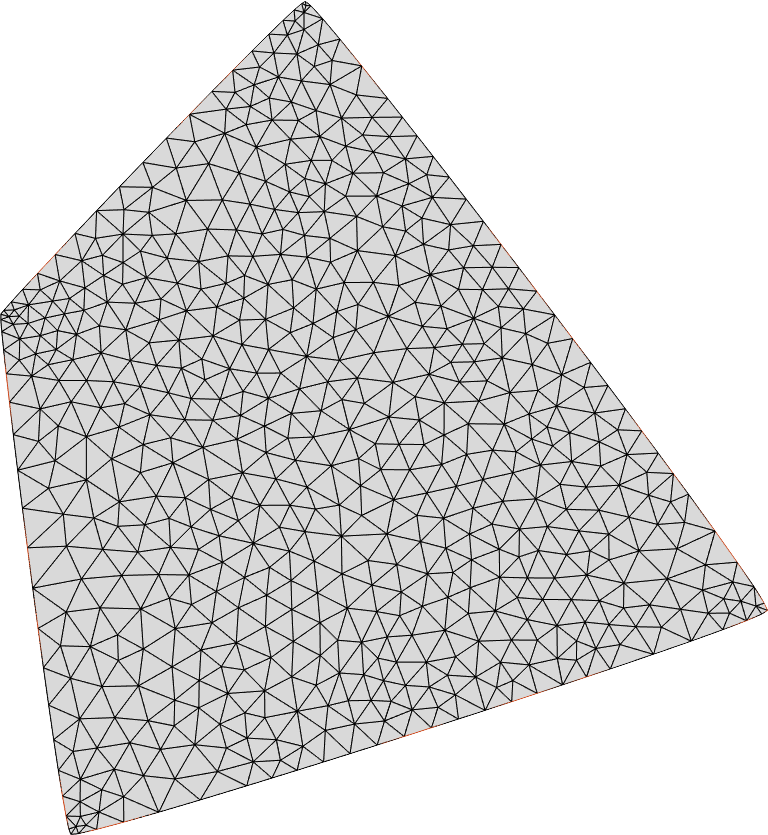}	&
		\includegraphics[width=\w]{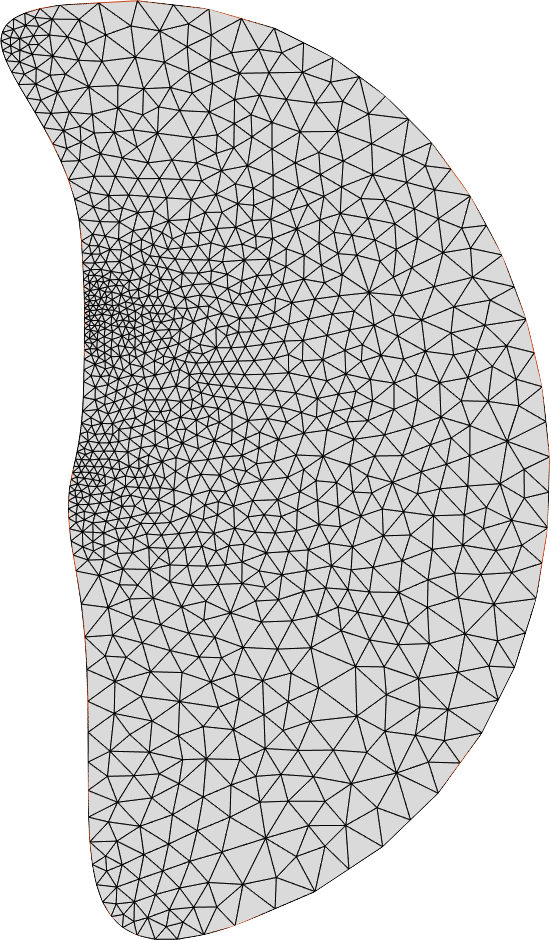}	&
		\includegraphics[width=\w]{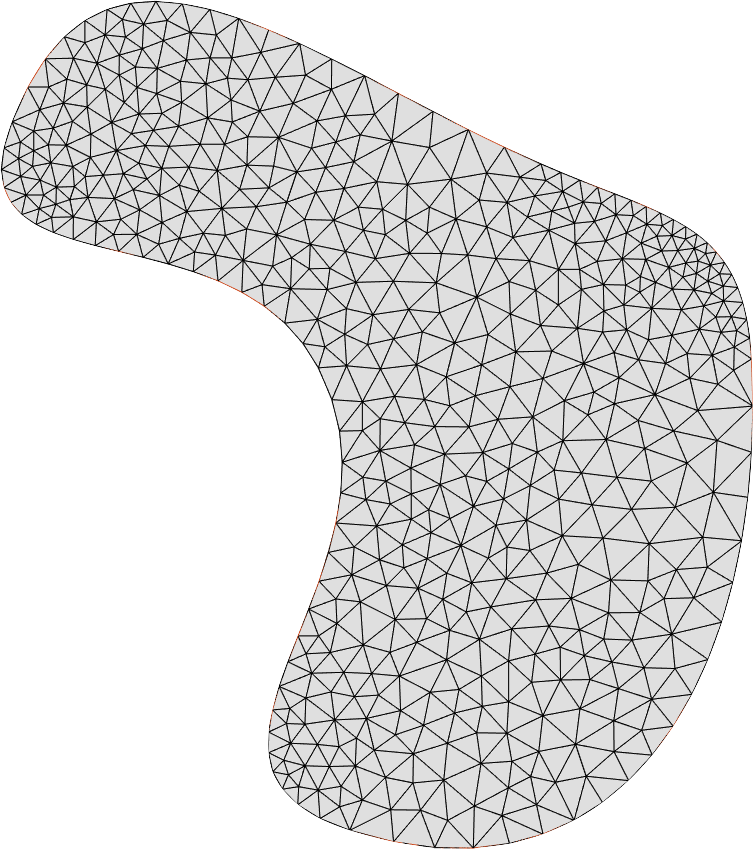}	&
		\includegraphics[width=\w]{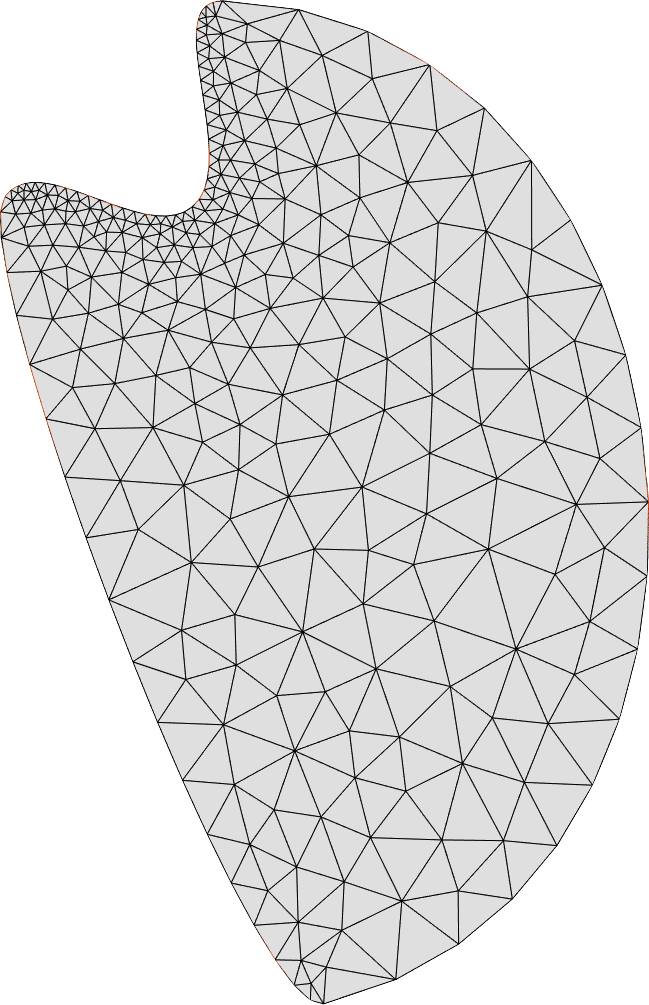}\\
		\includegraphics[width=\w]{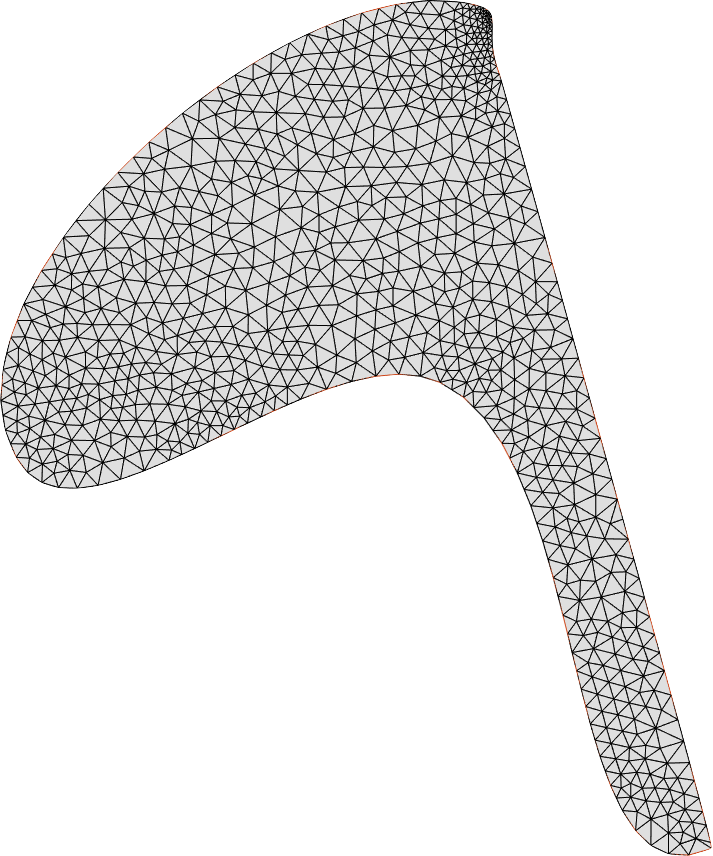} 	& 
		\includegraphics[width=\w]{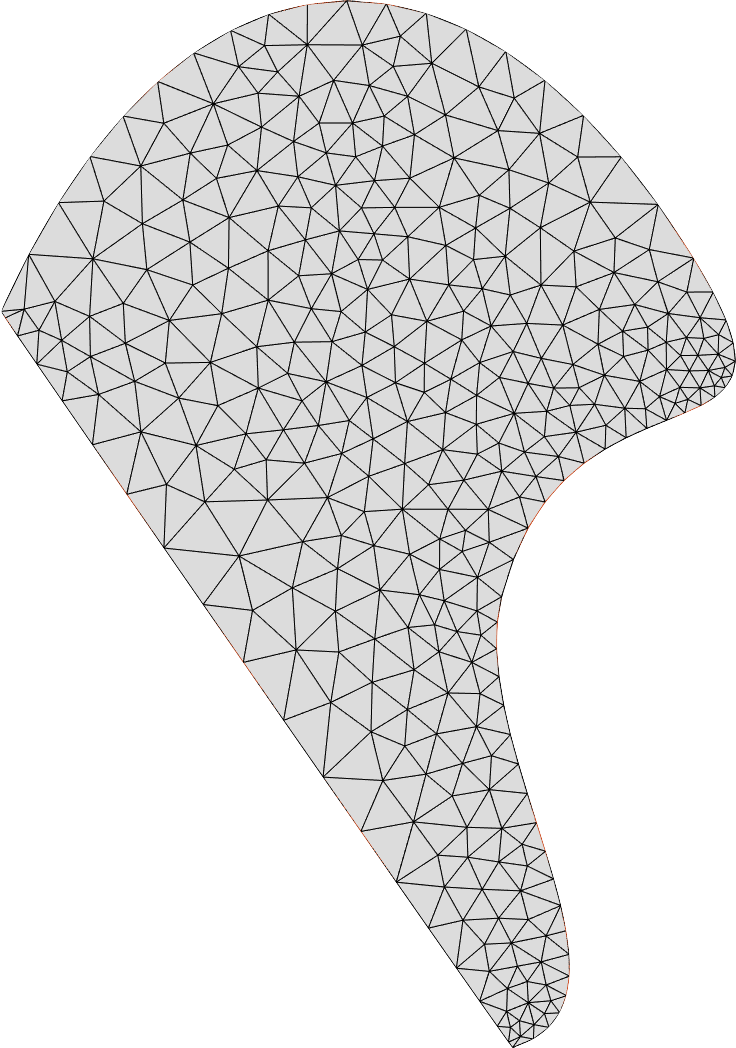}	&
		\includegraphics[width=\w]{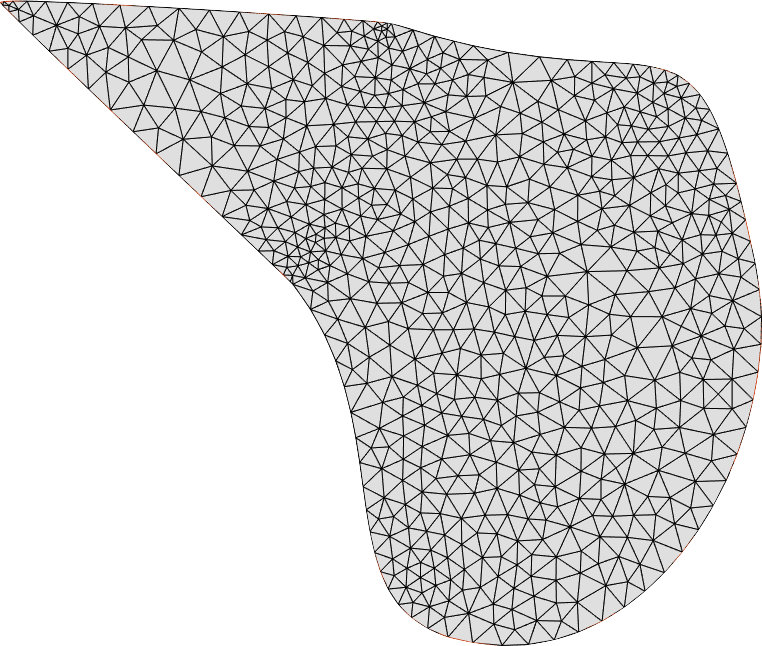}	&
		\includegraphics[width=\w]{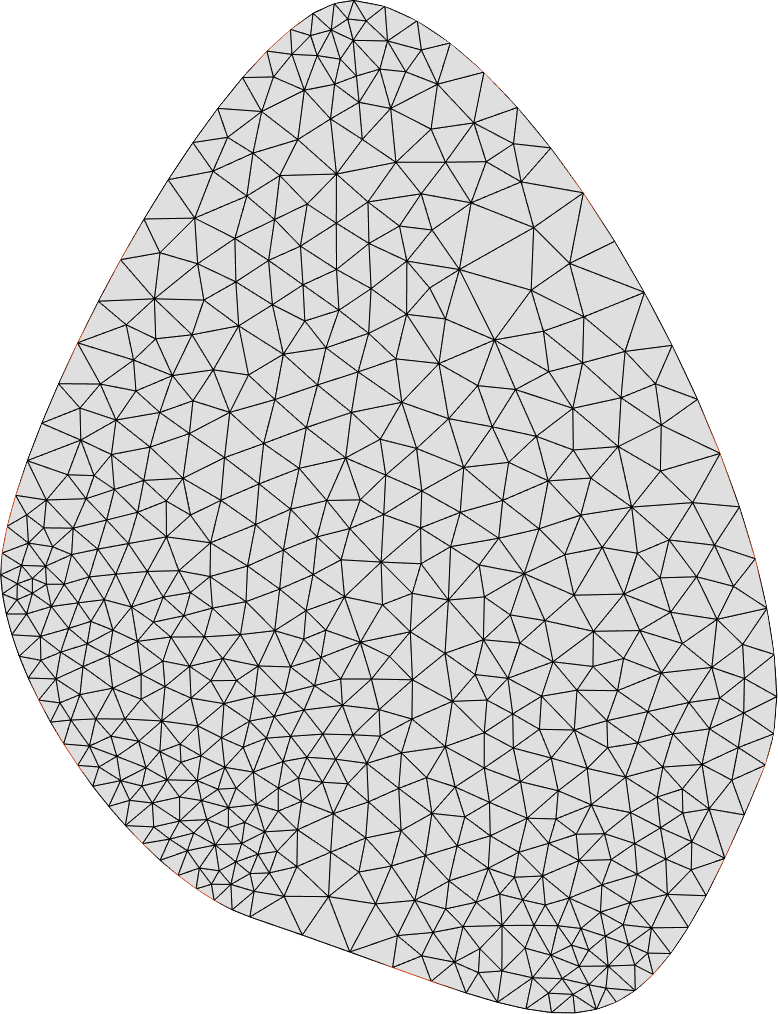}	&
		\includegraphics[width=\w]{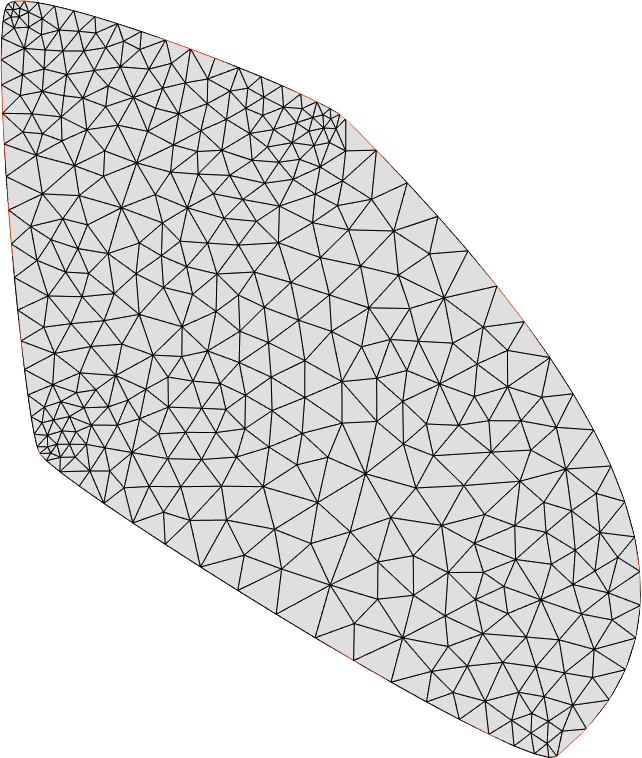}	&
		\includegraphics[width=\w]{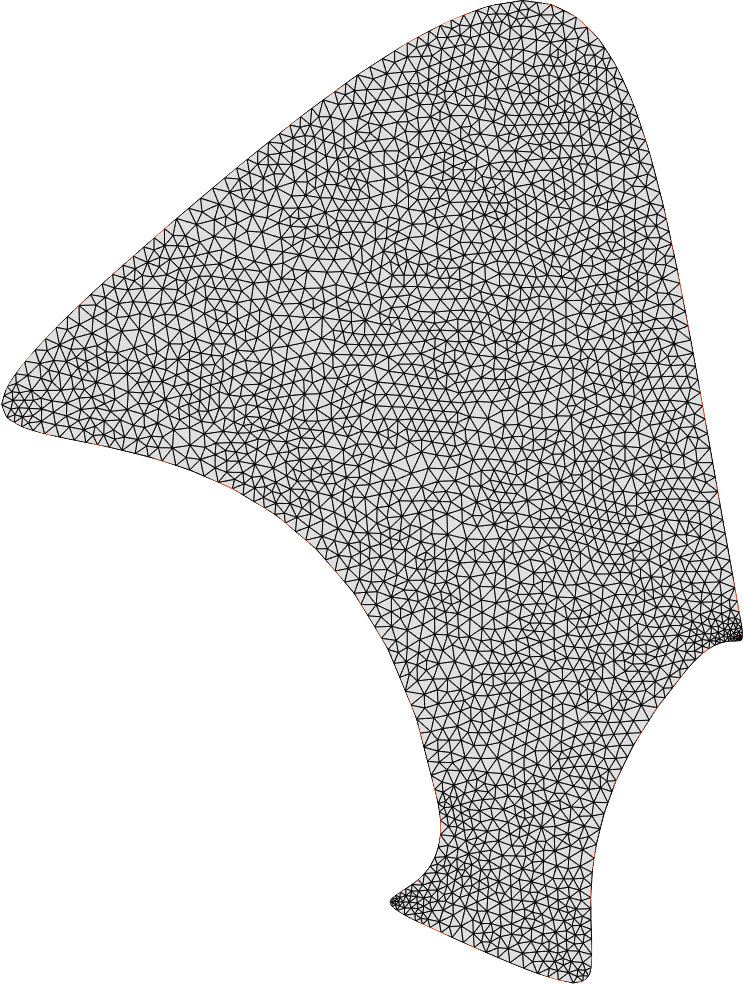}	&
		\includegraphics[width=\w]{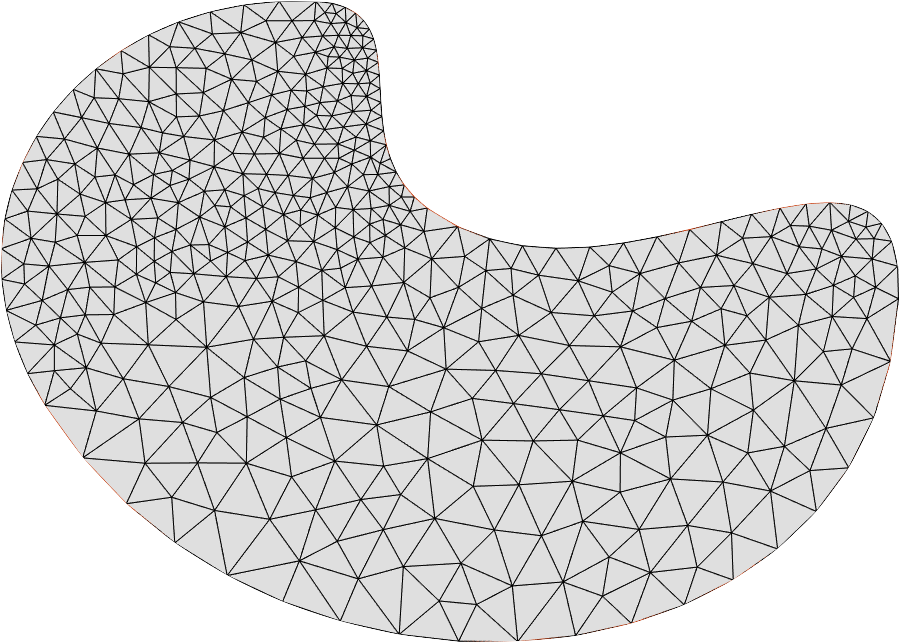}\\
		\includegraphics[width=\w]{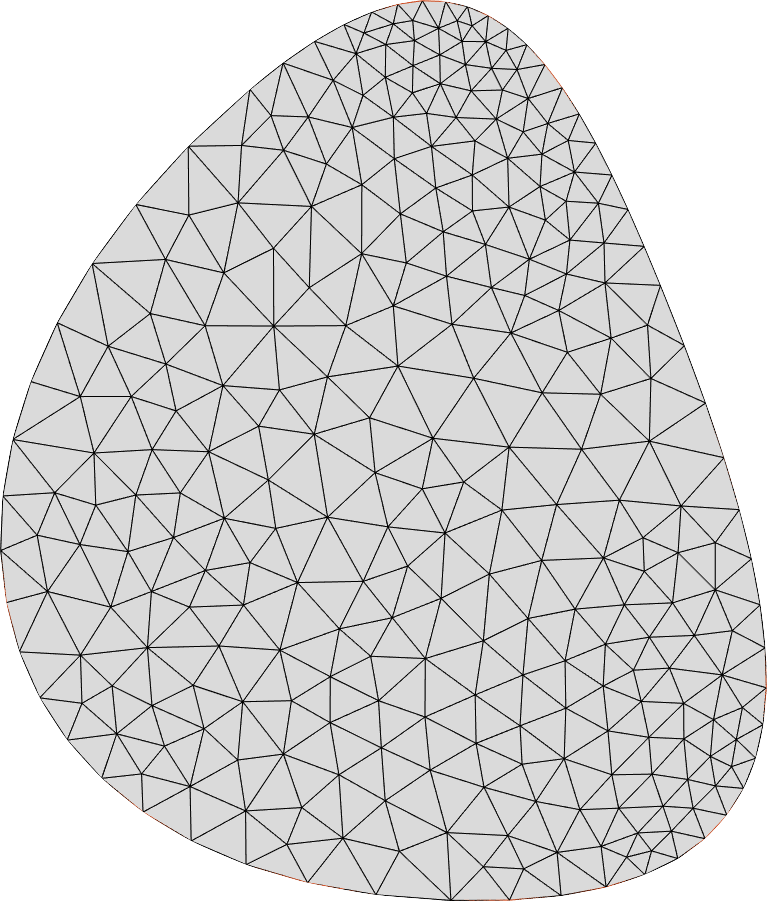} 	& 
		\includegraphics[width=\w]{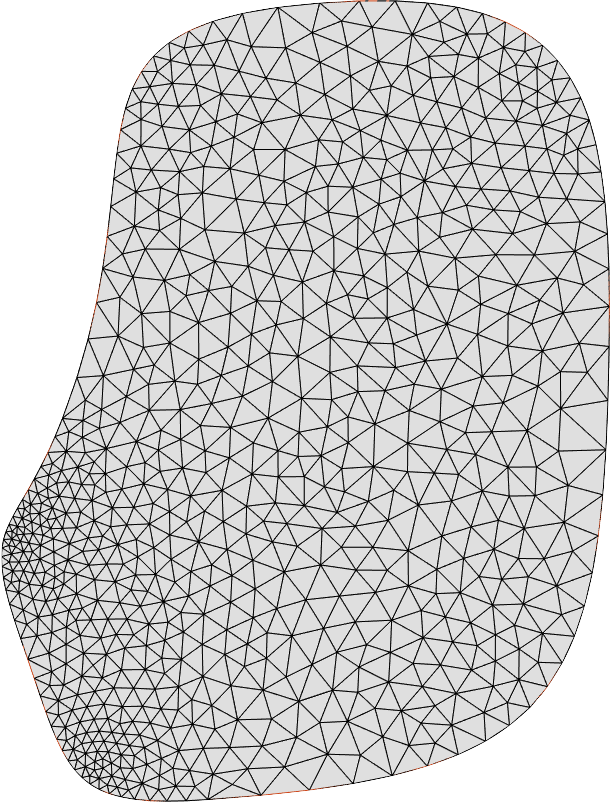}	&
		\includegraphics[width=\w]{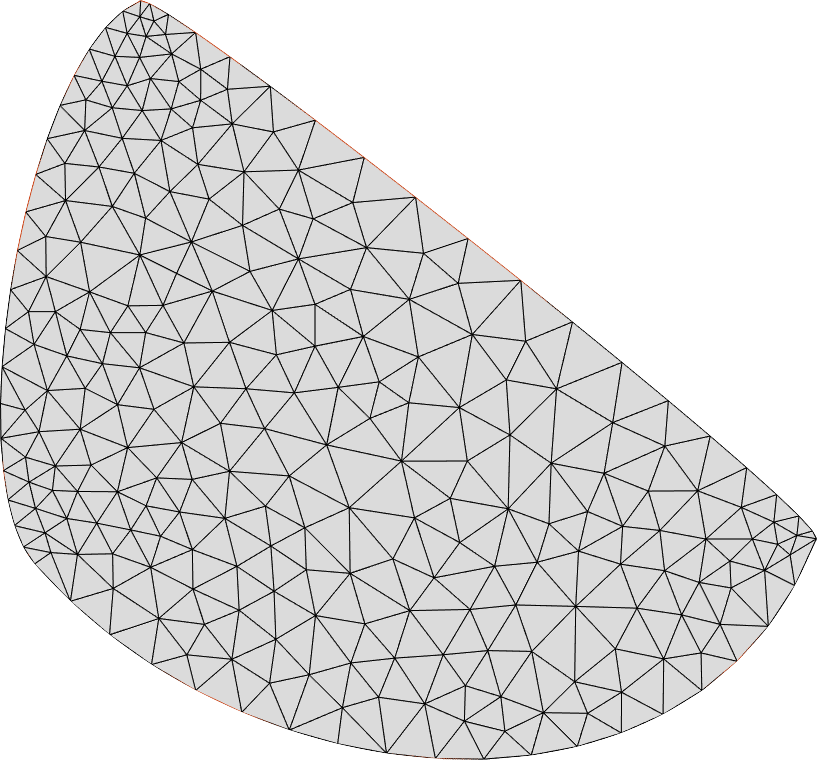}	&
		\includegraphics[width=\w]{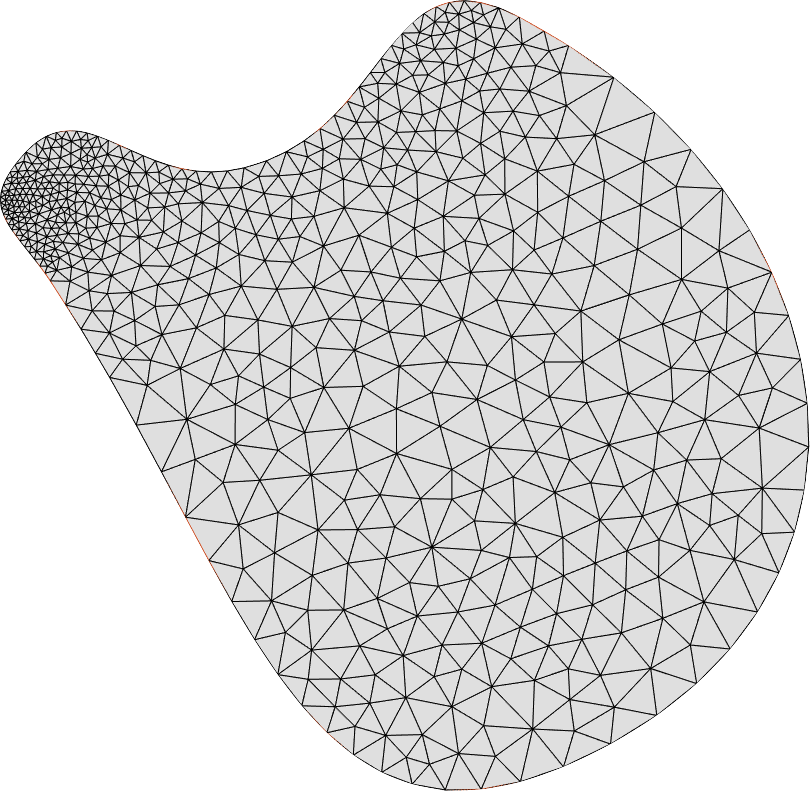}	&
		\includegraphics[width=\w]{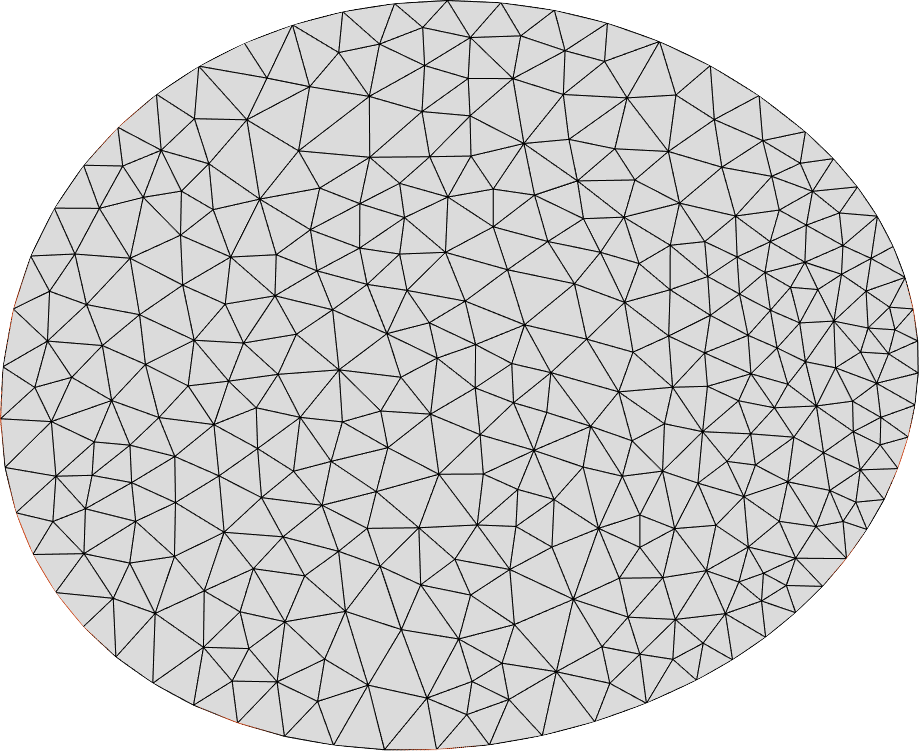}	&
		\includegraphics[width=\w]{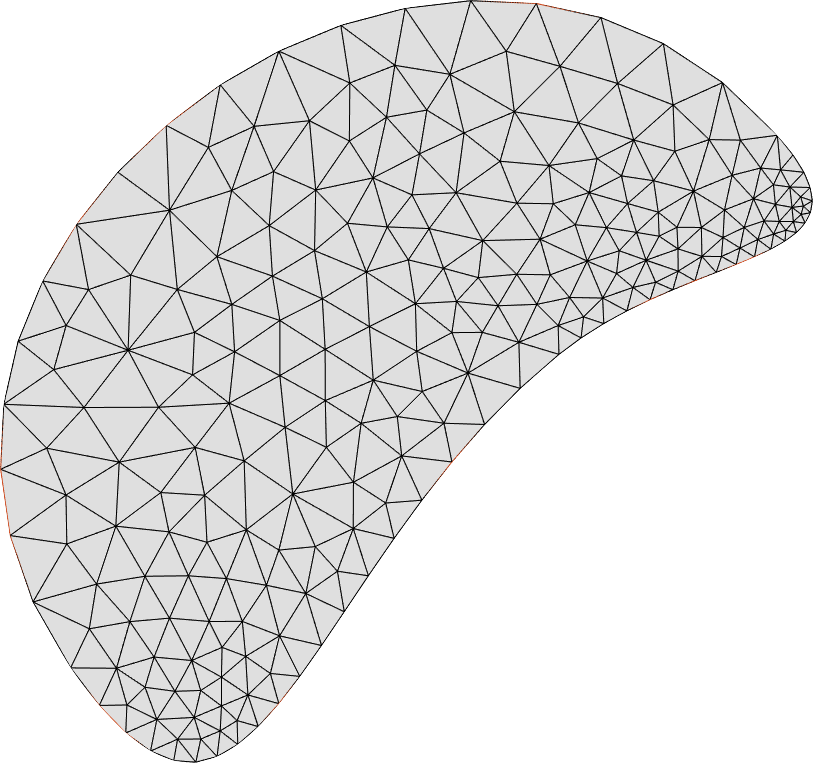}	&
		\includegraphics[width=\w]{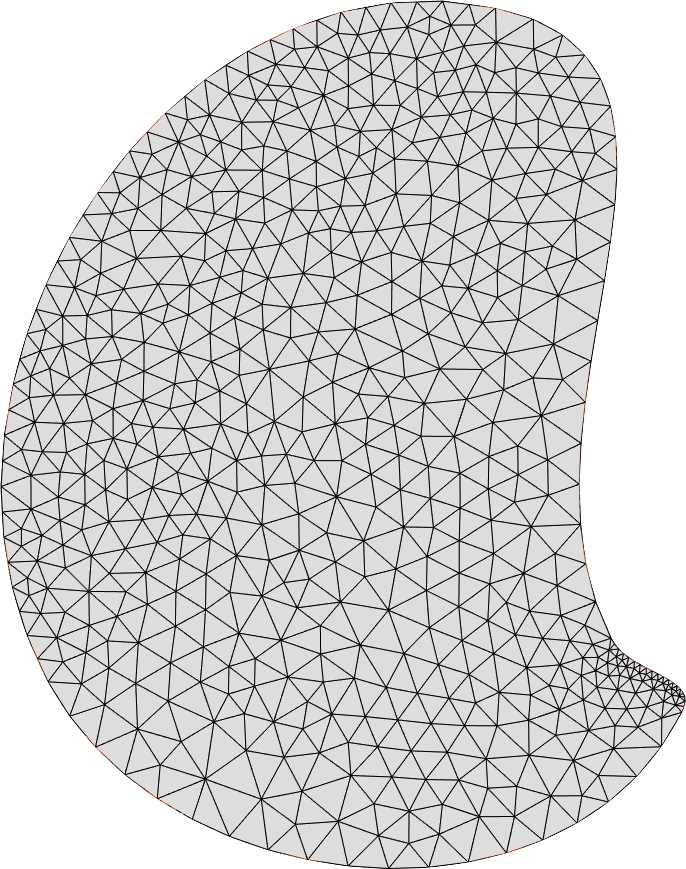}%
\end{tabular}}

\caption{\textbf{Shape examples drawn from the dataset.} A wide variety of shape is obtained using a restrained number of points ($n_s \in \left[ 4, 6 \right]$), as well as a local curvature $r$ and averaging parameter $\alpha$.}
\label{fig:shape_examples}
\end{figure}

In the first step, $n_s$ random points are drawn in $\left[ 0, 1 \right]^2$, and translated so their center of mass is positioned in $(0,0)$. An ascending trigonometric angle sort is then performed (see figure \ref{fig:shape_generation_1}), and the angles between consecutive random points are then computed. An average angle is then computed around each point (see figure \ref{fig:shape_generation_2}) using:

\begin{equation*}
	\theta^*_i = \alpha \theta_{i-1,i} + (1 - \alpha) \theta_{i,i+1},
\end{equation*}

\noindent with $\alpha \in \left[0,1\right]$. The averaging parameter $\alpha$ allows to alter the sharpness of the curve locally, maximum smoothness being obtained for $\alpha = 0.5$. Then, each pair of points is joined using a cubic B\'ezier curve, defined by four points: the first and last points, $p_i$ and $p_{i+1}$, are part of the curve, while the second and third ones, $p^*_i$ and $p^{**}_i$, are control points that define the tangent of the curve at $p_i$ and $p_{i+1}$. The tangents at $p_i$ and $p_{i+1}$ are respectively controlled by $\theta^*_i$ and $\theta^*_{i+1}$ (see figure \ref{fig:shape_generation_3}). A final sampling of the successive B\'ezier curves leads to a boundary description of the shape (figure \ref{fig:shape_generation_4}). Using this method, a wide variety of shapes can be attained, as shown in figure \ref{fig:shape_examples}.

\subsection{Numerical resolution of the Navier-Stokes equations}
\label{section:NS}

The flow motion of incompressible newtonian fluids is described by the Navier-Stokes (NS) equations: 

\begin{equation} \label{eq:ns_equation1}
	\left\{
	\begin{aligned}
		\rho\ (\partial_{t} \V{v} + \V{v} \cdot \nabla \V{v}) -\nabla \cdot \left( 2 \eta \GV{\epsilon}(\V{v}) - p \V{I} \right)  & = \V{f}, \\
		\nabla \cdot \V{v} &= 0,
	\end{aligned}
	\right.
\end{equation}

\noindent where $t \in [0,T]$ is the time, $\V{v}(x,t)$ the velocity, $p(x,t)$ the pressure, $\rho$ the fluid density, $\eta$ the dynamic viscosity and $\V{I}$ the identity tensor. In order to efficiently construct the dataset, an immersed volume method is used for resolution instead of the usual body-fitted method, avoiding a systematic re-meshing of the whole domain for each shape. This method rely on a unified fluid-solid Eulerian formulation based on level-set description of the geometry \cite{Bruchon2009}, and leads to the following set of modified equations: 

\begin{equation} \label{eq:ns_equation2}
	\left\{
	\begin{aligned}
		\rho^* (\partial_{t} \V{v} + \V{v} \cdot \nabla \V{v}) -\nabla \cdot \left( 2 \eta \GV{\epsilon}(\V{v}) + \GV{\tau} - p \V{I} \right)  & = \V{f}, \\
		\nabla \cdot \V{v} &= 0,
	\end{aligned}
	\right.
\end{equation}

\noindent where we have introduced the following mixed quantities:

\begin{equation*}
	\begin{aligned}
		\GV{\tau} & = H(\alpha) \GV{\tau}_{\text{s}},\\
		\rho^* & = H(\alpha) \rho_{\text{s}} + (1-H(\alpha)) \rho_{\text{f}},
	\end{aligned}
\end{equation*}

\noindent where the subscripts $f$ and $s$ respectively refer to the fluid and the solid, and $H(\alpha)$ is the Heaviside function:

\begin{equation} \label{eq:heavyside31}
	H(\alpha) = \left\{
	\begin{aligned}
		1 & \text{ if}\ \alpha > 0,\\
		0 & \text{ if}\ \alpha < 0.
	\end{aligned}
	\right.
\end{equation}

\noindent The reader is referred to \cite{Hachem2013} for additional details about formulation (\ref{eq:ns_equation2}). Eventually, the modified equations (\ref{eq:ns_equation2}) are cast into a stabilized finite element formulation, and solved using a variational multi-scale (VMS) solver \cite{Tezduyar1992, Bazilevs2007, Takizawa2018, Otoguro2019, Otoguro2020}.

\subsection{Dataset}

\begin{figure}
\centering

\setlength{\fboxsep}{0pt}%
\setlength{\fboxrule}{1pt}%

\begin{subfigure}[t]{.3\textwidth}
	\centering
	\fbox{\includegraphics[width=\linewidth]{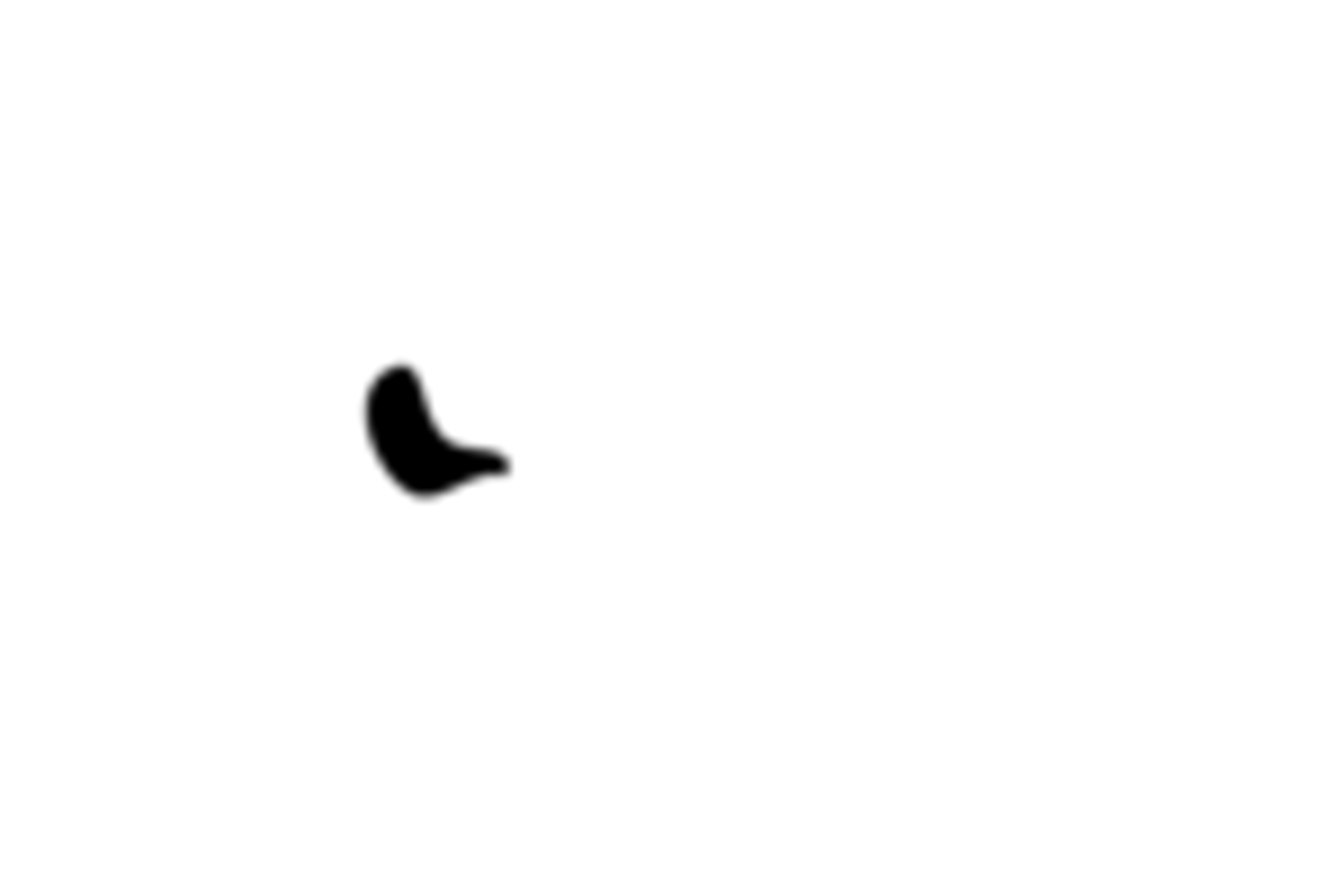}} 
	\caption{Network input}
\end{subfigure} \quad
\begin{subfigure}[t]{.3\textwidth}
	\centering
	\fbox{\includegraphics[width=\linewidth]{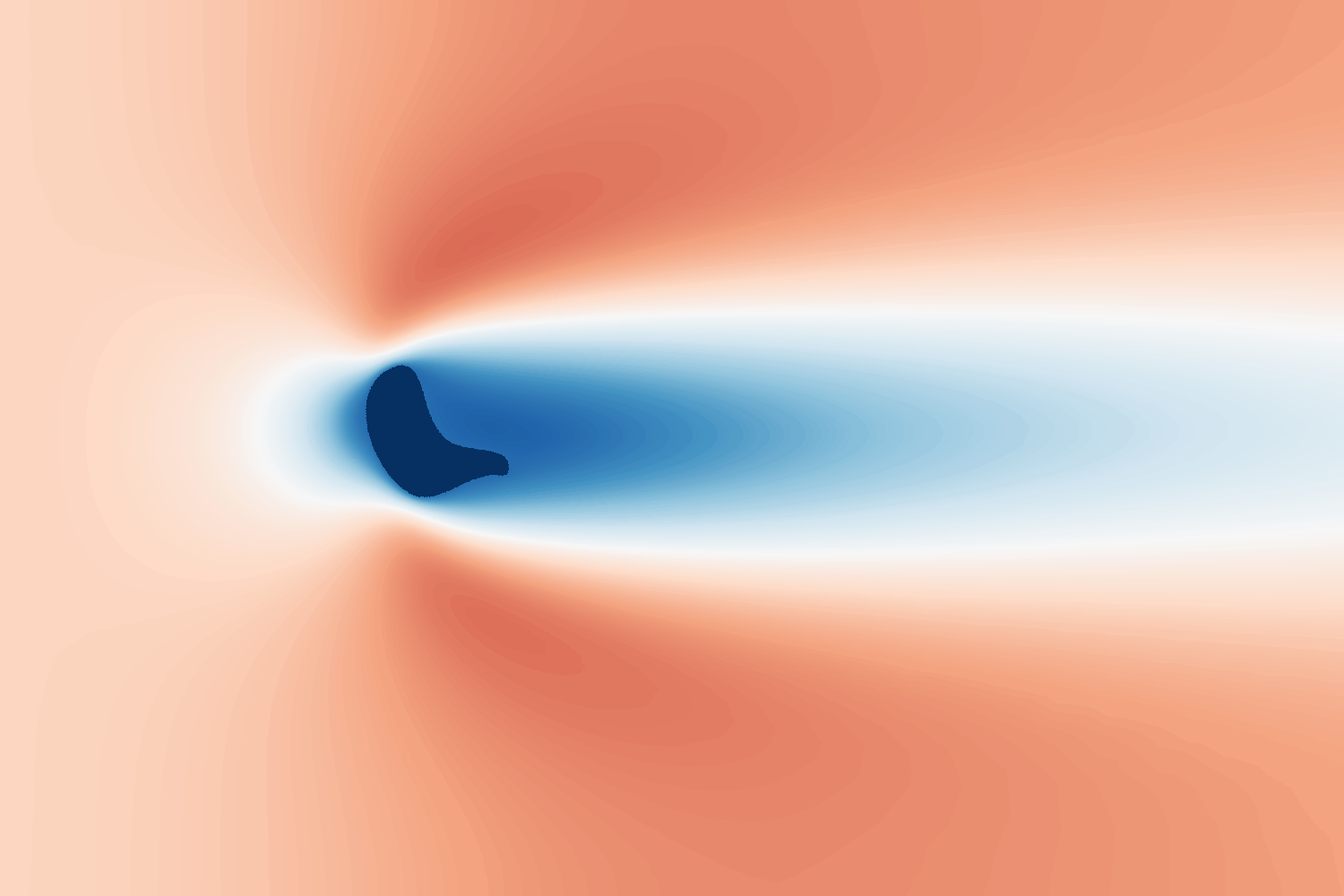}}
	\caption{Velocity field - $x$ component}
\end{subfigure}
\begin{subfigure}[t]{.1\textwidth}
	\centering
	\raisebox{-2.3mm}{\colorbar{0}{1}{3.15}}
\end{subfigure}

\medskip
\medskip

\begin{subfigure}{.3\textwidth}
	\centering
	\fbox{\includegraphics[width=\linewidth]{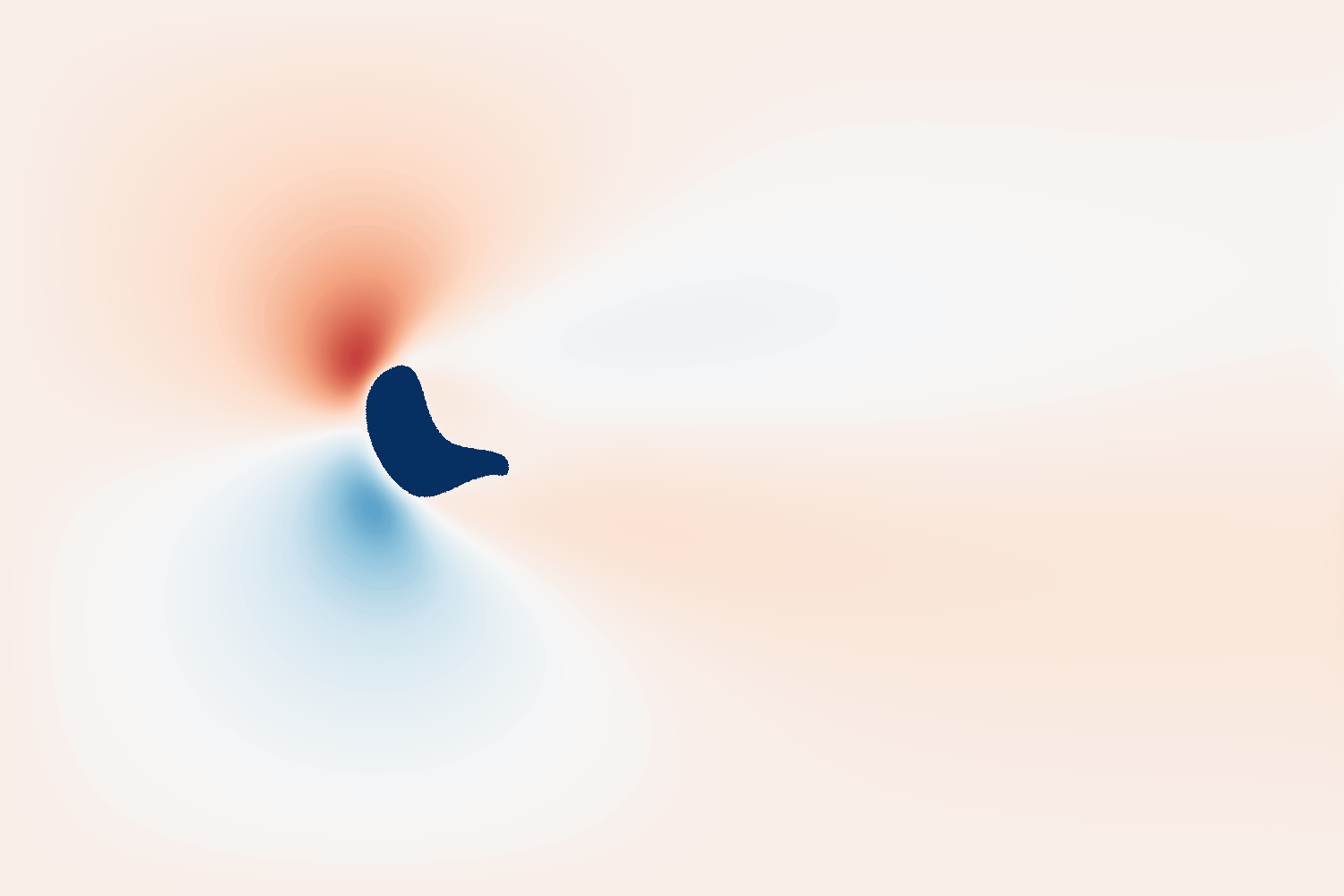}}
	\caption{Velocity field - $y$ component}
\end{subfigure} \quad
\begin{subfigure}{.3\textwidth}
	\centering
	\fbox{\includegraphics[width=\linewidth]{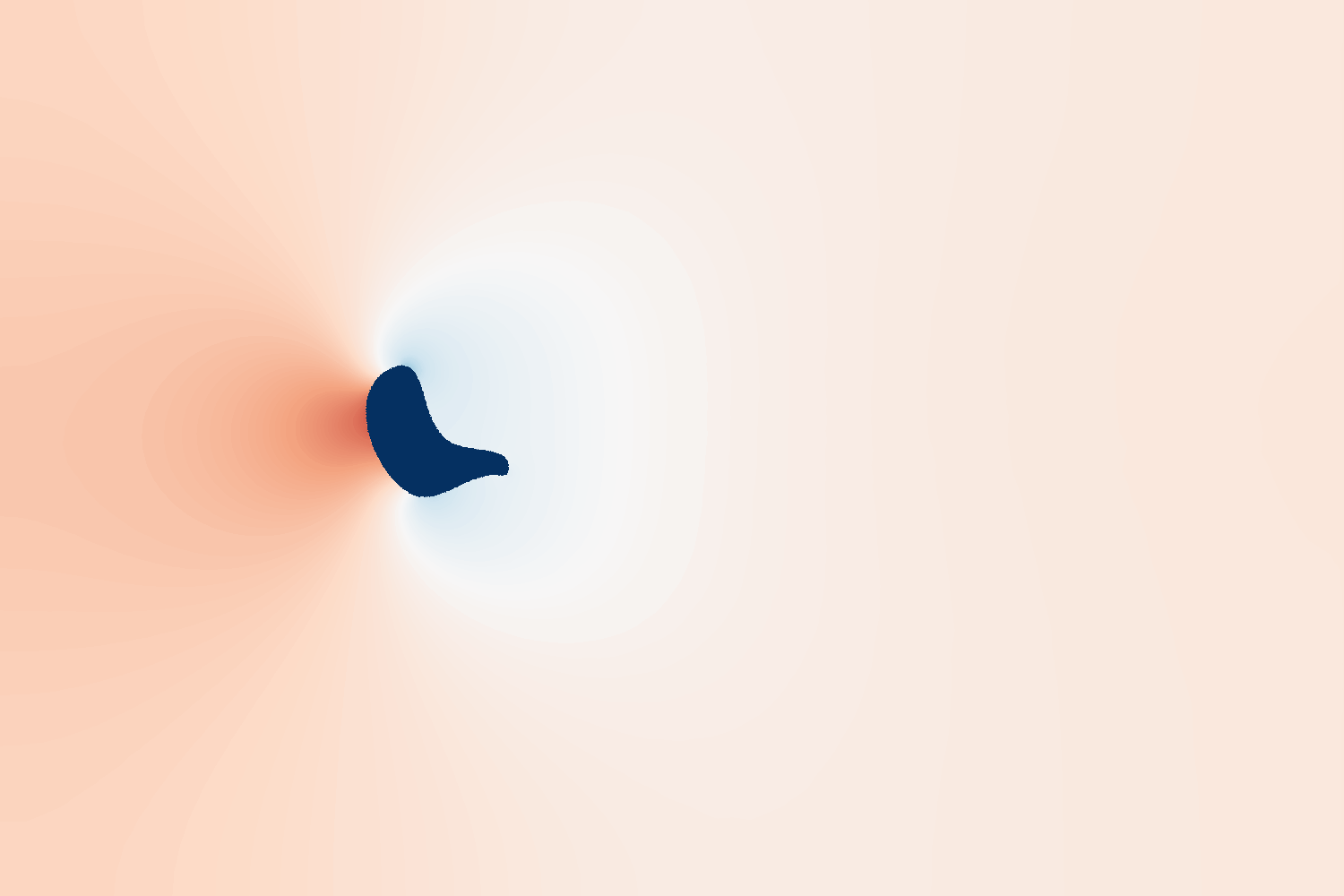}}
	\caption{Pressure field}
\end{subfigure}
\begin{subfigure}[t]{.1\textwidth}
	\centering
	\raisebox{-14.3mm}{\colorbar{0}{1}{3.15}}
\end{subfigure}

\caption{\textbf{Network input, velocity field and pressure field for a dataset element.} The shape is shown in its computational domain (upper left), along with the computed velocity field (top right, lower left) and pressure field (lower right).}
\label{fig:dataset_example}
\end{figure}

The dataset is composed of 12.000 shapes, along with their steady-state velocity and pressure fields at $Re=10$ (see figure \ref{fig:dataset_example}). All the labels were computed using CimLib-CFD \cite{Hachem2013}, following the methods exposed in section \ref{section:NS}. The input fields are resized to 2D $100\times 150$ arrays before being provided to the network. \red{As is customary in neural networks training,} a channel-wise normalization is applied, mapping all the pixels' value into $[0,1]$. In following sections we use a color scale to visualize the velocity and pressure fields, but each of them is always a 2D array. For additional details about the distribution of the elements in the dataset, the reader is referred to \cite{Viquerat2020} (section 3.5).

\section{Network architecture and training}
\label{section:autoencoder}

\subsection{Twin-decoder architecture}
\label{section:architecture}

\begin{figure}
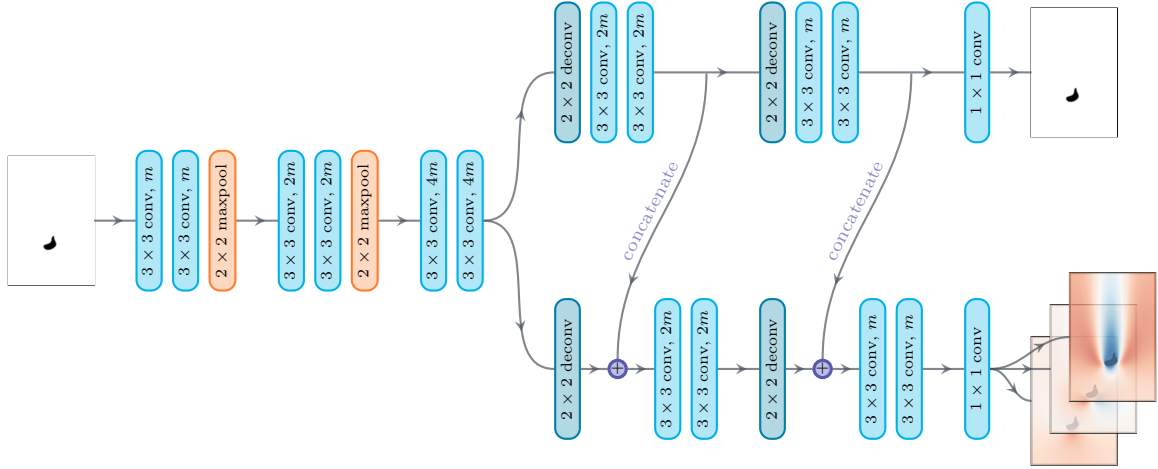

\centering

\begin{tikzpicture}[scale=0.85, node distance=0.15cm, rotate=90, transform shape] 
\tikzstyle{layer}		=[rectangle, rounded corners, thick, text centered]
\tikzstyle{cct}		=[circle, draw=mypurple1,		fill=mypurple4,		inner sep=0pt, very thick, text width=0.2cm]
\tikzstyle{skp}		=[inner sep=0pt, 			outer sep=0pt]
\tikzstyle{inp}		=[rectangle, thick, text centered, inner sep=0pt, 		outer sep=0pt,		text width=2cm, draw=black]
\tikzstyle{outp}		=[rectangle, thick, text centered, inner sep=0pt, 		outer sep=0pt,		text width=2cm, opacity=0.75, draw=black]
\tikzstyle{cv}		=[layer, draw=myblue1,		fill=myblue4, 		text width=2cm]
\tikzstyle{pl}		=[layer, draw=myorange1, 	fill=myorange4,		text width=2cm]
\tikzstyle{dcv}		=[layer, draw=mybluegray1, 	fill=mybluegray4,	text width=2cm]
\tikzstyle{connection}	=[thick,draw=mygray1,opacity=0.7,decoration={markings, mark=at position 0.7 with {\arrow{stealth}}},postaction={decorate}]
\tikzstyle{rconnection}=[thick,draw=mygray1,opacity=0.7,decoration={markings, mark=at position 0.25 with {\arrow[xscale=-1]{stealth}}},postaction={decorate}]
\scriptsize
\tikzstyle{space_small}=[yshift=-0.5cm]
\tikzstyle{space_med}=[yshift=-1cm]
\tikzstyle{space_large}=[yshift=-1.5cm]

\node[inp] 								(inp)  at (0,0)	{\includegraphics[width=\textwidth]{fig/shape_50.png}};

\node[cv, below=of inp, space_small] 		(c11) 		{$3\times3$ conv, $m$};
\node[cv, below=of c11] 					(c12) 		{$3\times3$ conv, $m$};
\node[pl,  below=of c12]					(p1) 			{$2\times2$ maxpool};

\node[cv, below=of p1, space_small] 		(c21) 	 	{$3\times3$ conv, $2m$};
\node[cv, below=of c21] 					(c22) 		{$3\times3$ conv, $2m$};
\node[pl,  below=of c22]					(p2) 			{$2\times2$ maxpool};

\node[cv, below=of p2, space_small] 		(c31) 	 	{$3\times3$ conv, $4m$};
\node[cv, below=of c31] 					(c32) 		{$3\times3$ conv, $4m$};

\node[dcv, below right=of c32, space_med] 	(dc1) 	 	{$2\times2$ deconv};
\node[cv, below=of dc1] 					(c41) 	 	{$3\times3$ conv,  	$2m$};
\node[cv, below=of c41] 					(c42) 	 	{$3\times3$ conv,  	$2m$};

\node[dcv, below=of c42, space_large] 		(dc2) 	 	{$2\times2$ deconv};
\node[skp] at ($(c42)!0.5!(dc2)$)			(sk1) 		{};
\node[cv, below=of dc2] 					(c51) 	 	{$3\times3$ conv,  	$m$};
\node[cv, below=of c51] 					(c52) 	 	{$3\times3$ conv,  	$m$};

\node[cv, below=of c52, space_large] 		(c61) 	 	{$1\times1$ conv};
\node[skp] at ($(c52)!0.5!(c61)$) 			(sk2)			{};

\node[inp, below=of c61, space_small] 		(outp1)  		{\includegraphics[width=\textwidth]{fig/shape_50.png}};

\node[dcv, below left=of c32, space_med] 		(dc3) 	 	{$2\times2$ deconv};
\node[cv, below=of dc3, space_med]			(c71) 	 	{$3\times3$ conv,  	$2m$};
\node[cct]	at ($(dc3)!0.5!(c71)$)			(cct1)		{$+$};
\node[cv, below=of c71] 					(c72) 	 	{$3\times3$ conv,  	$2m$};

\node[dcv, below=of c72, space_small] 		(dc4) 	 	{$2\times2$ deconv};
\node[cv, below=of dc4,space_med]			(c81) 	 	{$3\times3$ conv,  	$m$};
\node[cct]	at ($(dc4)!0.5!(c81)$)			(cct2)		{$+$};
\node[cv, below=of c81] 					(c82) 	 	{$3\times3$ conv,  	$m$};

\node[cv, below=of c82, space_small] 		(c91) 	 	{$1\times1$ conv};

\node[outp, below=of c91, yshift=-0.5cm, xshift=-0.5cm] 		(outp2)  		{\includegraphics[width=\textwidth]{fig/shape_50_p.png}};
\node[outp, below=of c91, yshift=-0.8cm, xshift=0cm] 		(outp3)  		{\includegraphics[width=\textwidth]{fig/shape_50_v.png}};
\node[outp, below=of c91, yshift=-1.1cm, xshift=0.5cm] 		(outp4)  		{\includegraphics[width=\textwidth]{fig/shape_50_u.png}};

\draw[connection]  (inp.south) -- (c11.north);
\draw[connection]  (p1.south) -- (c21.north);
\draw[connection]  (p2.south) -- (c31.north);
\draw[connection]  (c32.south)   to[out=-90,in=90] (dc1.north);
\draw[connection]  (c32.south)   to[out=-90,in=90] (dc3.north);
\draw[connection]  (c42.south) -- (dc2.north);
\draw[connection]  (c52.south) -- (c61.north);
\draw[connection]  (c61.south) -- (outp1.north);
\draw [connection]  (dc3.south) -- (cct1.north);
\draw [connection]  (cct1.south) -- (c71.north);
\draw [connection]  (c72.south) -- (dc4.north);
\draw [connection]  (dc4.south) -- (cct2.north);
\draw [connection]  (cct2.south) -- (c81.north);
\draw [connection]  (c82.south) -- (c91.north);
\draw [connection]  (c91.south) to[out=-90,in=90] (outp2.north);
\draw [connection]  (c91.south) to[out=-90,in=90] (outp3.north);
\draw [connection]  (c91.south) to[out=-90,in=90] (outp4.north);
\def\myshift#1{\raisebox{1ex}}
\draw[rconnection,postaction={decorate,decoration={text along path,text align=center,text color=mypurple1,
										text={|\scriptsize\myshift|concatenate}}}] (cct1.east) to[out=0,in=180] (sk1.west);
\draw[rconnection,postaction={decorate,decoration={text along path,text align=center,text color=mypurple1,
										text={|\scriptsize\myshift|concatenate}}}] (cct2.east) to[out=0,in=180] (sk2.west);

\end{tikzpicture}

\caption{\textbf{Proposed twin-decoder architecture.} The encoder is based on a pattern made of two convolutional layers followed by a max-pooling layer. At each occurence of the pattern, the image size is divided by two, while the number of filters, noted $m$, is doubled. In both decoder paths, a transposed convolution step is first applied to the input, while the number of filters is halved. The output of this layer in the flow decoder is then concatenated with its mirror counterpart in the shape decoder. Finally, two convolution layers are applied. At the end of the last layer, a $1 \times 1$ convolution is applied on each decoder to obtain a final 3D tensor with 4 channels. Every channel has the same dimension as the input.}
\label{fig:dualdecoder}
\end{figure}

The general autoencoder architecture with twin-decoder proposed in this contribution is shown in figure \ref{fig:dualdecoder}. Its input consists in a boolean 1-channel 2D tensor containing the computing domain and the obstacle. Its outputs are (i) a 1-channel 2D tensor containing the reconstructed input, and (ii) a 3-channels tensor containing the predicted velocity components and pressure fields. The encoder branch consists in stacked convolution-convolution-max-pooling blocks using $3 \times 3$ kernel size, stride size equal to 1, and zero-padding. The convolutional layers exploit rectified linear unit (ReLU) as activation functions. After every pooling operation, the number of kernels used for convolutional layers is doubled, until reaching the bottleneck.

The decoder is composed of two branches, hereafter denoted by ``shape decoder'' and ``flow decoder''. Both decoder branches are composed of deconvolution-convolution-convolution blocks, and share similar structures. The deconvolution layers use $2 \times 2$ kernel size, stride size equal to 2, zero padding and ReLU activation. Symmetrically to the encoder branch, the number of kernels is halved after each block in the decoder branches. Finally, a convolutional layer with a $1\times 1$ kernel is applied to set the final number of channels (1 for the shape decoder, and 3 for the flow decoder). The output of our network hence contains 1-channel tensor representing reconstructed input, and a 3-channel tensor representing the velocity and pressure.

The key ingredient of the proposed architecture lies in the skip connections that link the shape decoder and the flow decoder. The output of each shape decoder block is concatenated (along the channel axis, to preserve shape) to the output of the deconvolution layer of each flow decoder block. This idea is similar to that of U-net, except that low-level features do not originate from the encoder branch, but from the reconstruction of the input. Forcing such dependence between the two decoder branches is expected to induce a strong correlation between their performance levels. In essence, by enforcing a relation between the shape reconstruction error and the flow prediction error, the proposed method allows to reject possible outliers based on the reconstruction error. As the latter can be computed for an arbitrary input, the end-user can be warned whether the network prediction can be trusted or not. More details on the acceptance/rejection procedure are provided in section \ref{section:trust}.

\subsection{Training procedure}
\label{section:training}

The loss function used to train the twin-decoder architecture is a weighted sum of the shape and the flow decoder losses. Both decoders use the regular mean squared error (MSE) as loss function:

\begin{equation}
\label{eq:loss_mse}
	L = \frac{1}{h w n_c} \sum_{d,i,j}(y_{d,i,j} - \hat{y}_{d,i,j})^2,
\end{equation}

where $h$, $w$ and $n_c$ represent respectively channel height, width and the number of channels. Expression (\ref{eq:loss_mse}) represents the average squared error over all the pixels of a 3D tensor. The final loss function used for training is:

\begin{equation}
\label{eq:loss_beta}
	L_{\text{twin}} = L_{\text{flow}} + \beta L_{\text{shape}},
\end{equation}

where $\beta$ is a weighting parameter that remains to be tuned (see section \ref{section:results}). The network is trained with the Adam optimizer using an initial learning rate of \num{1e-3}, which is reduced to \num{1e-4} after 600 epochs. To prevent overfitting, the validation loss is monitored, and early stopping is used to determine the end of the training. The network parameters are initialized using a truncated Guassian distribution, following \cite{UnetRonneberger}. Training is performed using a Tesla V100 GPU, using mini-batches of size 128 to limit the required computational resources. As different models are evaluated and caompared, their training times are given in section \ref{section:results}.

\subsection{Trust level based on shape reconstruction}
\label{section:trust}

Given a trained twin-decoder neural network, it is feasible to evaluate the trust level of flow prediction by input reconstruction. As the error levels of the twin decoders are strongly correlated, a qualitative and a quantitative trust-level methods are proposed, both based on the reconstruction error. Their general concepts are proposed in the following sub-sections, while their applications on trained networks are presented in section \ref{section:results_trust_level}. In the following, the MSE for flow and shape reconstructions are respectively denoted $e_f$ and $e_s$.

\subsubsection{Qualitative method}
\label{section:qualitative}

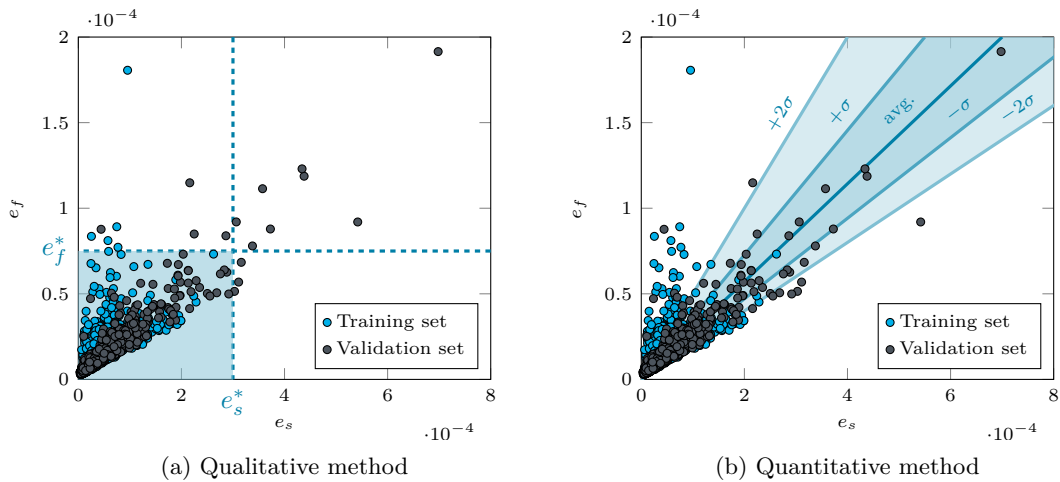
\begin{figure}
\centering

\begin{subfigure}[b]{.45\textwidth}
\centering
	\begin{tikzpicture}[trim axis left, trim axis right]
		\begin{axis}[	scale=0.8,transform shape, 
					label style={font=\scriptsize}, tick label style={font=\scriptsize}, legend style={font=\scriptsize},
					ymin=0, ymax=0.0002, xmin=0, xmax=0.0008,
					xlabel=$e_s$,ylabel=$e_f$,
					scaled y ticks=base 10:4,
					legend pos=south east,
					legend cell align={left},
					clip=false
					]
			\legend{Training set, Validation set}
			\addplot+[only marks,mark=*,mark options={draw=black,fill=myblue1},mark size=1.5pt] table[x index=1,y index=2] {fig/mse_train.csv};
			\addplot+[only marks,mark=*,mark options={draw=black,fill=mygray1},mark size=1.5pt] table[x index=1,y index=2] {fig/mse_valid.csv};
			\draw[mybluegray1, very thick, dash pattern=on 2pt] (axis cs:0.0003,\pgfkeysvalueof{/pgfplots/ymax}) -- (axis cs:0.0003,\pgfkeysvalueof{/pgfplots/ymin}) node[anchor=north]{$e_s^*$};
			\draw[mybluegray1, very thick, dash pattern=on 2pt] (axis cs:\pgfkeysvalueof{/pgfplots/xmax},0.000075) -- (axis cs:\pgfkeysvalueof{/pgfplots/xmin},0.000075) node[anchor=east]{$e_f^*$};
			\fill[mybluegray3, opacity=0.5] (axis cs:\pgfkeysvalueof{/pgfplots/xmin},\pgfkeysvalueof{/pgfplots/ymin}) rectangle (axis cs:0.0003,0.000075);
		\end{axis}	
	\end{tikzpicture}
	
	\caption{Qualitative method}
	\label{fig:qualitative}
\end{subfigure} \quad
\begin{subfigure}[b]{.45\textwidth}
\centering
	\begin{tikzpicture}[trim axis left, trim axis right]
		\begin{axis}[	scale=0.8,transform shape, 
					label style={font=\scriptsize}, tick label style={font=\scriptsize}, legend style={font=\scriptsize},
					ymin=0,ymax=0.0002,
					xmin=0,xmax=0.0008,
					xlabel=$e_s$,ylabel=$e_f$,
					scaled y ticks=base 10:4,
					legend pos=south east,
					legend cell align={left}
					]
			\legend{Training set, Validation set}
			\addplot+[only marks,mark=*,mark options={draw=black,fill=myblue1},mark size=1.5pt] table[x index=1,y index=2] {fig/mse_train.csv};
			\addplot+[only marks,mark=*,mark options={draw=black,fill=mygray1},mark size=1.5pt] table[x index=1,y index=2] {fig/mse_valid.csv};
			\addplot+[mybluegray1, very thick,name path=avg,mark=none] coordinates {(0,0) (0.0007,0.0002)} node[sloped, pos=0.75, anchor=south,color=mybluegray1] {\scriptsize avg.};
			
			\addplot[mybluegray2, very thick,name path=std11] coordinates {(0,0) (0.00055,0.0002)} node[sloped, pos=0.75, anchor=south,color=mybluegray1] {\scriptsize $+\sigma$};
			\addplot[mybluegray2, very thick,name path=std12] coordinates {(0,0) (0.00085,0.0002)} node[sloped, pos=0.75, anchor=south,color=mybluegray1] {\scriptsize $-\sigma$};
			\addplot[mybluegray3,opacity=0.5,forget plot] fill between[of=std11 and std12];
			
			\addplot[mybluegray3, very thick,name path=std21] coordinates {(0,0) (0.0004,0.0002)} node[sloped, pos=0.75, anchor=south,color=mybluegray1] {\scriptsize $+2\sigma$};;
			\addplot[mybluegray3, very thick,name path=std22] coordinates {(0,0) (0.0010,0.0002)} node[sloped, pos=0.75, anchor=south,color=mybluegray1] {\scriptsize $-2\sigma$};;
			\addplot[mybluegray4,opacity=0.5,forget plot] fill between[of=std21 and std22];
		\end{axis}	
	\end{tikzpicture}
	
	\caption{Quantitative method}
	\label{fig:quantitative}
\end{subfigure} 
	
\caption{\textbf{Description of the qualitative and quantitative methods} on the scatter plots of $e_f$ versus $e_s$ on training and validation sets for a reference twin-decoder architecture. (Left) Qualitative method: given a threshold $e^*_f$, the corresponding optimal $e_s^*$ is found by solving problem (\ref{eq:qualitative}). Then, the end-user rejects predictions that produce shape reconstruction errors superior to $e_s^*$. Only the predictions falling within the bottom left quarter (in orange) are accepted. (Right) Quantitative method: $e_f$ is modeled as an affine function of $e_s$ with an uncertainty interval, as shown in relation (\ref{eq:quantitative_rel}). Based on $e_s$, the method provides an estimated $e_f$ level, along with a confidence interval for the prediction.}
\label{fig:methods}
\end{figure}

In this case, the end-user provides an acceptable MSE level $e_f^*$ on the flow prediction. The method consists in selecting the associated shape reconstruction error $e_s^*$ that minimizes the probability of taking wrong decisions when supposing that the two error levels are linearly correlated. The threshold shape reconstruction error $e_s^*$ is the solution to the following minimization problem:

\begin{equation}
\label{eq:qualitative}
	e_s^* = \frac{1}{N} \min_{e} \big[ \text{card} \left( e_s < e \text{ and } e_f > e_f^* \right) + \text{card} \left( e_s > e \text{ and } e_f < e_f^* \right) \big],
\end{equation}

where $N$ is the number of MSE scatter points $(e_s, e_f)$ taken into account. The numerator sums the amount of wrong decisions as the error of both decoders are assumed to be positively correlated. Such a formulation prevents the end-user from making mistakes when accepting or rejecting predictions. An illustration of the method is shown in figure \ref{fig:qualitative}.

\subsubsection{Quantitative method}
\label{section:quantitative}

With the quantitative method, $e_f$ and an associated confidence interval $\delta_f$ are directly estimated from $e_s$ using the following relation:

\begin{equation}
\label{eq:quantitative_rel}
	e_f = a e_s + b + \epsilon,
\end{equation}

where $\epsilon$ follows a normal law of the form:

\begin{equation}
\label{eq:normal}
    \epsilon \sim \mathcal{N}(0, (c e_s)^2).
\end{equation} 

The assumption is supported by the linear pattern of error scatter plots. In essence, indexing the standard deviation on $e_s$ in (\ref{eq:normal}) represents the growing uncertainty on flow prediction as shape reconstruction turns worse. Under this formulation, the likelihood of $e_f$ on $N$ scatter points is:

\begin{equation}
\label{eq:quantitative}
	\prod_{i=1}^{N}p(e_f^i|e_s^i;a,b,c) = \prod_{i=1}^{N} \frac{1}{\sqrt{2\pi (c e_s^i)^2}} \exp \left(-\frac{(e_f^i-a e_s^i - b)^2}{2(c e_s^i)^2}\right)
\end{equation}

The optimal parameters $a^*$, $b^*$ and $c^*$ are obtained by minimizing the negative log likelihood:

\begin{equation}
\label{eq:quantitative}
	(a^*,b^*,c^*) = \min_{a,b,c}  \frac{1}{2}\sum_i \left( \frac{a e_s^i e_f^i - b}{c e_s^i} \right)^2 + \sum_i \log(c e_s^i) + \frac{N}{2} \log(2 \pi),
\end{equation}

which is achieved using a gradient descent algorithm. An illustration of the method is shown in figure \ref{fig:quantitative}. The minimum is determined by the scatter points of the training and validation error, then evaluated on the test set. To the difference of the qualitative method, in which the predictions are either plainly accepted or rejected, such formulation allows one to estimate $e_f$ with a confidence interval $\delta_f$, without a need of pre-selecting a threshold accuracy, thus providing an additional flexibility. As an example, given a shape reconstruction error $e_s$, the associated flow prediction accuracy level $e_f$ would fall into $[(a - 2c) e_s + b, (a + 2c) e_s + b]$
with $95\%$ probability.

\section{Results}
\label{section:results}

In this section, the performance of the twin-decoder architecture is explored. First, a minimal hyper-parameter calibration is presented, that highlights the impact of the loss weighting parameter $\beta$, the number of convolutional blocks used in the decoder structure, and \red{the number of kernels used in the convolutional layers}. The resulting architecture is evaluated on the test set, showing good performance on unseen shapes. Then, the qualitative and quantitative methods respectively presented in sections \ref{section:qualitative} and \ref{section:quantitative} are put into practice. Finally, a comparison with two other AE-based architectures is proposed, revealing the contribution of skip connections between shape and flow decoders.

\subsection{Hyper-parameter calibration}
\label{section:calibration}

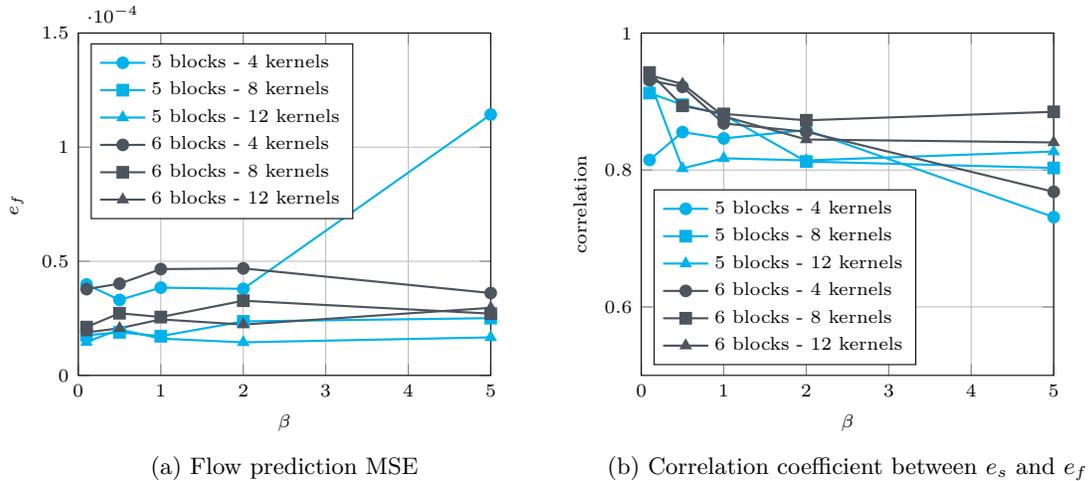
\begin{figure}
\centering

\begin{subfigure}[b]{.45\textwidth}
\centering
	\begin{tikzpicture}[trim axis left, trim axis right]
		\begin{axis}[	scale=0.8,transform shape,
					label style={font=\scriptsize}, tick label style={font=\scriptsize}, legend style={font=\scriptsize},
					ymin=0, ymax=0.00015, xmin=0, xmax=5,
					xlabel=$\beta$,ylabel=$e_f$,
					scaled y ticks=base 10:4,
					legend pos=north west,
					legend cell align={left},
					clip=false, grid=both
					]
			\legend{5 blocks - 4 kernels, 5 blocks - 8 kernels, 5 blocks - 12 kernels, 6 blocks - 4 kernels, 6 blocks - 8 kernels, 6 blocks - 12 kernels}
			\addplot[mark=*,thick,myblue1] table[x index=0,y index=1] {fig/calibration_1.csv};
			\addplot[mark=square*,thick,myblue1] table[x index=0,y index=2] {fig/calibration_1.csv};
			\addplot[mark=triangle*,thick,myblue1] table[x index=0,y index=3] {fig/calibration_1.csv};
			\addplot[mark=*,thick,mygray1] table[x index=0,y index=4] {fig/calibration_1.csv};
			\addplot[mark=square*,thick,mygray1] table[x index=0,y index=5] {fig/calibration_1.csv};
			\addplot[mark=triangle*,thick,mygray1] table[x index=0,y index=6] {fig/calibration_1.csv};
		\end{axis}	
	\end{tikzpicture}
	
	\caption{Flow prediction MSE}
\end{subfigure} \quad
\begin{subfigure}[b]{.45\textwidth}
\centering
	\begin{tikzpicture}[trim axis left, trim axis right]
		\begin{axis}[	scale=0.8,transform shape,
					label style={font=\scriptsize}, tick label style={font=\scriptsize}, legend style={font=\scriptsize},
					ymin=0.5, ymax=1.0, xmin=0, xmax=5,
					xlabel=$\beta$,ylabel=correlation,
					legend pos=south west,
					legend cell align={left},
					clip=false, grid=both
					]
			\legend{5 blocks - 4 kernels, 5 blocks - 8 kernels, 5 blocks - 12 kernels, 6 blocks - 4 kernels, 6 blocks - 8 kernels, 6 blocks - 12 kernels}
			\addplot[mark=*,thick,myblue1] table[x index=0,y index=1] {fig/calibration_2.csv};
			\addplot[mark=square*,thick,myblue1] table[x index=0,y index=2] {fig/calibration_2.csv};
			\addplot[mark=triangle*,thick,myblue1] table[x index=0,y index=3] {fig/calibration_2.csv};
			\addplot[mark=*,thick,mygray1] table[x index=0,y index=4] {fig/calibration_2.csv};
			\addplot[mark=square*,thick,mygray1] table[x index=0,y index=5] {fig/calibration_2.csv};
			\addplot[mark=triangle*,thick,mygray1] table[x index=0,y index=6] {fig/calibration_2.csv};
		\end{axis}	
	\end{tikzpicture}
	
	\caption{Correlation coefficient between $e_s$ and $e_f$}
\end{subfigure} 
	
\caption{\textbf{Hyper-parameter calibration.} The performance is evaluated on the validation set. To compare the accuracy of flow prediction between different architectures, the MSE is averaged over the entire dataset.}
\label{fig:calibration}
\end{figure}

In this section, the impact of three different parameters is explored: (i) the weighting parameter $\beta$, (ii) the number of convolutional blocks, and (iii) \red{the number of convolutional kernels}. In total, 30 configurations were tested, the network being evaluated not only on its flow field prediction performance, but also on the correlation coefficient between $e_f$ and $e_s$ on the validation set. As the different explorations are detailed below, the corresponding results can be found in figure \ref{fig:calibration}.

\paragraph{Convolution blocks} The number of convolution blocks in the encoder (or equivalently the number of deconvolution blocks in the decoders) proved to be determinant regarding the performance of the flow decoder. Based on previous experiments (not shown here), only architectures with 5 or 6 blocks were considered (less blocks showing low performance, while more blocks being too costly to train). Architectures with 5 blocks proved to outperform deeper ones in terms of flow prediction, probably due to a too high compression rate in the latent space when using 6 blocks. Adversely, architectures with 6 blocks presented a slightly better correlation coefficient between the two decoder errors.

\paragraph{Convolution kernels} The flow prediction accuracy was highly dependent on the number of convolution kernels used in each blocks. \red{Since the architecture is symmetric and scalable, we use the number of kernels in the first convolutional layer to represent this hyper-parameter}. Results show a clear advantage of using 8 kernels over 4, while increasing from 8 to 12 is not as beneficial. Hence, it is not necessary to use too many kernels, as a CNN complexity scales with the square of the kernel number. Similar conclusions hold for the correlation coefficient.

\paragraph{Weighting parameter} The $\beta$ parameter in the loss function (\ref{eq:loss_beta}) proved to have a crucial impact on correlation of the two decoder errors. Five different values were considered (from 0.1 to 5), with larger values giving more weight to the shape decoder during the training process. We found that small $\beta$ values were very beneficial to the correlation level, while alleviating the differences caused by different number of convolutional blocks. \red{With $\beta=0.1$}, a 5-block architecture with 12 kernels in the first convolutional layer obtained a correlation level of 93\%, which represents a strong linear relation. The impact of $\beta$ on the flow prediction accuracy is not as clear according to the obtained results.

\subsection{Prediction cost and accuracy}

\begin{figure}
\centering
\begin{tikzpicture}[trim axis left, trim axis right]
	\begin{axis}[	scale=0.8,transform shape,
				label style={font=\scriptsize}, tick label style={font=\scriptsize}, legend style={font=\scriptsize},
				ymin=0.0, ymax=0.25, xmin=0, xmax=350,
				xlabel=Training epochs,ylabel=Loss,ymode=log,
				legend pos=north east,
				legend cell align={left},
				clip=true, grid=both
				]
		\legend{Training set, Validation set}
		\addplot[mark=none,thick,myblue1] table[x index=0,y index=1] {fig/nllx.csv};
		\addplot[mark=none,thick,mygray1] table[x index=0,y index=2] {fig/nllx.csv};
	\end{axis}	
\end{tikzpicture}
\caption{\textbf{Training history of the model.} The training is ceased when the validation loss does not decrease for 10 successive epochs. A slight overfitting can be observed on the trained model.}
\label{fig:training_history}
\end{figure}
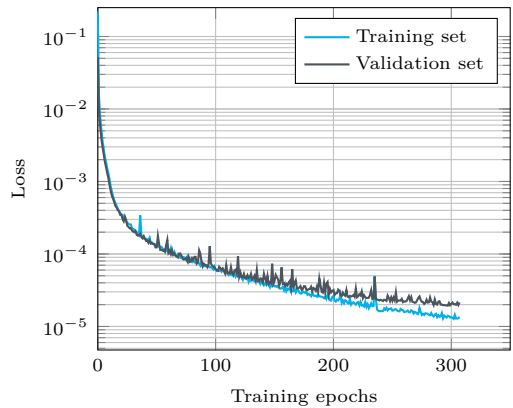

In order to evaluate the computational cost of the training, we provide the best performance attained by each combination with respect to the number of learnable parameters in table \ref{table:complexity}. Each point represents the smallest $e_f$ value obtained by tuning $\beta$. By using the 5-block architecture with 12 kernels, $e_f$ values as low as $1.4 \times 10^{-5}$ can be reached, requiring 1.4 million parameters, making it the architecture choice with the best cost-accuracy ratio. Learning time on a Tesla V100 GPU is approximately $0.7$ hours. The training curves for the training and validation subsets are presented in figure \ref{fig:training_history}.

\begin{table}
\footnotesize
\caption{\textbf{Flow prediction performance obtained for architectures of various complexities}. A good ratio must be found between the final achievable accuracy and the total number of learnable parameters, as training time rises dramatically with the network complexity. Here, best performance is obtained using a 5-block architecture and 12 initial kernels, with a total of 1.4 million parameters.}
\label{table:complexity}
\centering
\medskip
\begin{tabular}{lcc}
\toprule
Architecture		& Best $e_f$		& Nb. of parameters 	\\\midrule
5 blocks - 4 kernels	& \num{3.3063e-05}	& 157,208			\\\midrule
5 blocks - 8 kernels	& \num{1.7216e-05}	& 627,500			\\\midrule
\textbf{5 blocks - 12 kernels}	& \bm{$1.4491\times 10^{-05}$}	& \textbf{1,410,880}		\\\midrule
6 blocks - 4 kernels	& \num{3.6089e-05}	& 592,024			\\\midrule
6 blocks - 8 kernels	& \num{2.1178e-05}	& 2,365,484		\\\midrule
6 blocks - 12 kernels	& \num{1.8826e-05}	& 5,320,384		\\\bottomrule
\end{tabular}
\end{table}

\begin{figure}
\centering
\begin{subfigure}[b]{.3\linewidth}
	\centering
	\fbox{\includegraphics[width=.9\linewidth]{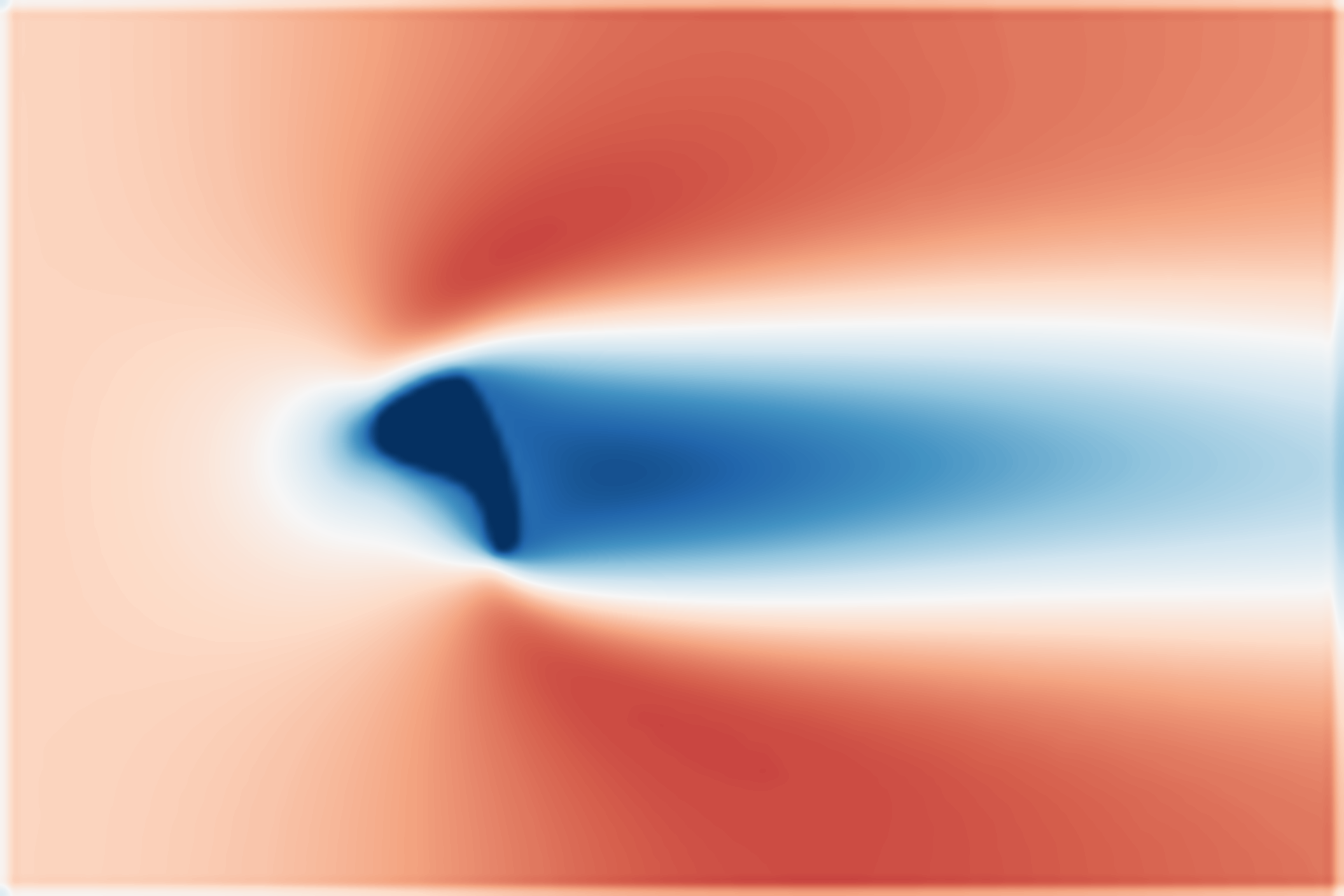}}
	\caption{$u$, reference}
	\label{fig:u0_ref}
\end{subfigure} \quad
\begin{subfigure}[b]{.3\linewidth}
	\centering
	\fbox{\includegraphics[width=.9\linewidth]{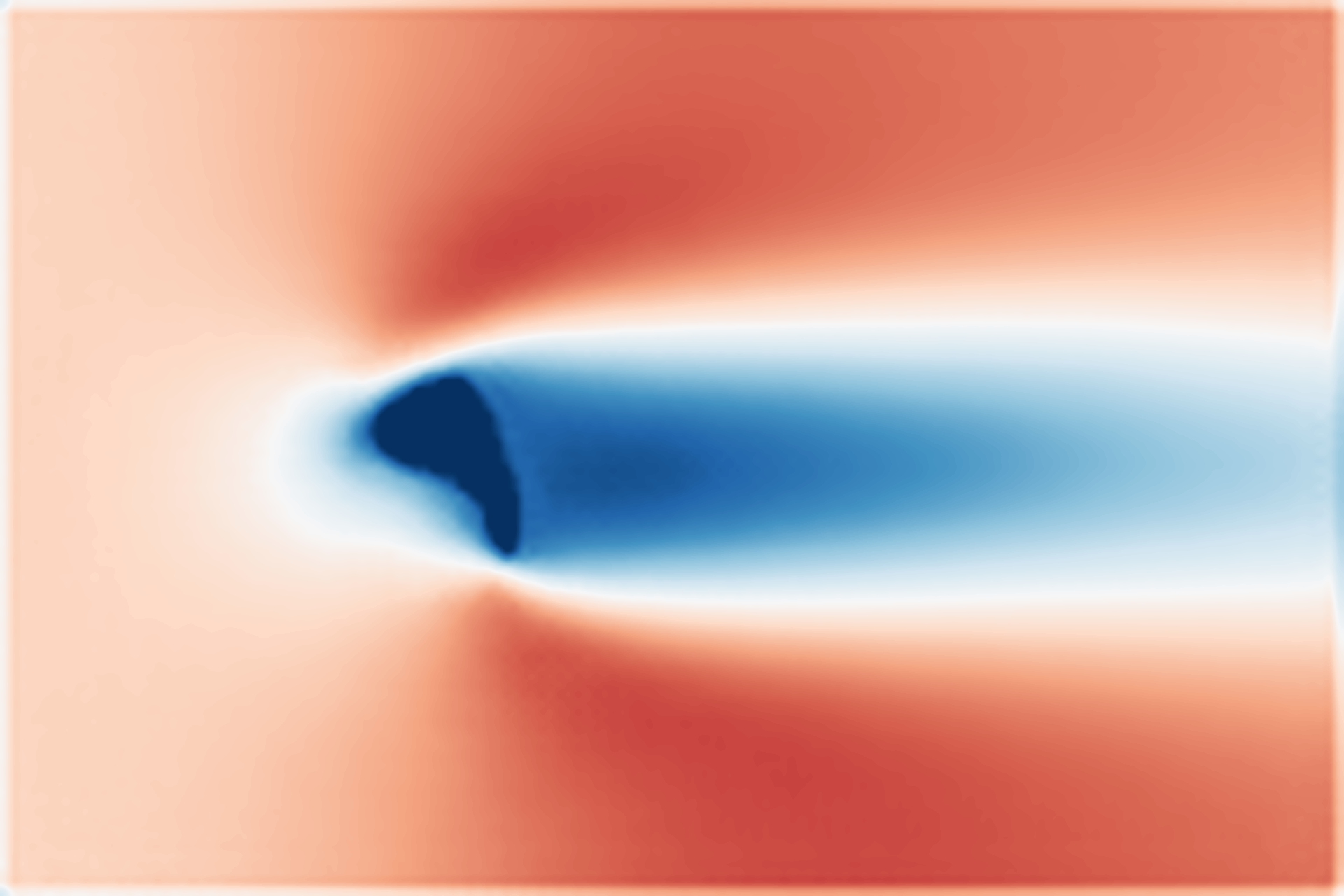}}
	\caption{$u$, predicted}
	\label{fig:u0_pred}
\end{subfigure} \quad
\begin{subfigure}[b]{.3\linewidth}
	\centering
	\fbox{\includegraphics[width=.9\linewidth]{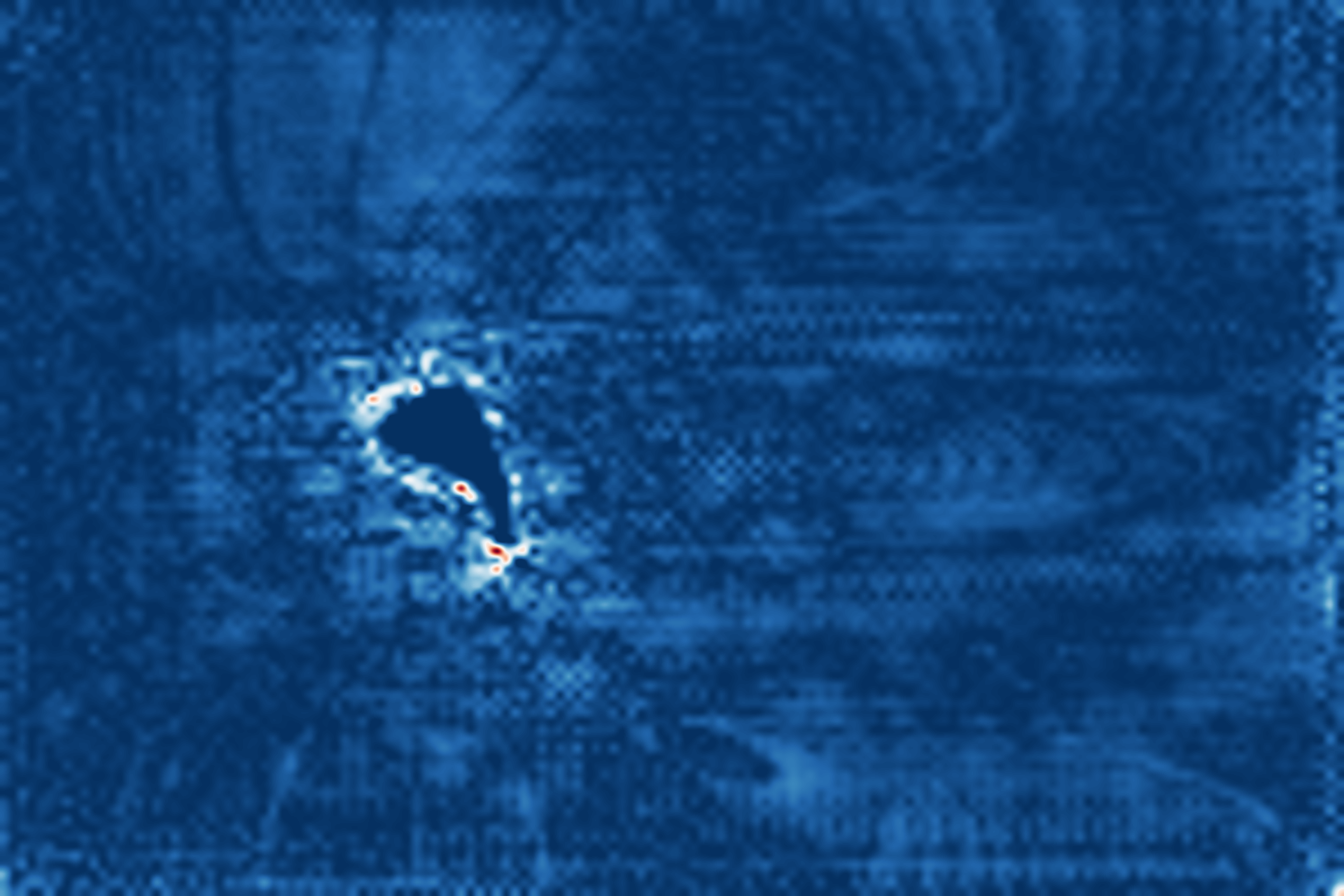}}
	\caption{$u$, absolute error}
	\label{fig:u0_error}
\end{subfigure}

\medskip

\begin{subfigure}[b]{.3\linewidth}
	\centering
	\fbox{\includegraphics[width=.9\linewidth]{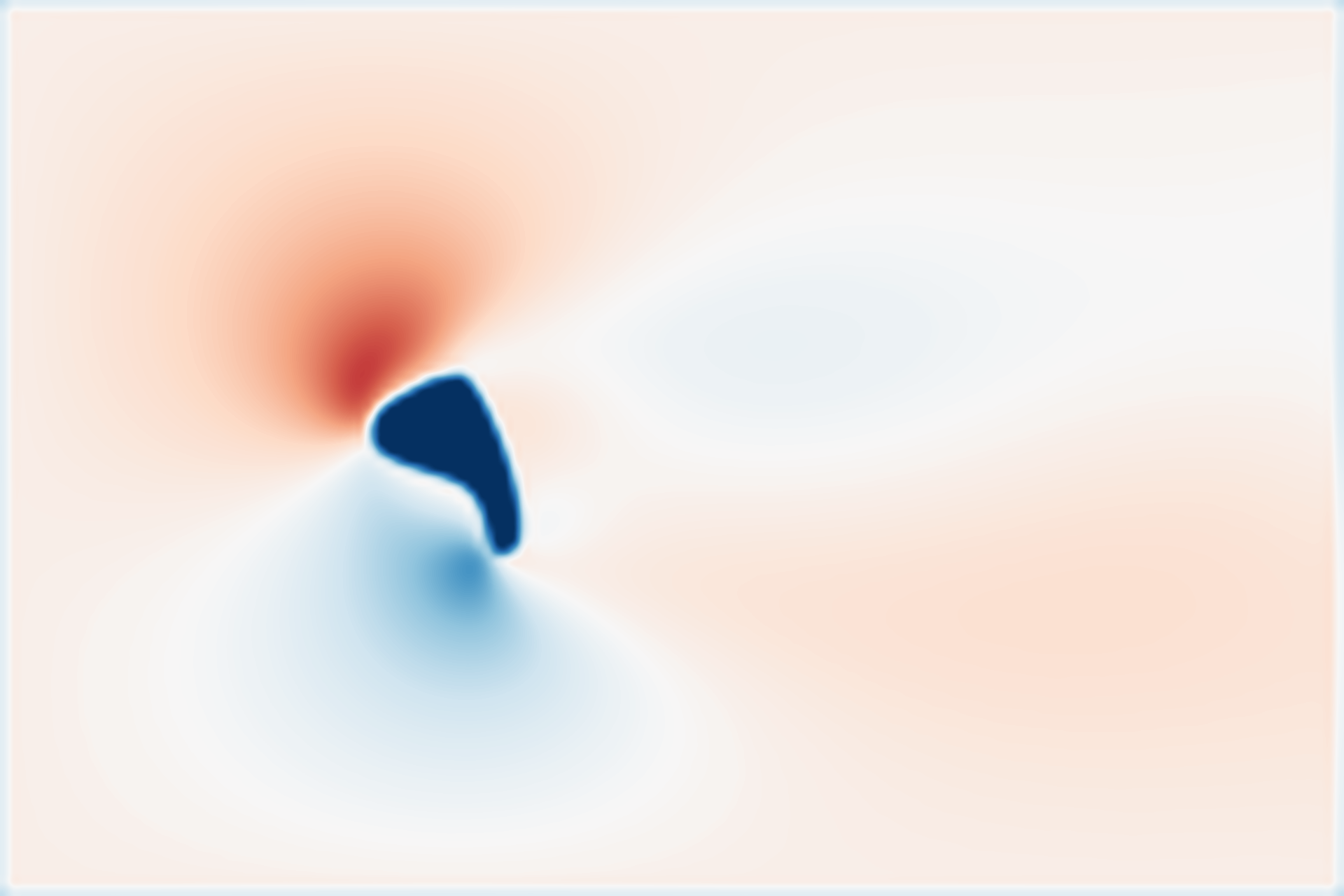}}
	\caption{$v$, reference}
	\label{fig:v0_ref}
\end{subfigure} \quad
\begin{subfigure}[b]{.3\linewidth}
	\centering
	\fbox{\includegraphics[width=.9\linewidth]{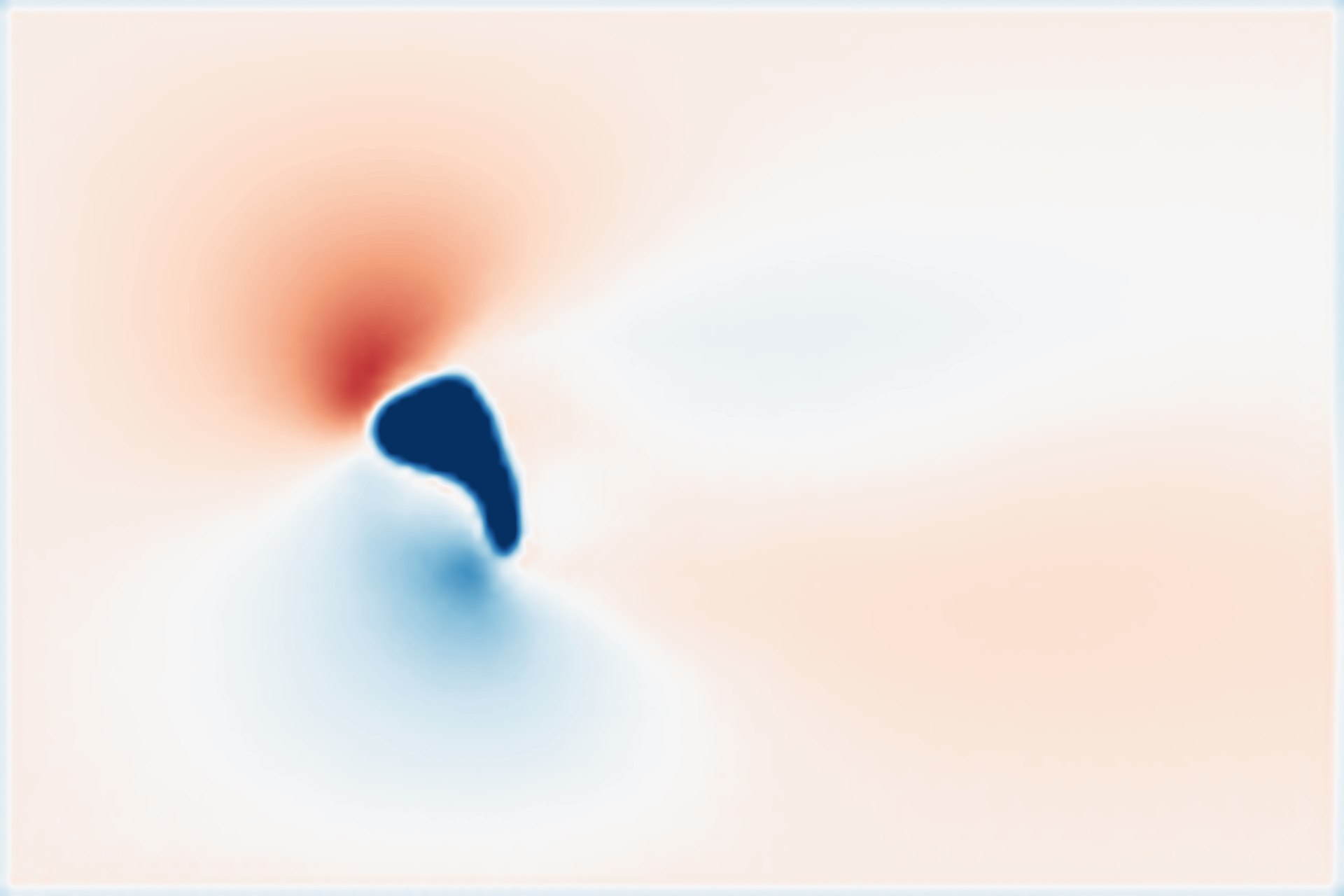}}
	\caption{$v$, predicted}
	\label{fig:v0_pred}
\end{subfigure} \quad
\begin{subfigure}[b]{.3\linewidth}
	\centering
	\fbox{\includegraphics[width=.9\linewidth]{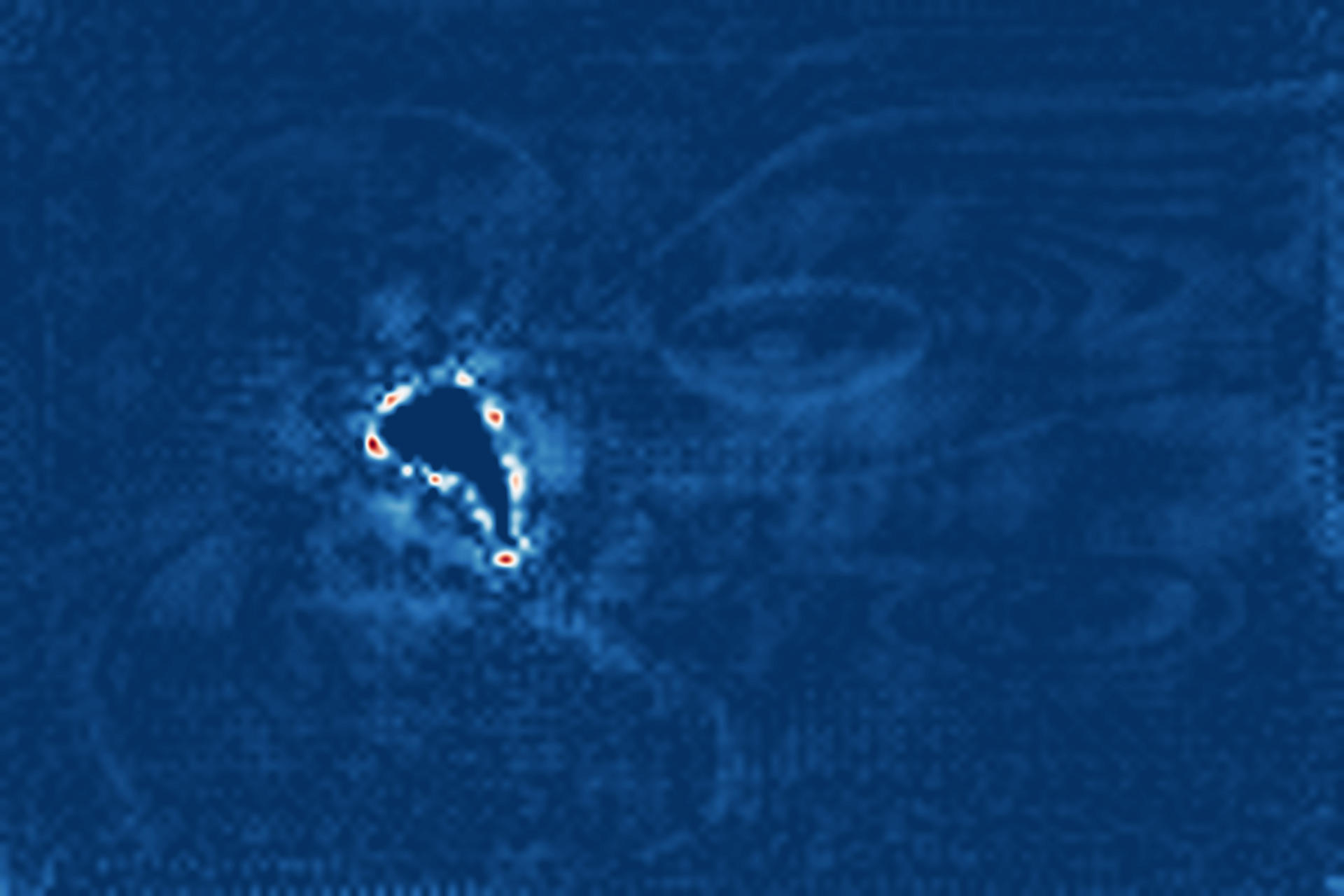}}
	\caption{$v$, absolute error}
	\label{fig:v0_error}
\end{subfigure}

\medskip

\begin{subfigure}[b]{.3\linewidth}
	\centering
	\fbox{\includegraphics[width=.9\linewidth]{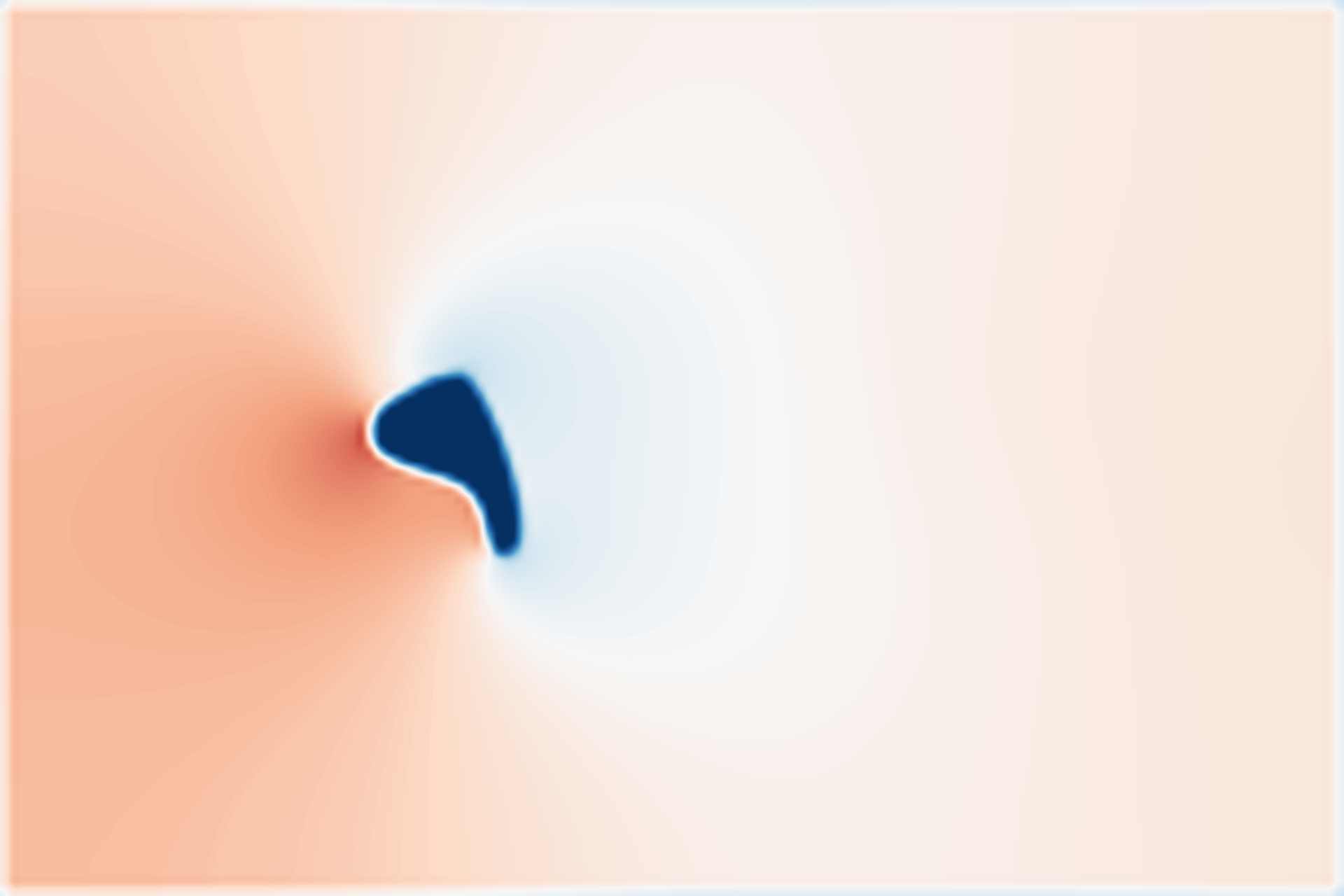}}
	\caption{$p$, reference}
	\label{fig:p0_ref}
\end{subfigure} \quad
\begin{subfigure}[b]{.3\linewidth}
	\centering
	\fbox{\includegraphics[width=.9\linewidth]{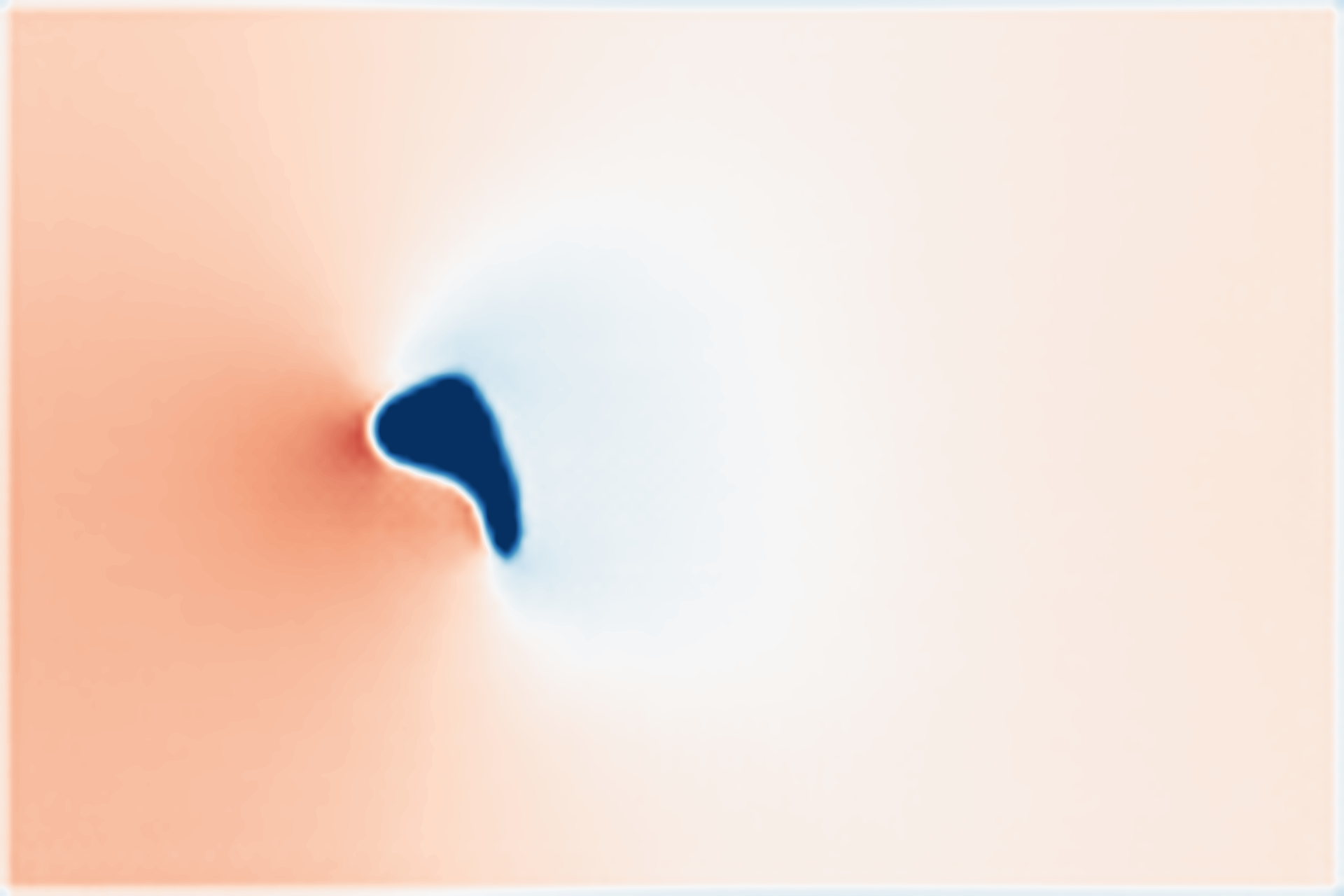}}
	\caption{$p$, predicted}
	\label{fig:p0_pred}
\end{subfigure} \quad
\begin{subfigure}[b]{.3\linewidth}
	\centering
	\fbox{\includegraphics[width=.9\linewidth]{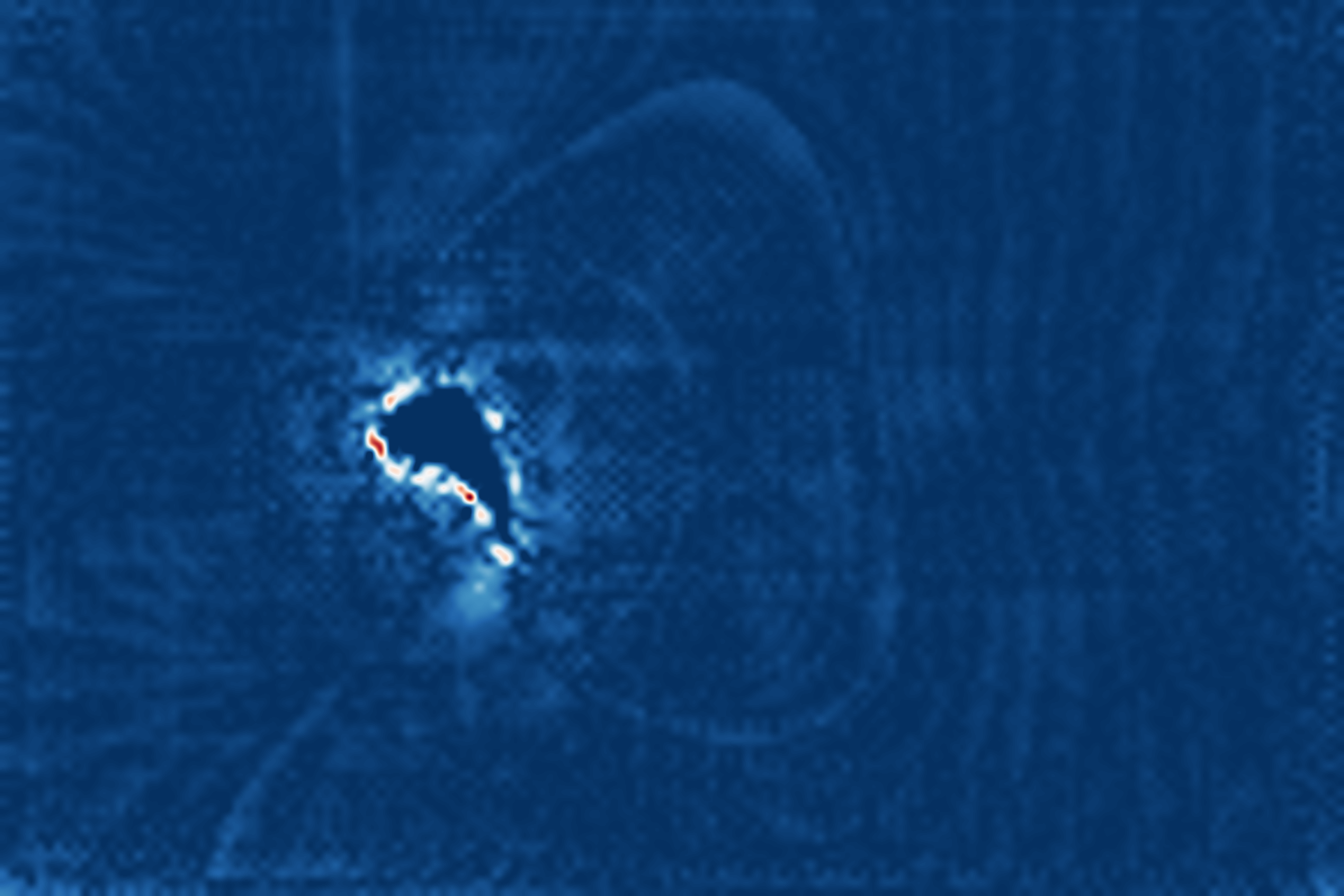}}
	\caption{$p$, absolute error}
	\label{fig:p0_error}
\end{subfigure}

\medskip

\begin{subfigure}[b]{.3\linewidth}
	\centering
	\fbox{\includegraphics[width=.9\linewidth]{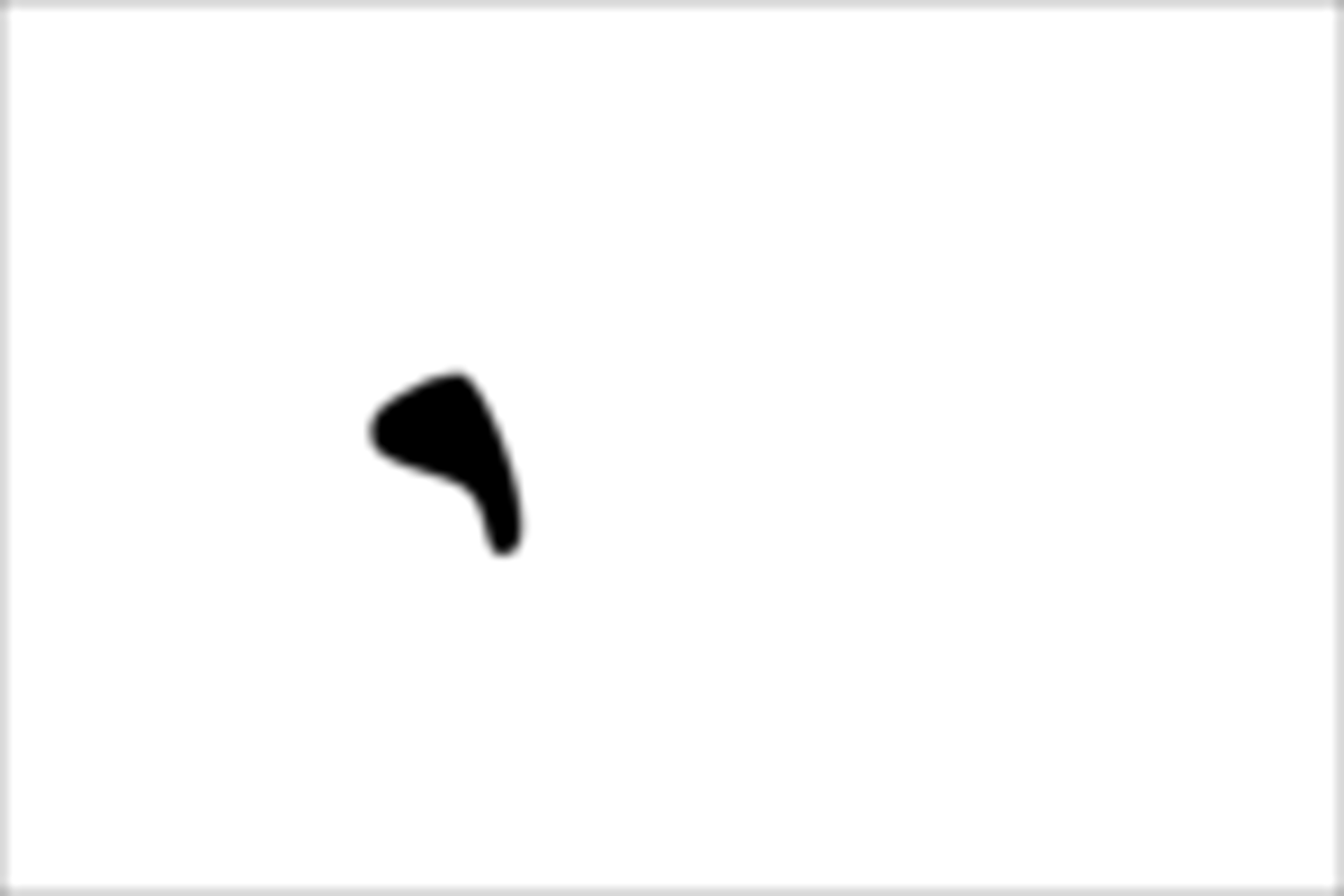}}
	\caption{Shape, reference}
	\label{fig:shape_ref}
\end{subfigure} \quad
\begin{subfigure}[b]{.3\linewidth}
	\centering
	\fbox{\includegraphics[width=.9\linewidth]{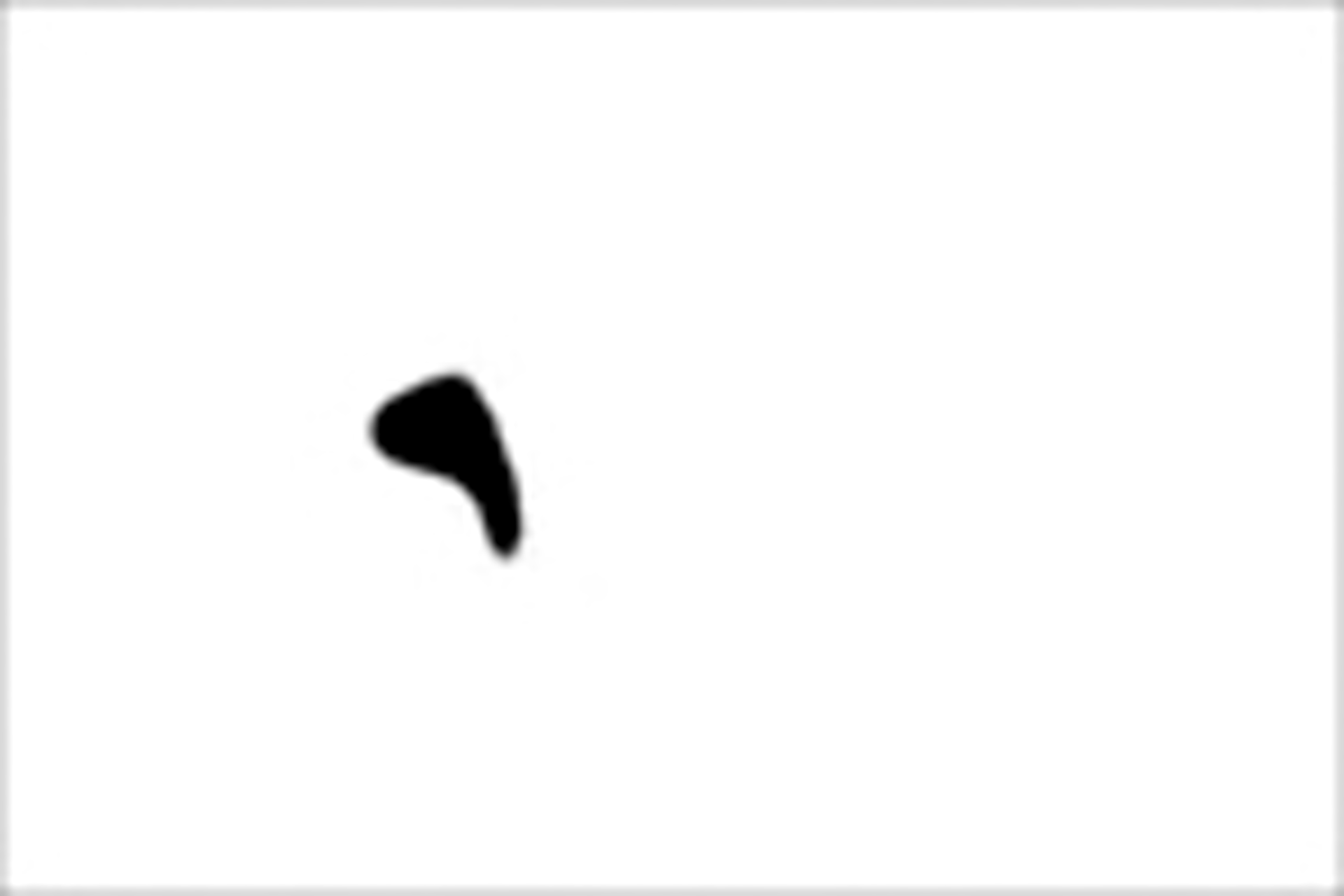}}
	\caption{Shape, predicted}
	\label{fig:shape_pred}
\end{subfigure} \quad
\begin{subfigure}[b]{.3\linewidth}
	\centering
	\fbox{\includegraphics[width=.9\linewidth]{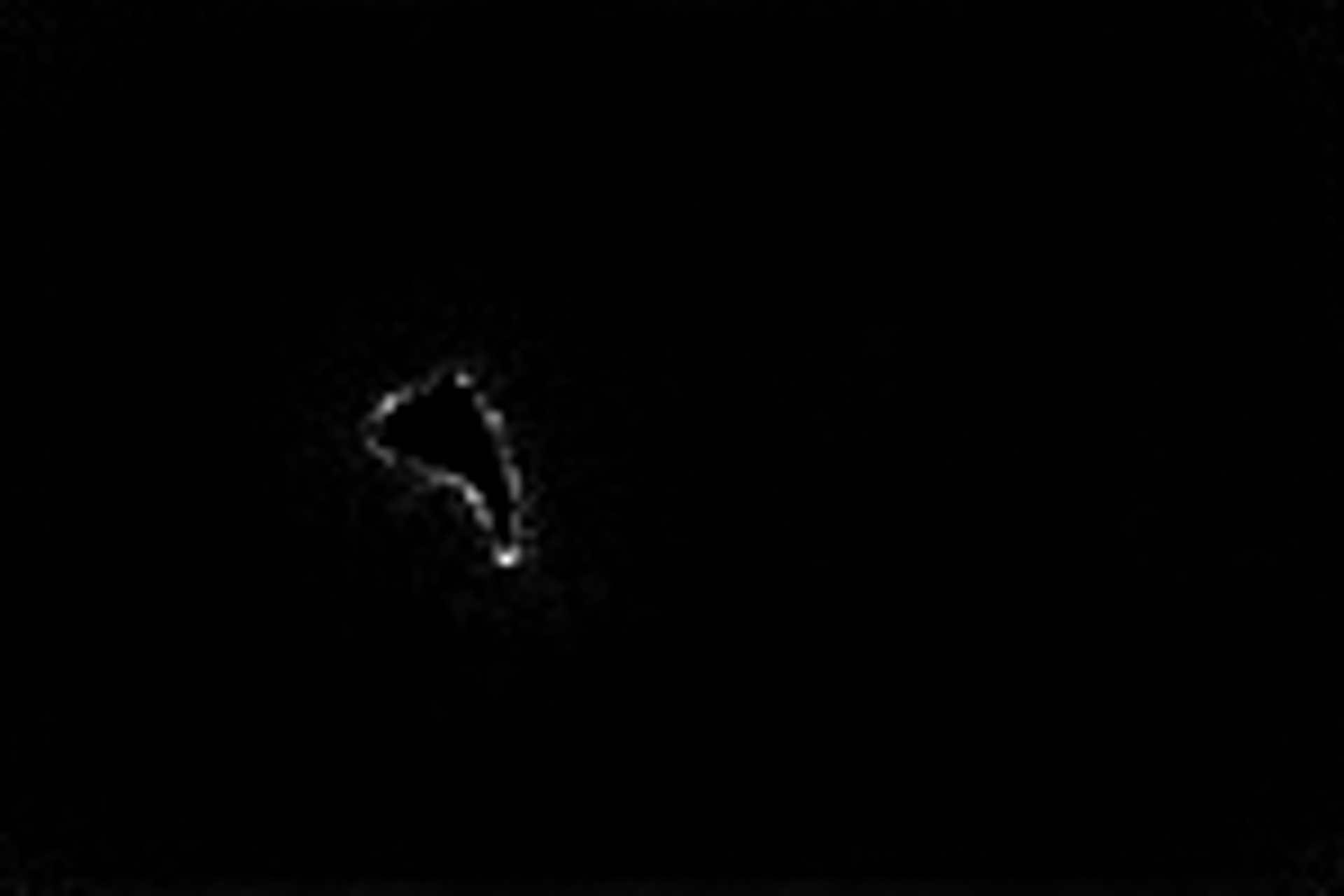}}
	\caption{Shape, absolute error}
	\label{fig:shape_error}
\end{subfigure}

\caption{\textbf{Flow and shape predictions around an obstacle from the test set.} On this instance, the flow prediction error is $e_f = \num{1.57e-5}$, with most of the error concentrated on the boundary of the shape, \ie in the area of large pressure and velocity gradients. For the $u$, $v$ and $p$ predictions, the pixels' value range is still $[0,1]$. An RGB colormap is used for better visualization}.
\label{fig:example}
\end{figure}

\begin{figure}
\centering

\setlength{\fboxsep}{0pt}%
\setlength{\fboxrule}{1pt}%

\begin{subfigure}[b]{.3\linewidth}
	\centering
	\fbox{\includegraphics[width=.9\linewidth]{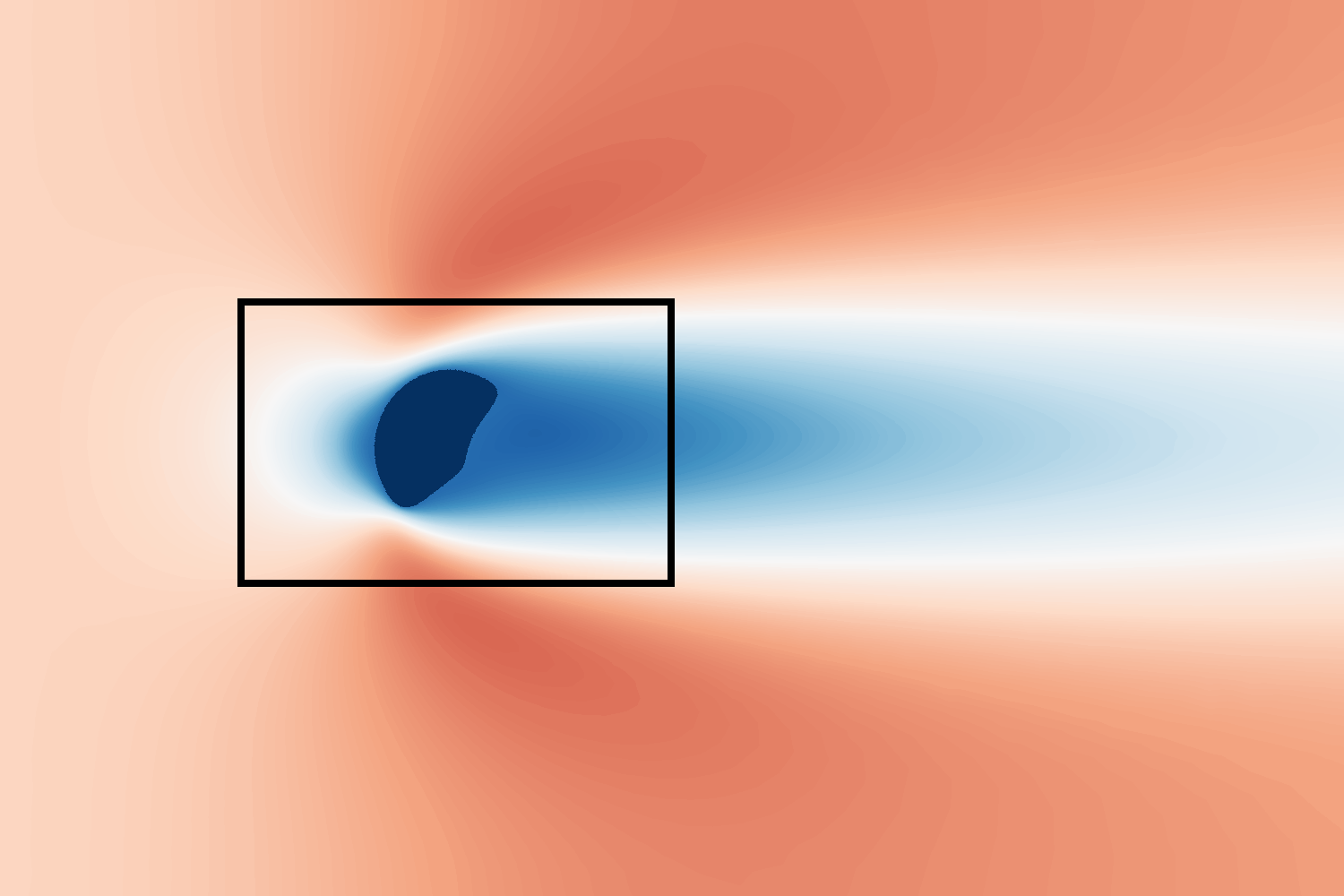}}
	\caption{Area taken into account to compute relative error histograms.}
	\label{fig:rectangle}
\end{subfigure}

\medskip

\begin{subfigure}[b]{.3\linewidth}
	\centering
	\begin{tikzpicture}[trim axis left, trim axis right]
		\begin{axis}[	scale=0.6,
					xmin=0, xmax=0.1, ybar interval, 
					xtick={0,0.02,0.04,0.06,0.08,0.1},
					xticklabels={0,0.02,0.04,0.06,0.08,0.1}]
		\addplot[draw=mybluegray1,fill=mybluegray4] table [x index=0, y index=2] {fig/histo_u.csv};
		\end{axis}
	\end{tikzpicture}
	\caption{Relative error for $u$.}
	\label{fig:u_histo}
\end{subfigure} \quad
\begin{subfigure}[b]{.3\linewidth}
	\centering
	\begin{tikzpicture}[trim axis left, trim axis right]
		\begin{axis}[	scale=0.6,
					xmin=0, xmax=0.1, ybar interval, 
					xtick={0,0.02,0.04,0.06,0.08,0.1},
					xticklabels={0,0.02,0.04,0.06,0.08,0.1}]
		\addplot[draw=mybluegray1,fill=mybluegray4] table [x index=0, y index=2] {fig/histo_v.csv};
		\end{axis}
	\end{tikzpicture}
	\caption{Relative error for $v$.}
	\label{fig:v_histo}
\end{subfigure} \quad
\begin{subfigure}[b]{.3\linewidth}
	\centering
	\begin{tikzpicture}[trim axis left, trim axis right]
		\begin{axis}[	scale=0.6, ybar interval, 
					xmin=0, xmax=0.1, ymin=0, ymax=160,
					xtick={0,0.02,0.04,0.06,0.08,0.1},
					xticklabels={0,0.02,0.04,0.06,0.08,0.1}]
		\addplot[draw=mybluegray1,fill=mybluegray4] table [x index=0, y index=2] {fig/histo_p.csv};
		\end{axis}
	\end{tikzpicture}
	\caption{Relative error for $p$.}
	\label{fig:p_histo}
\end{subfigure}
\caption{\textbf{Relative error for flow predictions} over test set. The black rectangle around the obstacle indicates the area on which the $u$, $v$ and $p$ relative errors are computed. The histograms indicate the error levels obtained when comparing predictions to labels on the $1200$ elements of the test set.}
\label{fig:rel_errors}
\end{figure}

On the test set, the proposed model reaches a flow reconstruction accuracy $e_f = \num{1.38 e-5}$, thus showing good generalization capabilities on unseen data. In figure \ref{fig:example}, a prediction example from the test set is shown, along with the associated exact solution and a pixel-wise error map. As can be seen, the velocity and pressure fields are well recovered by the network, the error being concentrated in the vicinity of the obstacle, \ie in the area of large pressure and velocity gradients. As the obstacle itself only represents a small portion of the predicted field, a second metric is proposed: for each element of the test set, the mean pixel-wise relative error is computed on a smaller area surrounding the obstacle. The corresponding error distributions are plotted in figure \ref{fig:rel_errors}. Overall, the average relative error is 3.92\% for horizontal velocity, 3.57\% for vertical velocity and 3.55\% for pressure, with very few elements exceeding 6\% of relative error, showing again decent generalization capabilities.

\subsection{Correlation levels}
\label{section:correlation_levels}

In this section, the correlation levels obtained between the shape reconstruction and the flow prediction errors are commented. To further show the interest of the autoencoder architecture with twin-decoder (twin-AE), two close network architectures are considered, namely the dual autoencoder (dual-AE), and the U-net dual autoencoder (U-dual-AE). The dual-AE architecture is simply obtained by removing the skip connections of the twin-AE, while the U-dual-AE exploits skip connections coming from the encoder path instead of the reconstruction path. To ensure a fair comparison, the same configuration is used for all three architectures. A concatenation with constant tensor is applied to the flow decoder of the dual-AE, so its number of parameters is equal to that of the U-dual-AE and the twin-AE.

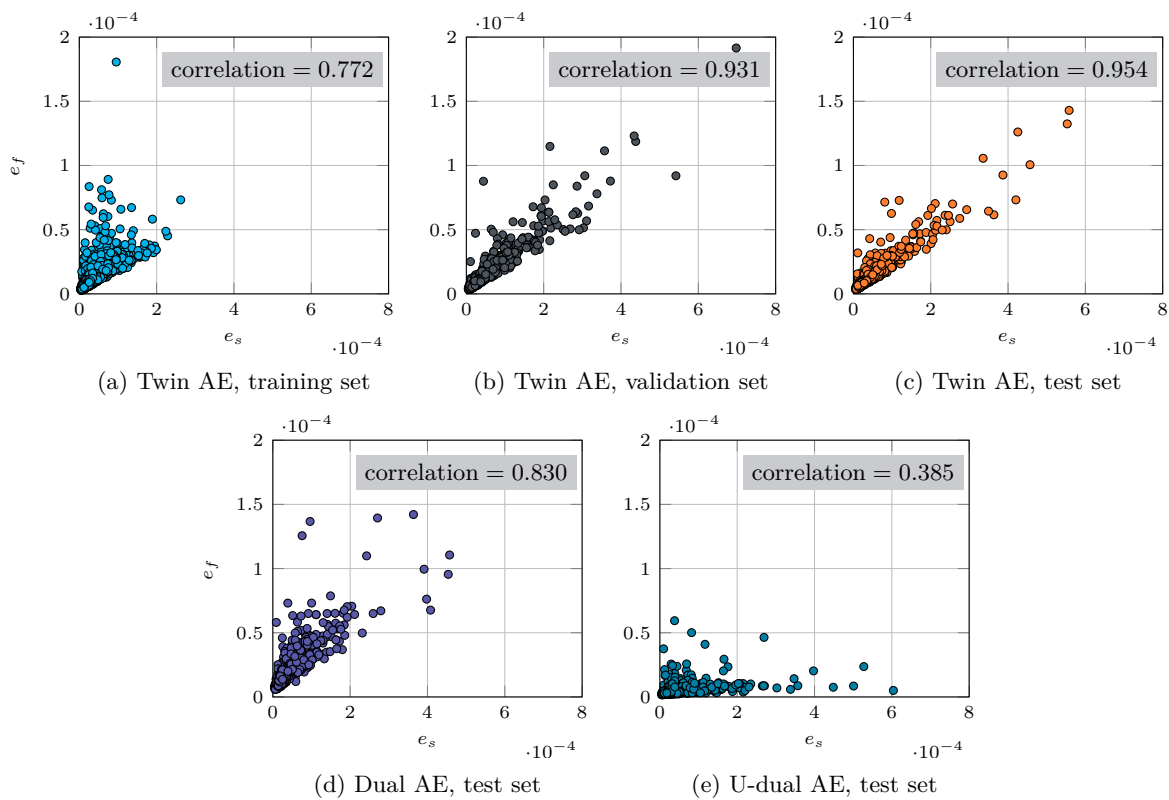
\begin{figure}
\centering

\begin{subfigure}[b]{.3\textwidth}
\centering
	\begin{tikzpicture}[trim axis left, trim axis right]
		\begin{axis}[	scale=0.6,transform shape, 
					label style={font=\scriptsize}, tick label style={font=\scriptsize}, legend style={font=\scriptsize},
					ymin=0, ymax=0.0002, xmin=0, xmax=0.0008,
					xlabel=$e_s$,ylabel=$e_f$,
					scaled y ticks=base 10:4,
					legend pos=south east,
					legend cell align={left},
					clip=true, grid=both
					]
			\addplot+[only marks,mark=*,mark options={draw=black,fill=myblue1},mark size=1.5pt] table[x index=1,y index=2] {fig/mse_train.csv};
			\node[fill=mygray4] at (0.0005,0.000175) {\small $\text{correlation}=0.772$};
		\end{axis}	
	\end{tikzpicture}
	
	\caption{Twin AE, training set}
	\label{fig:train_set}
\end{subfigure} \quad
\begin{subfigure}[b]{.3\textwidth}
\centering
	\begin{tikzpicture}[trim axis left, trim axis right]
		\begin{axis}[	scale=0.6,transform shape, 
					label style={font=\scriptsize}, tick label style={font=\scriptsize}, legend style={font=\scriptsize},
					ymin=0, ymax=0.0002, xmin=0, xmax=0.0008,
					xlabel=$e_s$,
					scaled y ticks=base 10:4,
					legend pos=south east,
					legend cell align={left},
					clip=true, grid=both
					]
			\addplot+[only marks,mark=*,mark options={draw=black,fill=mygray1},mark size=1.5pt] table[x index=1,y index=2] {fig/mse_valid.csv};
			\node[fill=mygray4] at (0.0005,0.000175) {\small  $\text{correlation}=0.931$};
		\end{axis}	
	\end{tikzpicture}
	
	\caption{Twin AE, validation set}
	\label{fig:validation_set}
\end{subfigure} \quad
\begin{subfigure}[b]{.3\textwidth}
\centering
	\begin{tikzpicture}[trim axis left, trim axis right]
		\begin{axis}[	scale=0.6,transform shape, 
					label style={font=\scriptsize}, tick label style={font=\scriptsize}, legend style={font=\scriptsize},
					ymin=0, ymax=0.0002, xmin=0, xmax=0.0008,
					xlabel=$e_s$,
					scaled y ticks=base 10:4,
					legend pos=south east,
					legend cell align={left},
					clip=true, grid=both
					]
			\addplot+[only marks,mark=*,mark options={draw=black,fill=myorange1},mark size=1.5pt] table[x index=1,y index=2] {fig/mse_test.csv};
			\node[fill=mygray4] at (0.0005,0.000175) {\small  $\text{correlation}=0.954$};
		\end{axis}	
	\end{tikzpicture}
	
	\caption{Twin AE, test set}
	\label{fig:test_set}
\end{subfigure} 

\medskip

\begin{subfigure}[b]{.3\textwidth}
\centering
	\begin{tikzpicture}[trim axis left, trim axis right]
		\begin{axis}[	scale=0.6,transform shape, 
					label style={font=\scriptsize}, tick label style={font=\scriptsize}, legend style={font=\scriptsize},
					ymin=0, ymax=0.0002, xmin=0, xmax=0.0008,
					xlabel=$e_s$,ylabel=$e_f$,
					scaled y ticks=base 10:4,
					legend pos=south east,
					legend cell align={left},
					clip=true, grid=both
					]
			\addplot+[only marks,mark=*,mark options={draw=black,fill=mypurple1},mark size=1.5pt] table[x index=1,y index=2] {fig/mse_test_dual.csv};
			\node[fill=mygray4] at (0.0005,0.000175) {\small  $\text{correlation}=0.830$};
		\end{axis}	
	\end{tikzpicture}
	
	\caption{Dual AE, test set}
	\label{fig:dual_ae_set}
\end{subfigure} \quad
\begin{subfigure}[b]{.3\textwidth}
\centering
	\begin{tikzpicture}[trim axis left, trim axis right]
		\begin{axis}[	scale=0.6,transform shape, 
					label style={font=\scriptsize}, tick label style={font=\scriptsize}, legend style={font=\scriptsize},
					ymin=0, ymax=0.0002, xmin=0, xmax=0.0008,
					xlabel=$e_s$,
					scaled y ticks=base 10:4,
					legend pos=south east,
					legend cell align={left},
					clip=true, grid=both
					]
			\addplot+[only marks,mark=*,mark options={draw=black,fill=mybluegray1},mark size=1.5pt] table[x index=1,y index=2] {fig/mse_test_udual.csv};
			\node[fill=mygray4] at (0.0005,0.000175) {\small  $\text{correlation}=0.385$};
		\end{axis}	
	\end{tikzpicture}
	
	\caption{U-dual AE, test set}
	\label{fig:udual_ae_set}
\end{subfigure} 

\caption{\textbf{Comparison of scatter plots of $e_f$ versus $e_s$ for different sets and architectures}. Top row: the correlation levels obtained with the twin AE on the training, validation and test sets are respectively 0.772, 0.931 and 0.954. Bottom row: dual AE and U-dual AE architectures show weaker correlation levels on the test set, with respective levels of 0.830 and 0.385.}
\label{fig:correlation}
\end{figure}

The scatter plots obtained with the twin-AE on the training, the validation and the test sets are respectively shown in figures \ref{fig:train_set}, \ref{fig:validation_set} and \ref{fig:test_set}. Associated correlation levels are 0.772, 0.931 and 0.954, meaning that strong linear relations are observed on the validation and test set. Regarding the training set, the weaker correlation is interpreted as a consequence of the slight overfitting observed during training (see figure \ref{fig:training_history}). In comparison, the obtained correlation level of the dual AE on the test set is 0.830, which is significantly weaker than that of the twin AE, thus proving the interest of the skip connections between the two decoder branches (see figure \ref{fig:dual_ae_set}). The twin-AE also has slightly lower relative errors than its dual-AE counterpart (3.92\%, 3.57\% and 3.55\% respectively for $u$, $v$, $p$, against 4.30\%, 4.05\% and 4.00\%). Finally, the U-dual-AE architecture exhibits almost no correlation between $e_s$ and $e_f$, with a computed correlation level of 0.385 (see figure \ref{fig:udual_ae_set}). Adversely, the relative error levels of the U-dual-AE are significantly lower (3.01\%, 1.94\% and 1.94\%), which is in line with results from the literature \cite{Chen2019}.

\subsection{Trust level based on input reconstruction}
\label{section:results_trust_level}

This section presents the application of the trust level methods detailed in section \ref{section:trust}. Again, the underlying concept is to take advantage of the strong correlation between shape and flow reconstruction error levels (see section \ref{section:correlation_levels}) to propose an uncertainty estimation (either qualitative or quantitative) along with the flow prediction.

\paragraph{Qualitative method}

\begin{figure}
\centering
\begin{subfigure}[b]{.45\linewidth}
	\centering	
	\begin{tikzpicture}[trim axis left, trim axis right]
		\begin{axis}[	scale=0.8,transform shape,
					label style={font=\scriptsize}, tick label style={font=\scriptsize}, legend style={font=\scriptsize},
					ymin=0, ymax=0.1, xmin=1e-4, xmax=3e-4,
					xlabel=$e_s^*$,ylabel=False classification rate,
					ytick={0,0.02,0.04,0.06,0.08,0.1},
					yticklabels={0,0.02,0.04,0.06,0.08,0.1},
					legend pos=north east,
					legend cell align={left},
					scaled x ticks=base 10:4,
					clip=true, grid=both
					]
			\legend{Training set, Validation set, Test set}
			\addplot[mark=none,smooth,thick,myblue1] table[x index=0,y index=1] {fig/qualitative_min.csv};
			\addplot[mark=none,smooth,thick,mygray1] table[x index=0,y index=2] {fig/qualitative_min.csv};
		\end{axis}	
	\end{tikzpicture}
	\caption{False classification rate as a function of $e_s^*$ for $e_f^* = \num{5e-5}$.}
	\label{fig:qualitative_min}
\end{subfigure} \quad
\begin{subfigure}[b]{.45\linewidth}
	\centering
	\begin{tikzpicture}[trim axis left, trim axis right]
		\begin{axis}[	scale=0.8,transform shape,
					label style={font=\scriptsize}, tick label style={font=\scriptsize}, legend style={font=\scriptsize},
					ymin=0, ymax=0.1, xmin=0, xmax=1e-4,
					xlabel=$e_f^*$,ylabel=False classification rate,
					ytick={0,0.02,0.04,0.06,0.08,0.1},
					yticklabels={0,0.02,0.04,0.06,0.08,0.1},
					legend pos=north east,
					legend cell align={left},
					scaled x ticks=base 10:4,
					clip=true, grid=both
					]
			\legend{Training set, Validation set, Test set}
			\addplot[mark=*,thick,myblue1] table[x index=0,y index=1] {fig/mistake_ef.csv};
			\addplot[mark=*,thick,mygray1] table[x index=0,y index=2] {fig/mistake_ef.csv};
			\addplot[mark=*,thick,myorange1] table[x index=0,y index=3] {fig/mistake_ef.csv};
		\end{axis}	
	\end{tikzpicture}
	\caption{False classification rate achieved on different subsets for different $e_f^*$ values.}
	\label{fig:mistake_ef}
\end{subfigure}

\caption{\textbf{Trust level identification based on the $e_s$ indicator.} (Left) The optimal threshold is obtained by minimizing the mistaken classification rate on the training and the validation sets. (Right) The mistake rate rises significantly on all three subsets when decreasing the threshold $e_f^*$ value. For the optimal $e_s^*$ value, the mistake rate on the validation and test sets is approximately 1\%.}
\label{fig:threshold}
\end{figure}
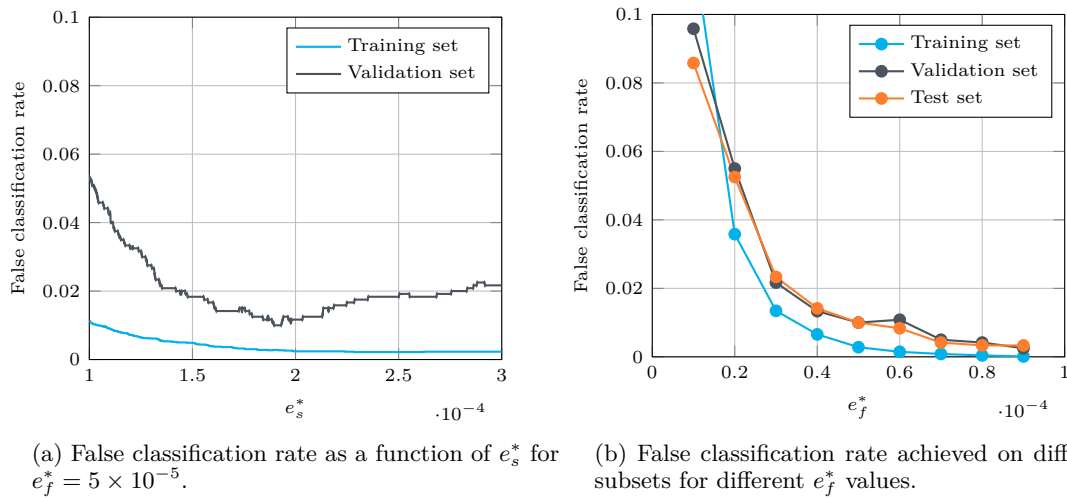

\begin{figure}
\centering
\begin{subfigure}[b]{.45\linewidth}
	\centering	
	\begin{tikzpicture}[trim axis left, trim axis right]
		\begin{axis}[	scale=0.8,transform shape, 
					label style={font=\scriptsize}, tick label style={font=\scriptsize}, legend style={font=\scriptsize},
					ymin=0, ymax=0.0002, xmin=0, xmax=0.0008,
					xlabel=$e_s$,ylabel=$e_f$,
					scaled y ticks=base 10:4,
					legend pos=south east,
					legend cell align={left},
					clip=true
					]
			\legend{Test set}
			\addplot+[only marks,mark=*,mark options={draw=black,fill=myorange1},mark size=1.5pt] table[x index=1,y index=2] {fig/mse_test.csv};
			\draw[mybluegray1, very thick, dash pattern=on 2pt] (axis cs:0.00019,\pgfkeysvalueof{/pgfplots/ymax}) -- (axis cs:0.00019,\pgfkeysvalueof{/pgfplots/ymin});
			\draw[mybluegray1, very thick, dash pattern=on 2pt] (axis cs:\pgfkeysvalueof{/pgfplots/xmax},0.00005) -- (axis cs:\pgfkeysvalueof{/pgfplots/xmin},0.00005);
			\fill[mybluegray3, opacity=0.5] (axis cs:\pgfkeysvalueof{/pgfplots/xmin},\pgfkeysvalueof{/pgfplots/ymin}) rectangle (axis cs:0.00019,0.00005);
		\end{axis}	
	\end{tikzpicture}
	\caption{Qualitative method: accepted prediction area when choosing $e_f^* = \num{5e-5}$ and selecting $e_s^*$ following (\ref{eq:qualitative}).}
	\label{fig:qualitative_accept}
\end{subfigure} \quad
\begin{subfigure}[b]{.45\linewidth}
	\centering
	\begin{tikzpicture}[trim axis left, trim axis right]
		\begin{axis}[	scale=0.8,transform shape, 
					label style={font=\scriptsize}, tick label style={font=\scriptsize}, legend style={font=\scriptsize},
					ymin=0,ymax=0.0002,
					xmin=0,xmax=0.0008,
					xlabel=$e_s$,ylabel=$e_f$,
					scaled y ticks=base 10:4,
					legend pos=south east,
					legend cell align={left}
					]
			\legend{Test set}
			\addplot+[only marks,mark=*,mark options={draw=black,fill=myorange1},mark size=1.5pt] table[x index=1,y index=2] {fig/mse_test.csv};
			\addplot+[mybluegray1, very thick,name path=avg,mark=none] coordinates {(0,0) (0.0008,0.0002079738)};
			
			\addplot[mybluegray3, very thick,name path=std21] coordinates {(0,0) (0.0008,0.0002926466)};
			\addplot[mybluegray3, very thick,name path=std22] coordinates {(0,0) (0.0008,0.000123301)};
			\addplot[mybluegray4,opacity=0.5,forget plot] fill between[of=std21 and std22];
		\end{axis}	
	\end{tikzpicture}
	\caption{Quantitative method: regression line with its $1 \sigma$ confidence interval obtained by solving problem (\ref{eq:quantitative}).}
	\label{fig:quantitative_accept}
\end{subfigure}
\caption{\textbf{Representations of the qualitative and quantitative methods} along with the test set scatter plot.}
\label{fig:err_interval}
\end{figure}
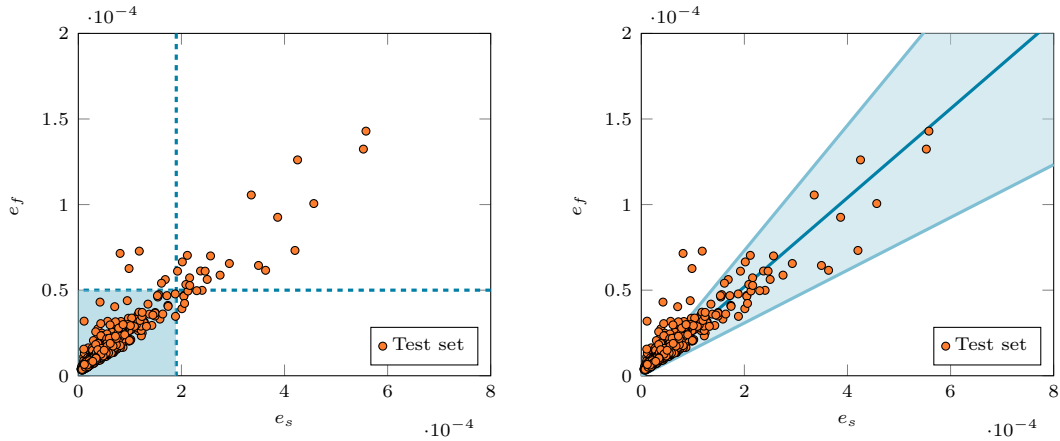

A threshold mean squared error tolerance for the flow reconstruction is provided by the user, and is here chosen to be $e_f^{*} = \num{5e-5}$. The corresponding threshold shape reconstruction error $e_s^*$ is obtained by solving the minimization problem (\ref{eq:qualitative}). As the dataset size is relatively limited, the problem is solved by an exhaustive search, \red{the results of which are shown} in figure \ref{fig:qualitative_min}. One observes that choosing $e_s^* = \num{1.9e-4}$ minimizes the risks of accepting bad predictions and rejecting good predictions. When testing the procedure on the elements of the validation and testing sets, it is observed that the mistake rate is close to 1\% in both cases. In figure \ref{fig:mistake_ef}, the false classification rate on the  three subsets is plotted as a function of $e_f^*$, showing that stricter $e_f^*$ choices inevitably lead to worse performances with this method. The area of accepted predictions is plotted in figure \ref{fig:qualitative_accept}, along with a representation of the elements of the test set.

\paragraph{Quantitative method}

A gradient descent algorithm is used to minimize the negative log-likelihood problem (\ref{eq:quantitative}) over the training set, in order to obtain the optimal parameter set $(a^*,b^*,c^*)$. With an initial value $(a_0, b_0, c_0) = (0.1, 0, 0.1)$, the algorithm converges after 6 iterations (see figure  \ref{fig:NLL}). The optimal parameters retained are $(a^*,b^*,c^*) = (0.257157, \num{2.24820e-6}, 0.105841)$, which minimize the \red{negative log likelihood} over the validation set. Hence, $e_f$ can be estimated from $e_s$ as:

\begin{equation}
\label{eq:estimate}
	e_f = 0.257157 \, e_s + \num{2.24820e-6} + \mathcal{N}(0, (0.105841 \, e_s)^2).
\end{equation}

As shown in figure \ref{fig:quantitative_accept}, the regression line and its $1 \sigma$ confidence interval matches well with the distribution of the test set, with only a handful of $(e_f, e_s)$ couples falling outside of the range. The fact that the confidence interval widens with larger values of $e_s$ translates the increasing scarcity of samples in the test set when $e_s$ rises. For the majority of predictions, though, it provides a good grasp of the flow prediction quality. The $1 \sigma$ (68\% probability) and $2 \sigma$ ($95\%$ probability) confidence intervals for sampled $e_s$ values are provided in table \ref{tab:err_interval}.

\begin{figure}
\centering
\begin{tikzpicture}[trim axis left, trim axis right]
	\begin{axis}[	scale=0.8,transform shape,
				label style={font=\scriptsize}, tick label style={font=\scriptsize}, legend style={font=\scriptsize},
				ymin=-11.6, ymax=-10.6, xmin=0, xmax=14,
				xlabel=BFGS iterations,ylabel=Negative log-likelihood,
				legend pos=north east,
				legend cell align={left},
				clip=true, grid=both
				]
		\legend{Training set, Validation set}
		\addplot[mark=*,thick,myblue1] table[x index=0,y index=1] {fig/nll.csv};
		\addplot[mark=*,thick,mygray1] table[x index=0,y index=2] {fig/nll.csv};
	\end{axis}	
\end{tikzpicture}
\caption{\textbf{Minimization of the negative log-likelihood problem (\ref{eq:quantitative}) using the BFGS algorithm.}}
\label{fig:NLL}
\end{figure}
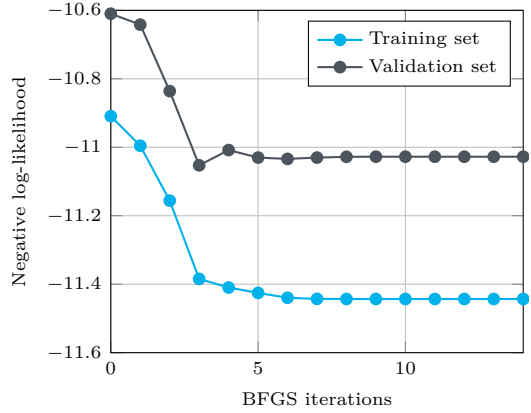

\begin{table}
\footnotesize
\caption{\textbf{Estimating $e_f$ for given $e_s$ values.}}
\label{tab:err_interval}
\centering
\begin{tabular}{cccc}
\toprule
&$e_s$      			&$e_f$ ($1\sigma$ interval)               		&$e_f$ ($2\sigma$ interval)				\\\midrule
&$1\times 10^{-5}$	 	&$[3.76\times 10^{-6},5.88\times 10^{-6}]$ 	&$[2.70\times 10^{-6},6.94\times 10^{-6}]$   	\\\midrule
&$2\times 10^{-5}$	 	&$[5.27\times 10^{-6},9.51\times 10^{-6}]$ 	&$[3.16\times 10^{-6},1.16\times 10^{-5}]$   	\\\midrule
&$4\times 10^{-5}$	 	&$[8.30\times 10^{-6},1.68\times 10^{-5}]$ 	&$[4.07\times 10^{-6},2.10\times 10^{-5}]$      	\\\midrule
&$8\times 10^{-5}$	 	&$[1.44\times 10^{-5},3.13\times 10^{-5}]$ 	&$[5.89\times 10^{-6},3.98\times 10^{-5}]$ 	\\\midrule
&$1.6\times 10^{-4}$ 	&$[2.65\times 10^{-5},6.03\times 10^{-5}]$ 	&$[9.52\times 10^{-6},7.73\times 10^{-5}]$ 	\\\midrule
&$3.2\times 10^{-4}$ 	&$[5.07\times 10^{-5},1.18\times 10^{-4}]$ 	&$[1.68\times 10^{-5},1.52\times 10^{-4}]$ 	\\\bottomrule
\end{tabular}
\end{table}

\subsection{Flow prediction on outliers}
\label{section:prediction}

\begin{figure}
\centering

\setlength{\fboxsep}{0pt}%
\setlength{\fboxrule}{1pt}%

\begin{subfigure}[t]{.3\textwidth}
	\centering
	\fbox{\includegraphics[width=.9\linewidth]{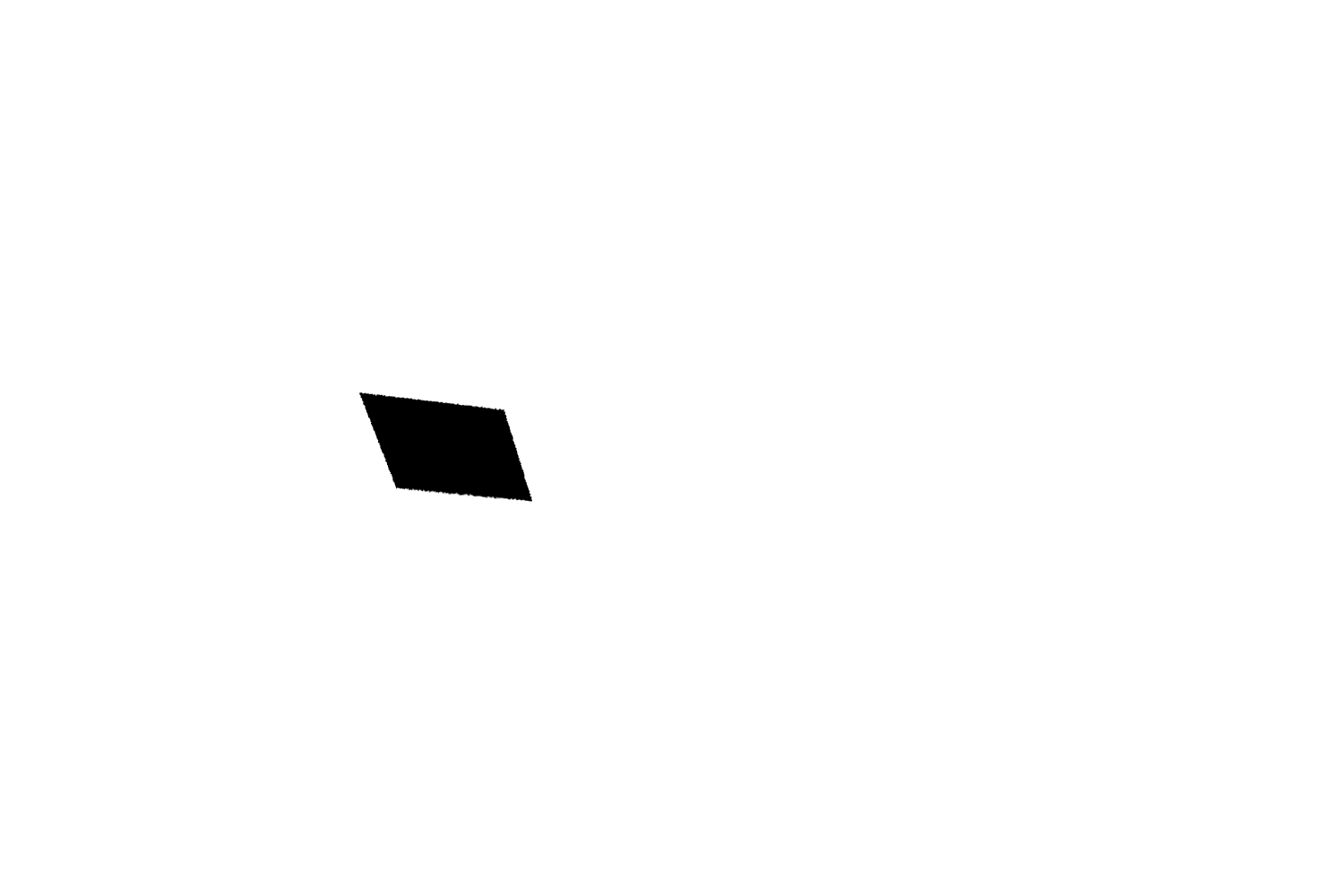}} 
	\caption{Polygon}
\end{subfigure} \quad
\begin{subfigure}[t]{.3\textwidth}
	\centering
	\fbox{\includegraphics[width=.9\linewidth]{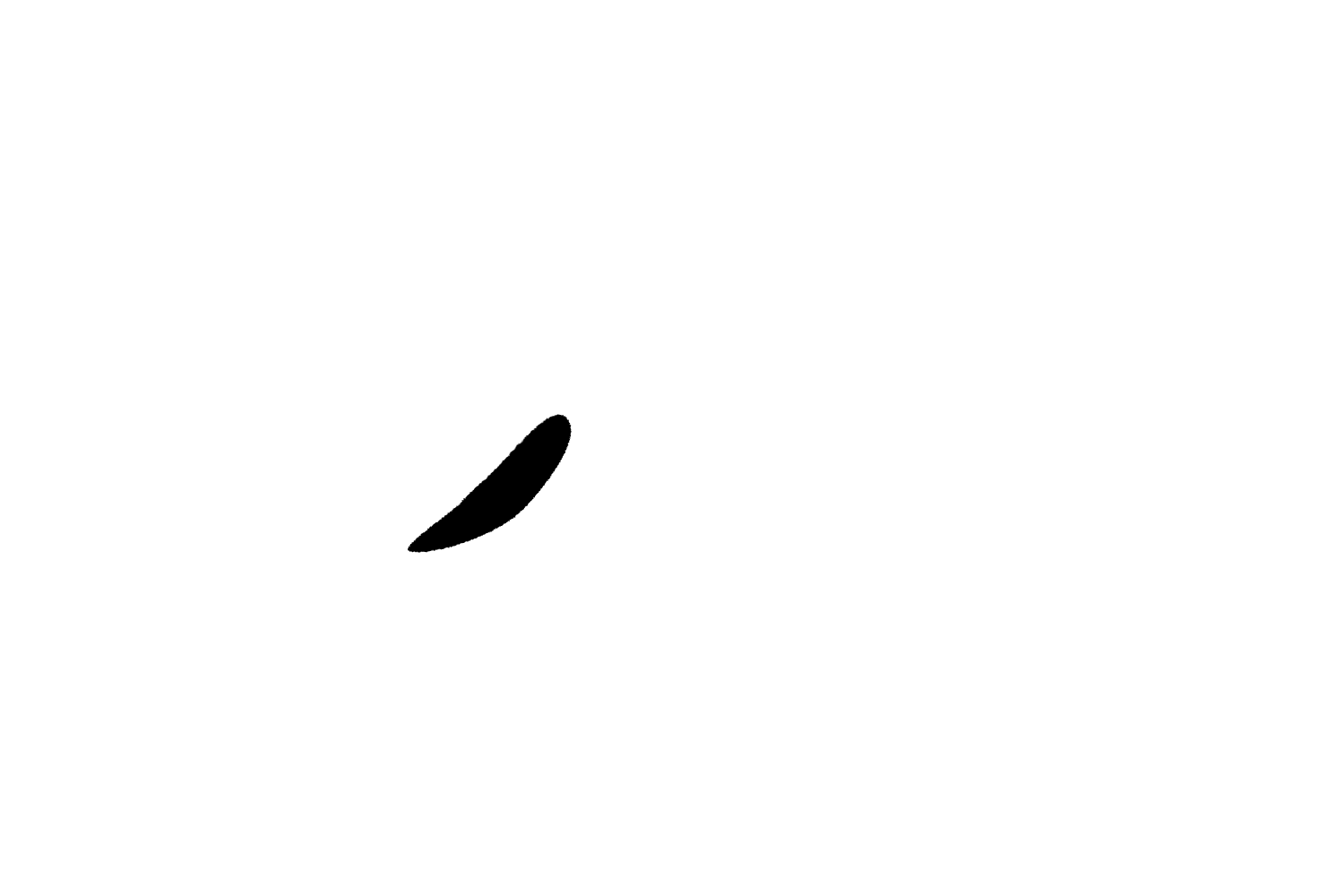}}
	\caption{Bezier shape misplaced by $(+0.5, -0.5)$}
\end{subfigure}

\medskip
\medskip

\begin{subfigure}[t]{.3\textwidth}
	\centering
	\fbox{\includegraphics[width=.9\linewidth]{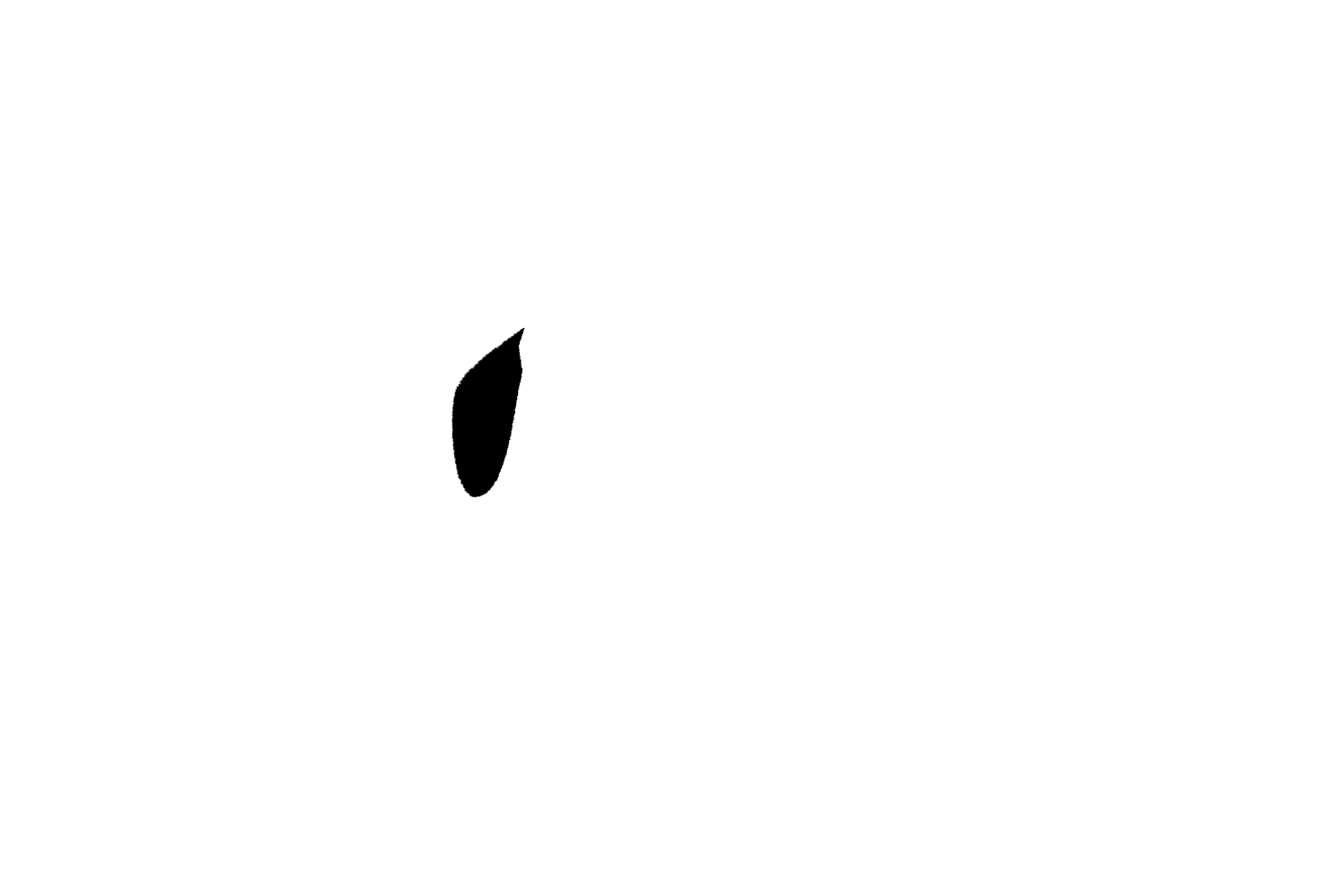}}
	\caption{Bezier shape misplaced by $(+0.5, +0.5)$}
\end{subfigure}\quad
\begin{subfigure}[t]{.3\textwidth}
	\centering
	\fbox{\includegraphics[width=.9\linewidth]{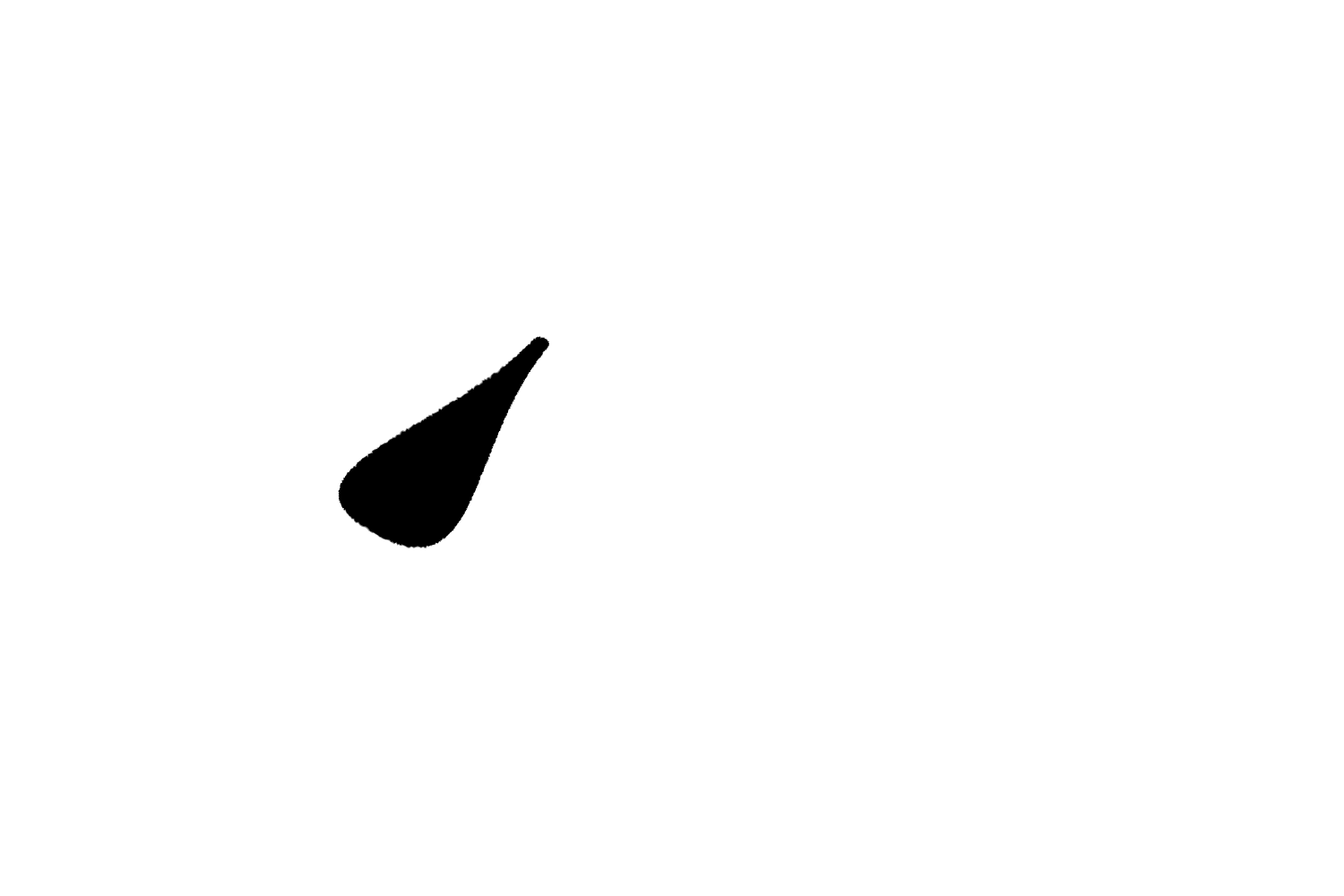}}
	\caption{Enlarged Bezier shape}
\end{subfigure}

\caption{\textbf{Outlier examples for model evaluation.} Polygons are generated with fewer sampling points and sharper edges than the dataset shapes. Misplaced and enlarged shapes have the same curve characteristics as the original data set, but are ill-positioned, or significantly larger than those of the dataset.}
\label{fig:outliers}
\end{figure}

\begin{figure}
\centering
\begin{subfigure}[b]{.3\linewidth}
	\centering
	\fbox{\includegraphics[width=.9\linewidth]{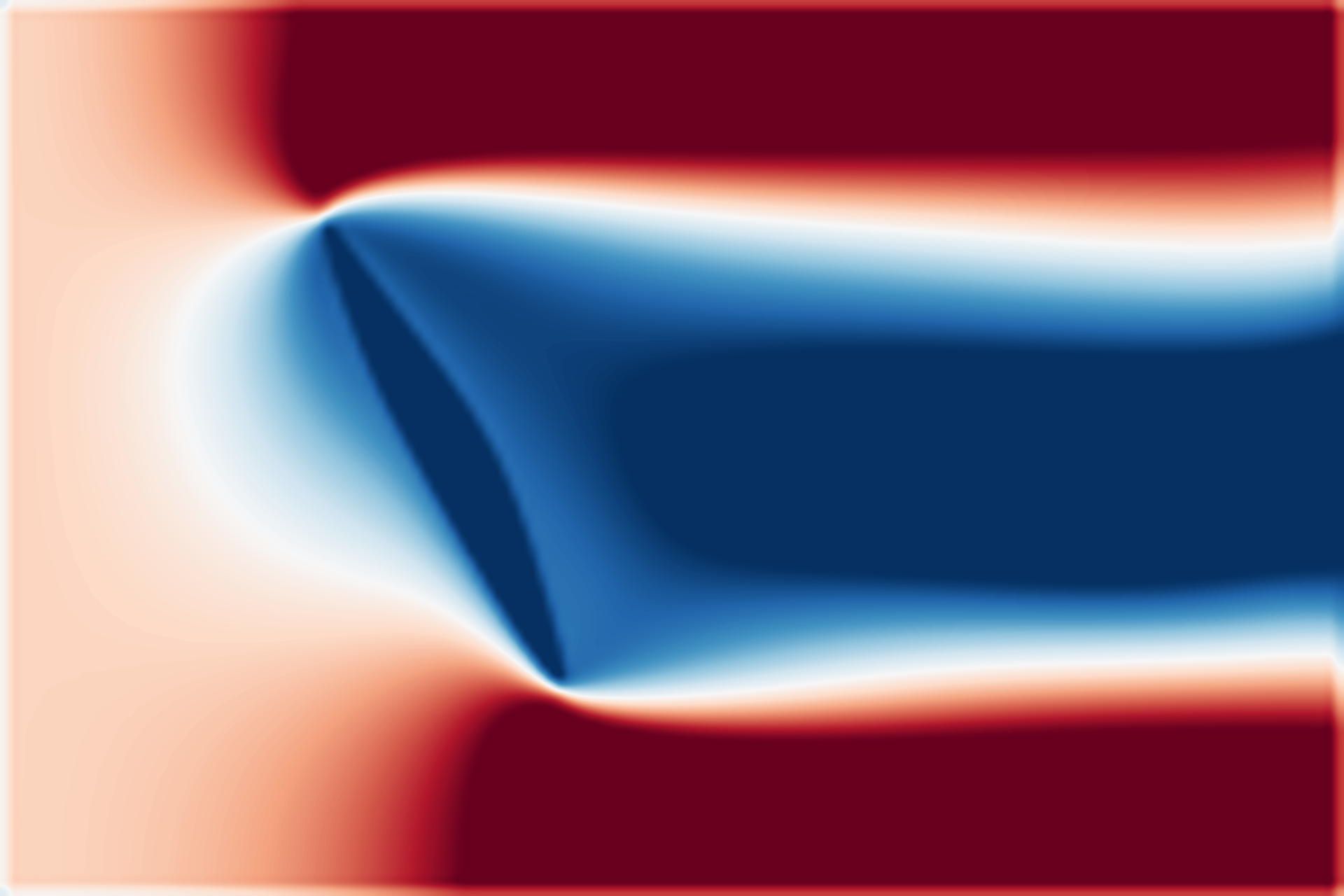}}
	\caption{$u$, reference}
	\label{fig:huge_u0_ref}
\end{subfigure} \quad
\begin{subfigure}[b]{.3\linewidth}
	\centering
	\fbox{\includegraphics[width=.9\linewidth]{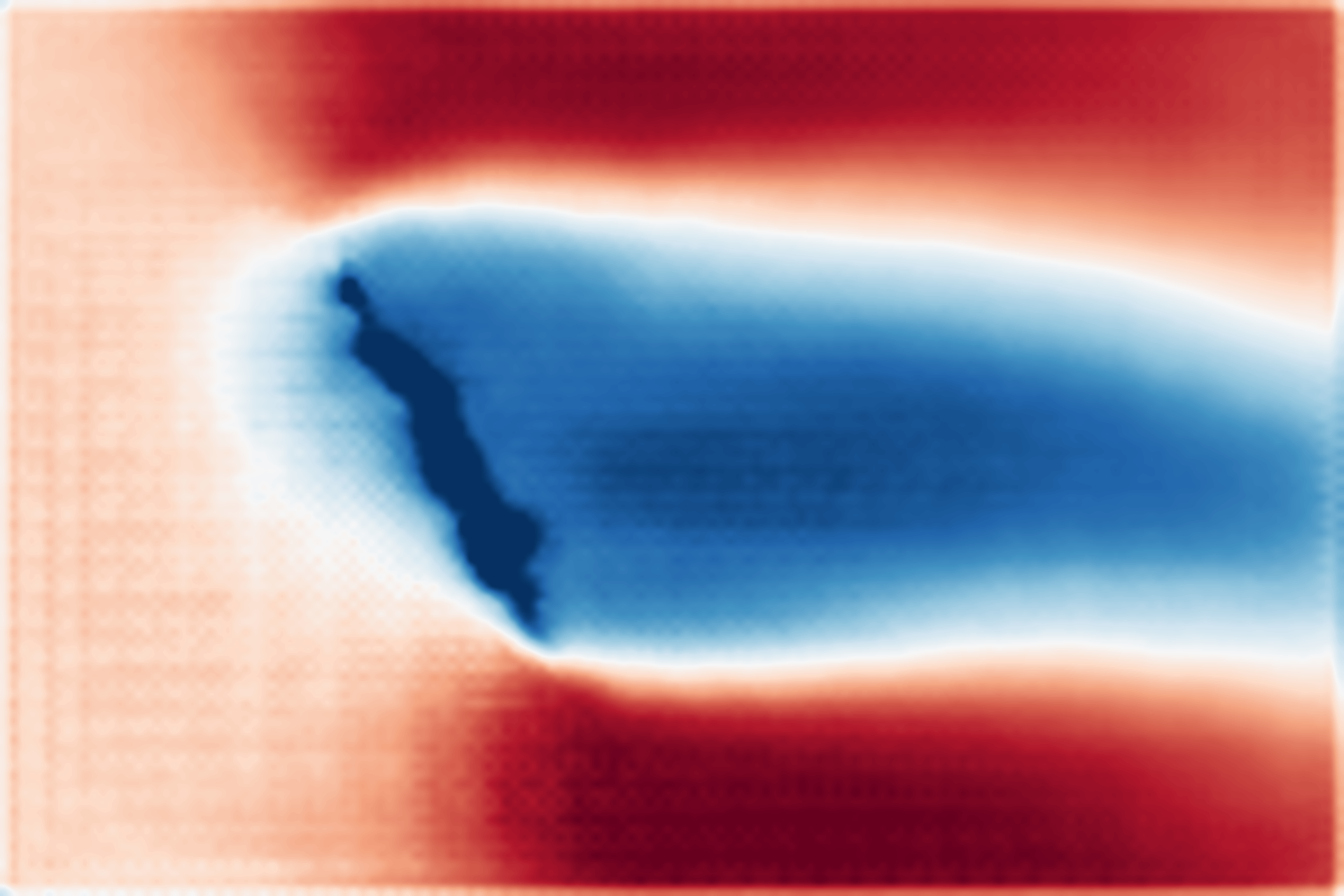}}
	\caption{$u$, predicted}
	\label{fig:huge_u0_pred}
\end{subfigure} \quad
\begin{subfigure}[b]{.3\linewidth}
	\centering
	\fbox{\includegraphics[width=.9\linewidth]{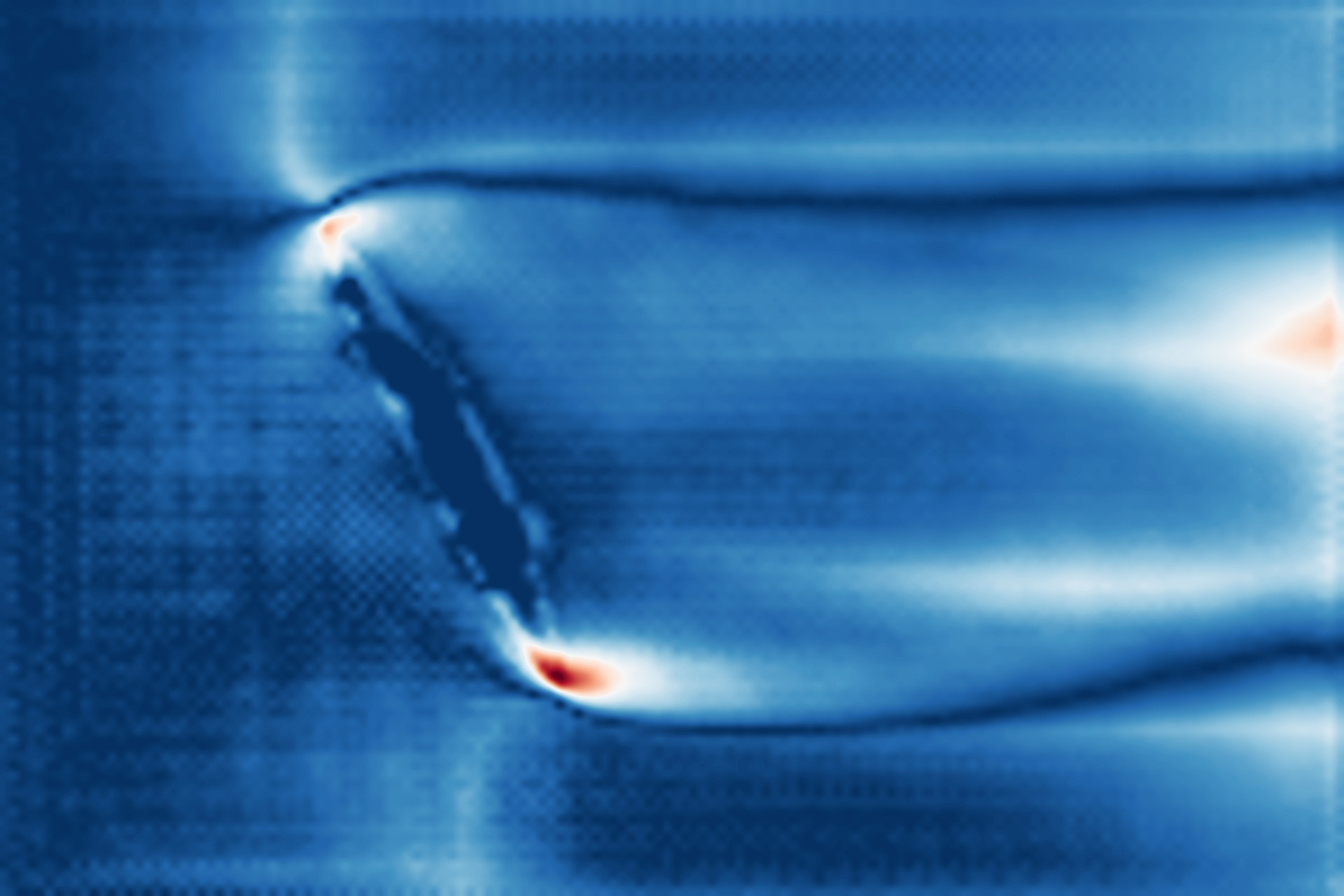}}
	\caption{$u$, absolute error}
	\label{fig:huge_u0_error}
\end{subfigure}

\medskip

\begin{subfigure}[b]{.3\linewidth}
	\centering
	\fbox{\includegraphics[width=.9\linewidth]{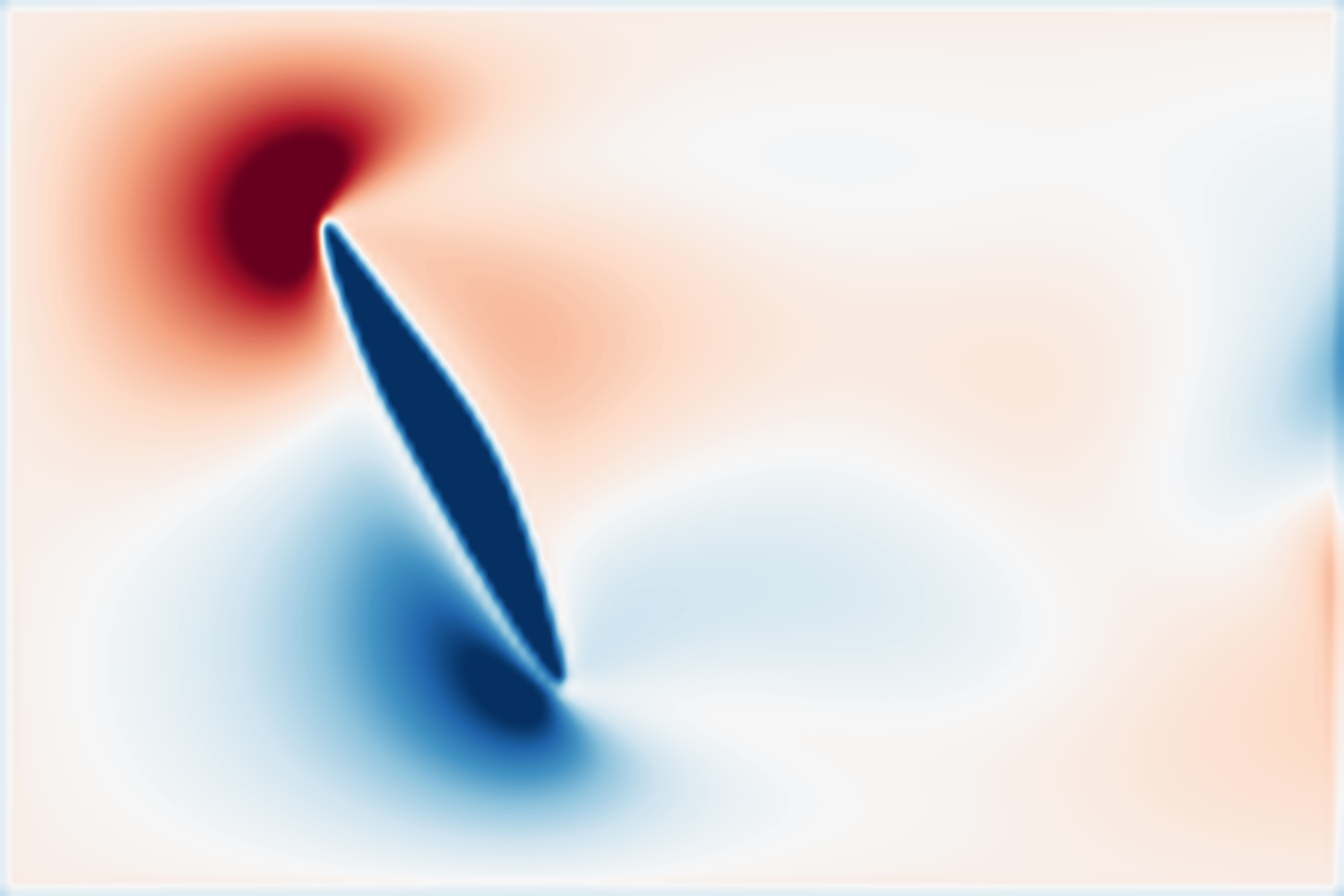}}
	\caption{$v$, reference}
	\label{fig:huge_v0_ref}
\end{subfigure} \quad
\begin{subfigure}[b]{.3\linewidth}
	\centering
	\fbox{\includegraphics[width=.9\linewidth]{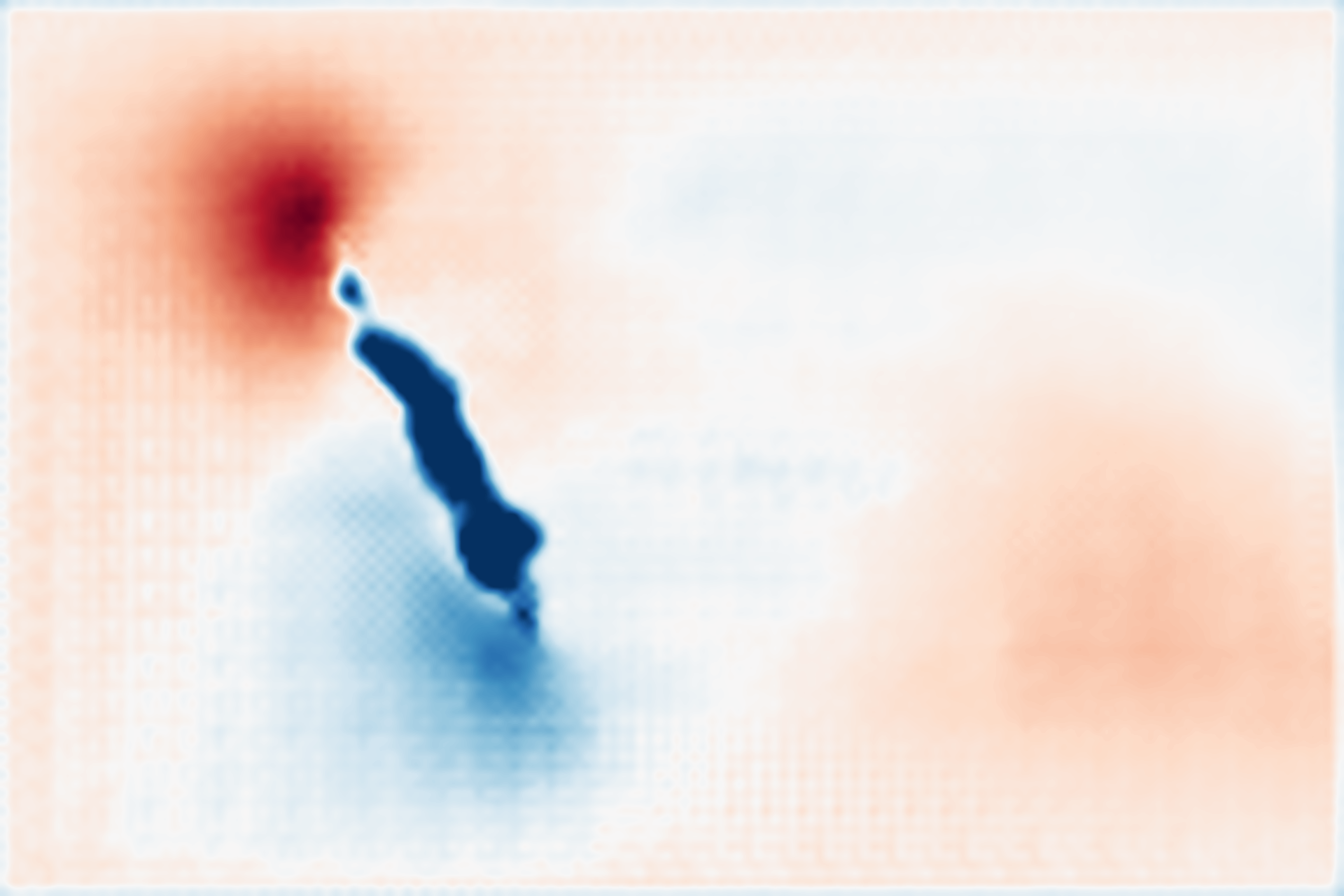}}
	\caption{$v$, predicted}
	\label{fig:huge_v0_pred}
\end{subfigure} \quad
\begin{subfigure}[b]{.3\linewidth}
	\centering
	\fbox{\includegraphics[width=.9\linewidth]{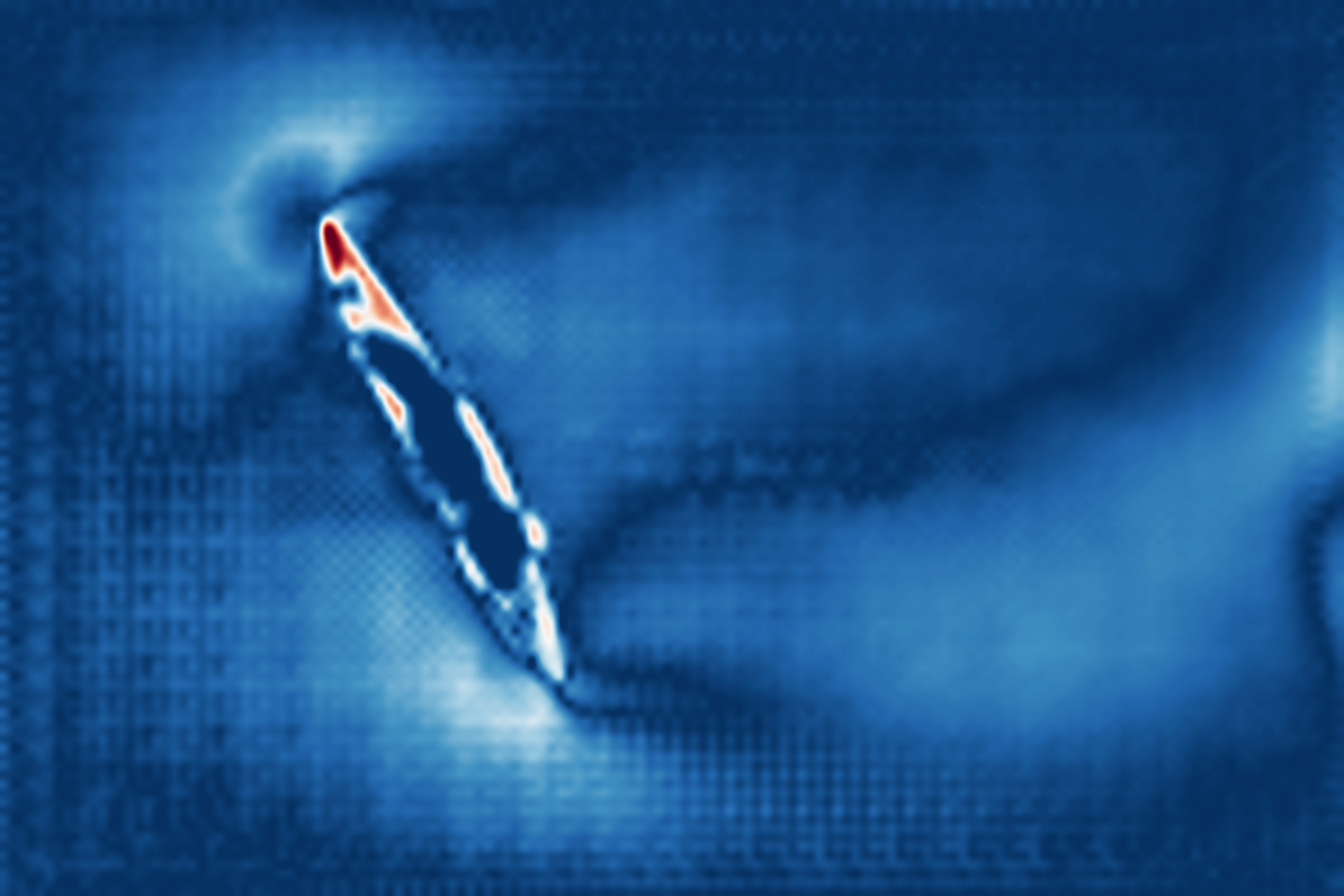}}
	\caption{$v$, absolute error}
	\label{fig:huge_v0_error}
\end{subfigure}

\medskip

\begin{subfigure}[b]{.3\linewidth}
	\centering
	\fbox{\includegraphics[width=.9\linewidth]{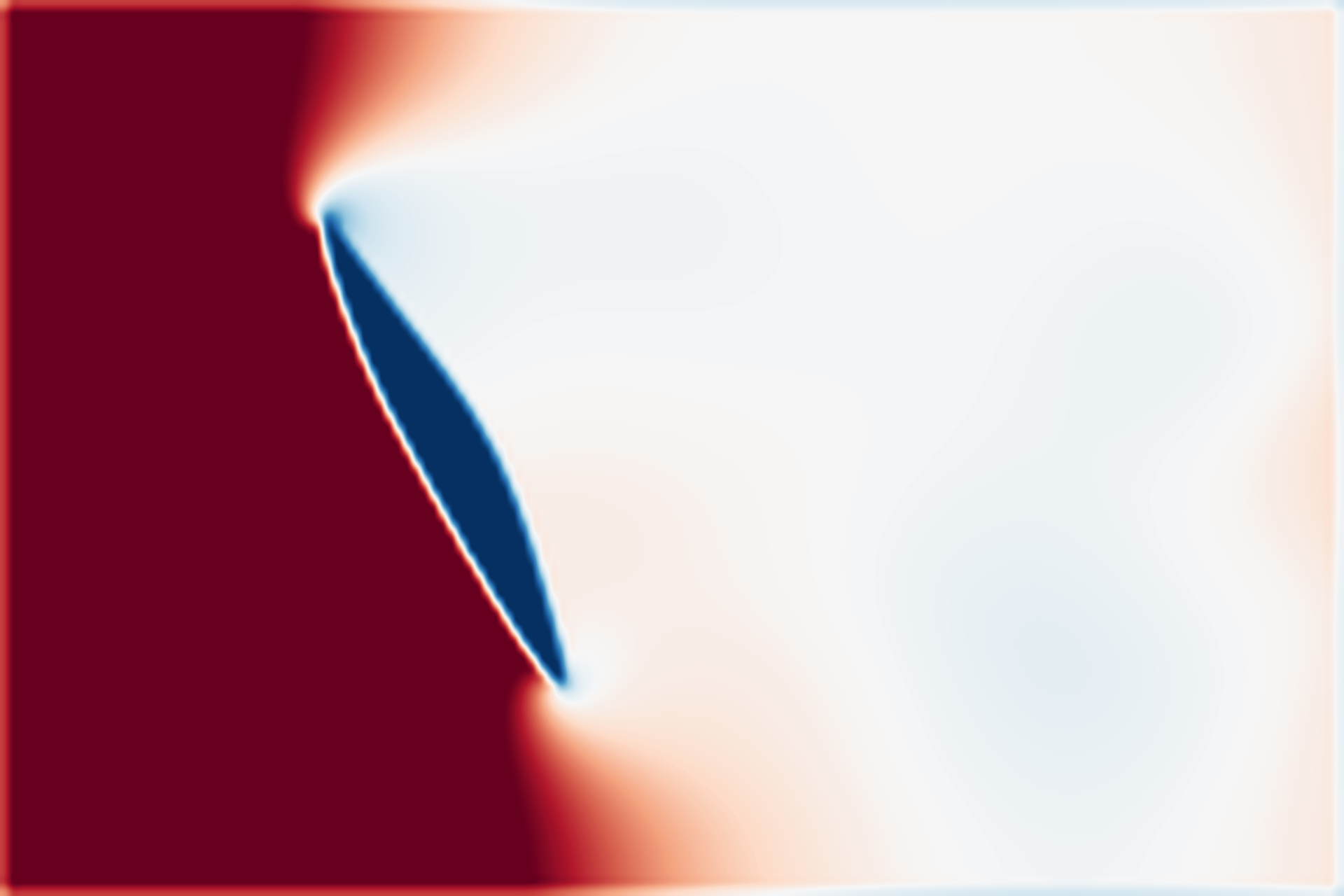}}
	\caption{$p$, reference}
	\label{fig:huge_p0_ref}
\end{subfigure} \quad
\begin{subfigure}[b]{.3\linewidth}
	\centering
	\fbox{\includegraphics[width=.9\linewidth]{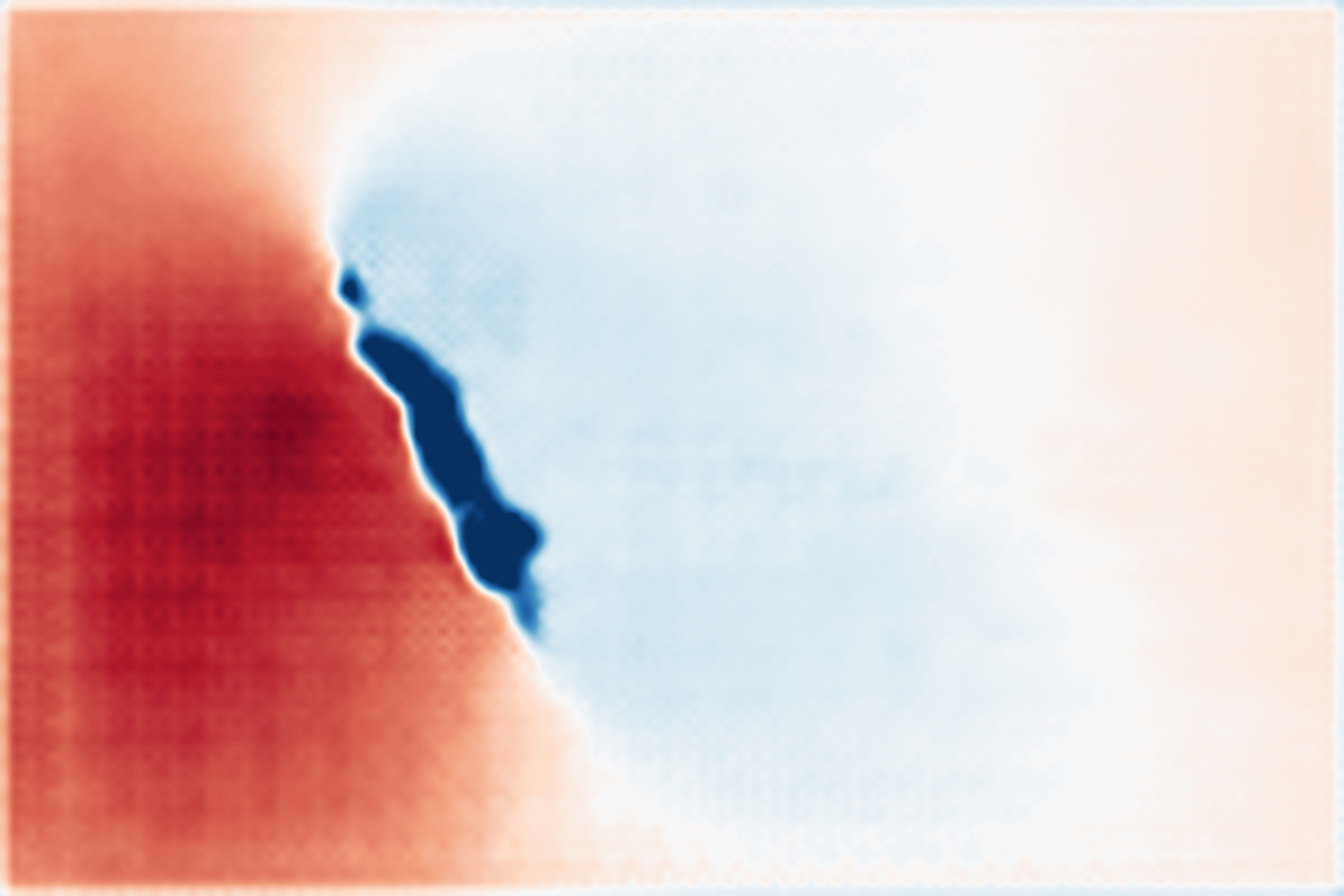}}
	\caption{$p$, predicted}
	\label{fig:huge_p0_pred}
\end{subfigure} \quad
\begin{subfigure}[b]{.3\linewidth}
	\centering
	\fbox{\includegraphics[width=.9\linewidth]{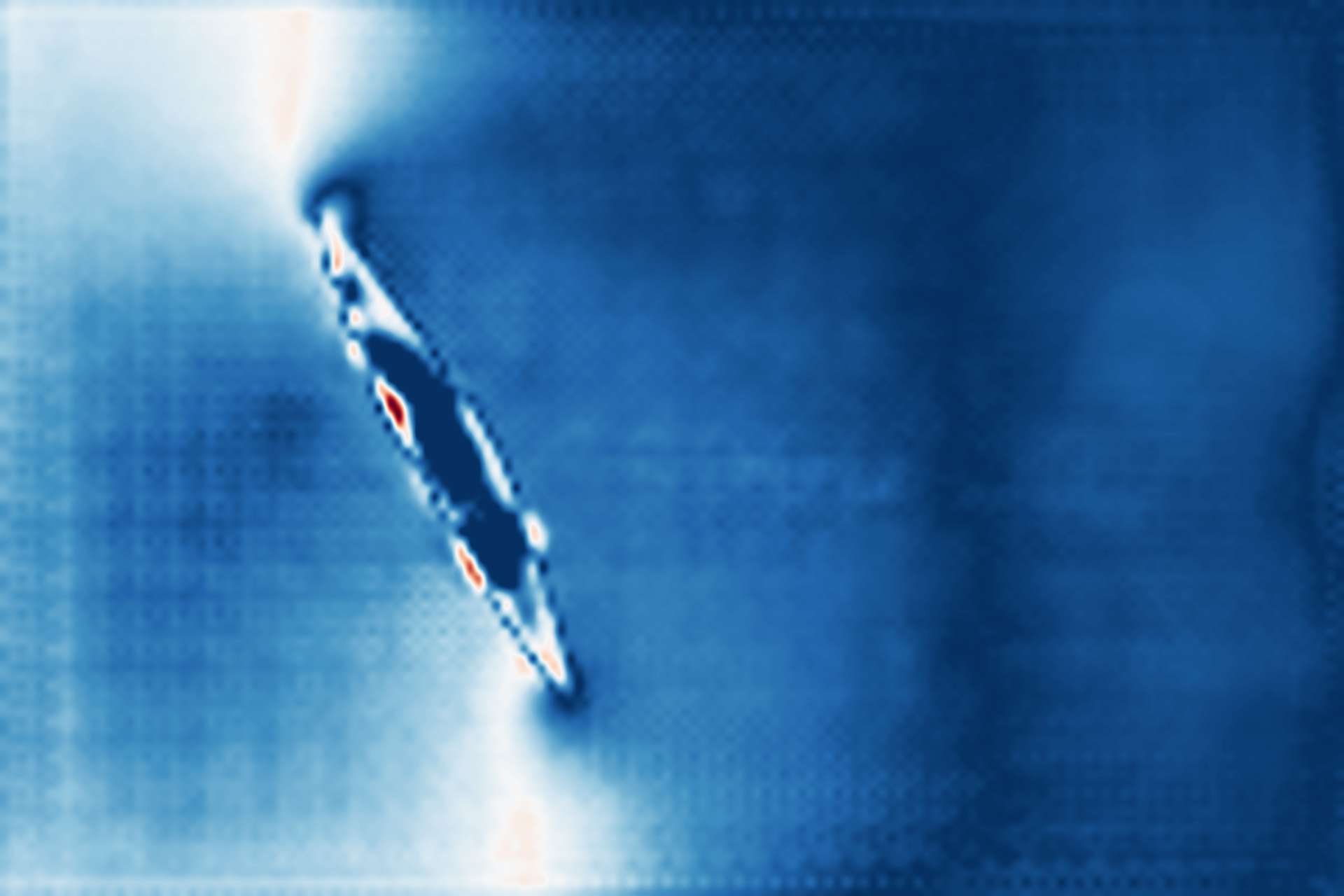}}
	\caption{$p$, absolute error}
	\label{fig:huge_p0_error}
\end{subfigure}

\medskip

\begin{subfigure}[b]{.3\linewidth}
	\centering
	\fbox{\includegraphics[width=.9\linewidth]{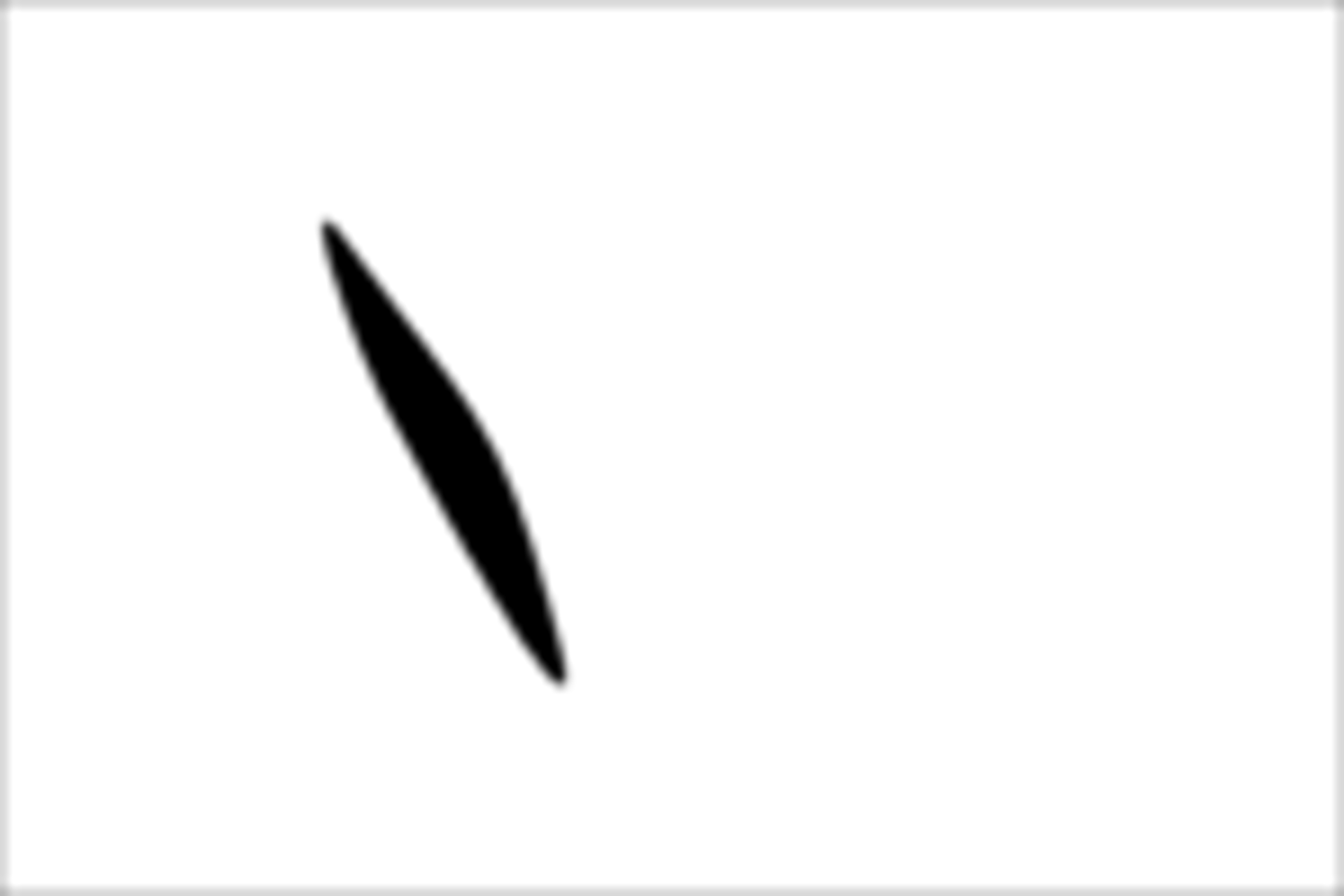}}
	\caption{Shape, reference}
	\label{fig:huge_shape_ref}
\end{subfigure} \quad
\begin{subfigure}[b]{.3\linewidth}
	\centering
	\fbox{\includegraphics[width=.9\linewidth]{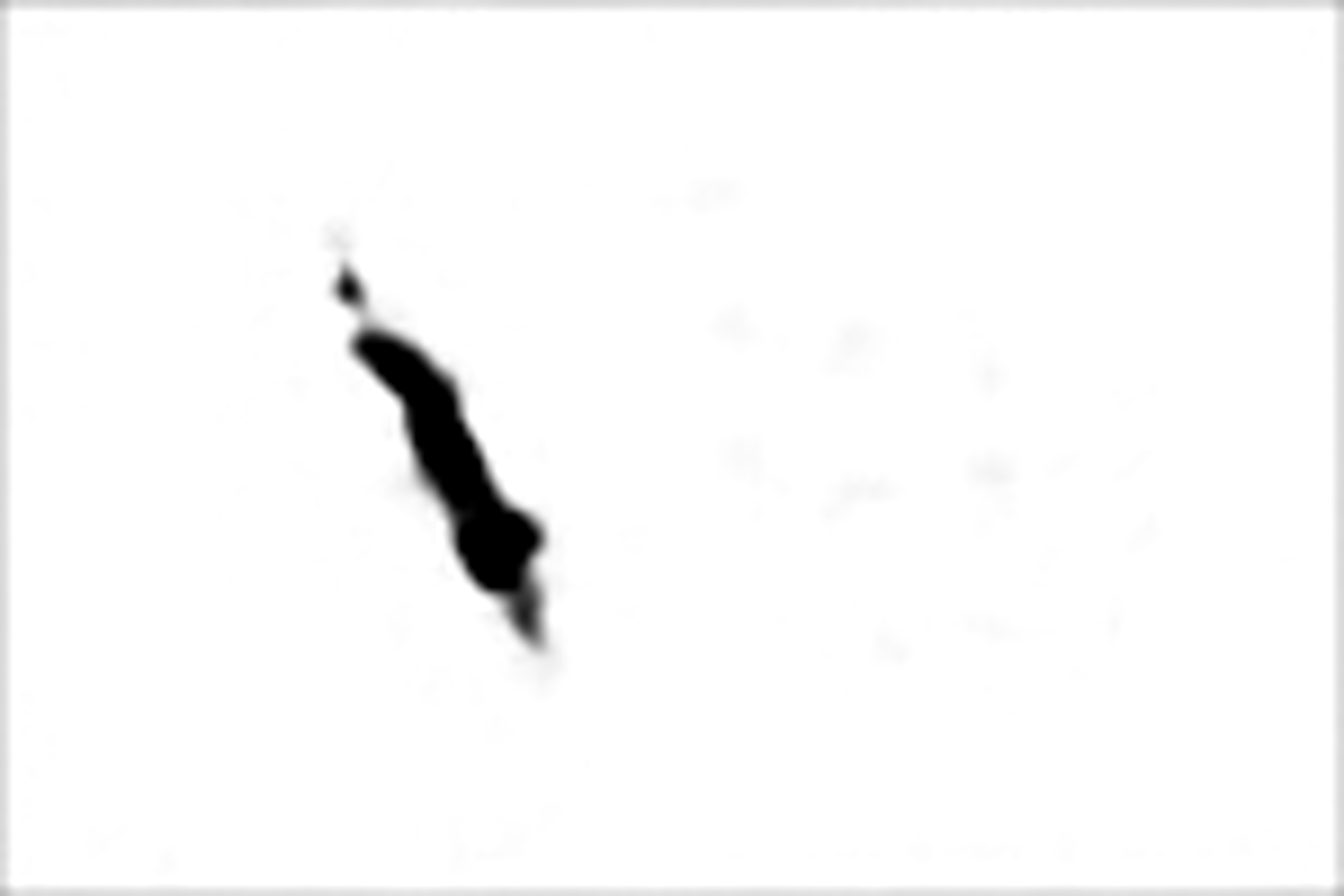}}
	\caption{Shape, predicted}
	\label{fig:huge_shape_pred}
\end{subfigure} \quad
\begin{subfigure}[b]{.3\linewidth}
	\centering
	\fbox{\includegraphics[width=.9\linewidth]{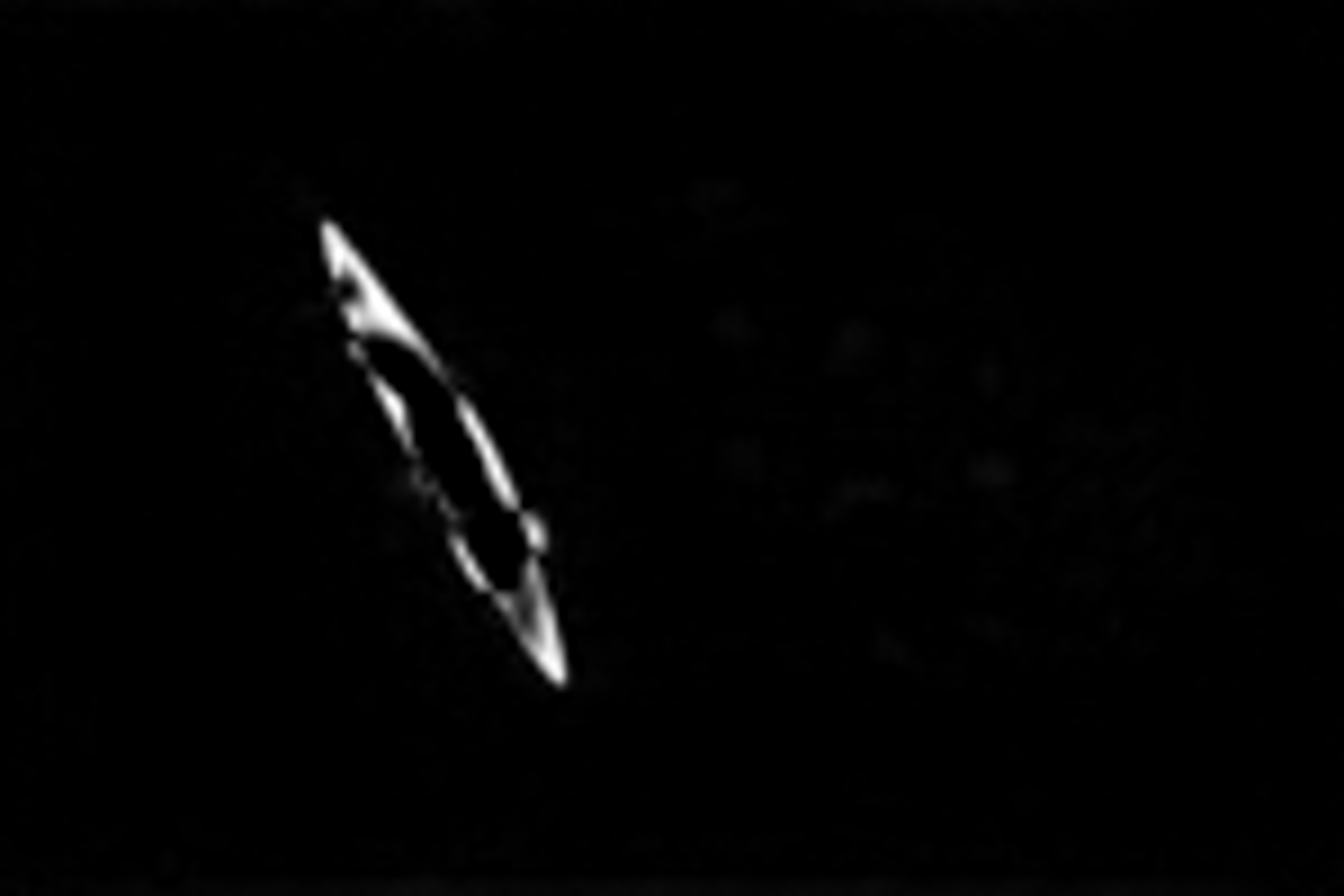}}
	\caption{Shape, absolute error}
	\label{fig:huge_shape_error}
\end{subfigure}

\caption{\textbf{Flow and shape predictions around an enlarged Bezier shape.} As this input shape is an outlier, the shape reconstruction is poor, and is associated with large prediction errors ($e_f = \num{1.316e-2}$). For the $u$, $v$ and $p$ predictions, the color scales are the same as for figure \ref{fig:dataset_example}.}
\label{fig:example_enlarged}
\end{figure}

In this section, the capabilities of the qualitative and quantitative methods to detect invalid inputs and outliers are evaluated. To do so, multiple shapes are generated that do not fit within the dataset, including polygons with sharp edges (see figure \ref{fig:outliers}), shapes included in the dataset but misplaced in the input domain (\ie moved away from the position used in the dataset), and enlarged shapes from the dataset. In total, 120 outliers are tested. In table \ref{tab:outliers}, one can see that the relative errors on the different classes of outliers are systematically larger than those obtained on the dataset shapes. While the prediction errors on polygons are only slightly higher that those from the test set, field prediction errors for misplaced shapes are systematically superior to 10\%. Enlarged shapes present extremely high reconstruction and prediction errors, higher than 100\%. An example of prediction on enlarged Bezier shape is shown in figure \ref{fig:example_enlarged}, illustrating the interest of incorporating uncertainty estimation and outlier detection processes in neural network architectures.

\begin{figure}
\centering
\begin{subfigure}[t]{.45\linewidth}
	\centering	
	\begin{tikzpicture}[trim axis left, trim axis right]
		\begin{axis}[	scale=0.8,transform shape, 
					label style={font=\scriptsize}, tick label style={font=\scriptsize}, legend style={nodes={scale=0.7, transform shape}, font=\scriptsize},
					ymin=0, ymax=0.0002, xmin=0, xmax=0.0008,
					xlabel=$e_s$,ylabel=$e_f$,
					scaled y ticks=base 10:4,
					legend pos=south east,
					legend cell align={left},
					clip=true
					]
		        \addplot+[only marks,mark=*,mark options={draw=black,fill=myorange1},mark size=1.5pt] table[x index=1,y index=2] {fig/mse_test.csv};
		        \addlegendentry{Test set}
		        \addplot+[only marks,mark=*,mark options={draw=black,fill=black},mark size=1.5pt] table[x index=1,y index=2] {fig/mse_polygons.csv};
		        \addlegendentry{Polygons}
		        \addplot+[only marks,mark=*,mark options={draw=black,fill=green},mark size=1.5pt] table[x index=1,y index=2] {fig/mse_lower_right.csv};
		        \addlegendentry{Lower right}
		        \addplot+[only marks,mark=*,mark options={draw=black,fill=yellow},mark size=1.5pt] table[x index=1,y index=2] {fig/mse_upper_right.csv};
		        \addlegendentry{Upper right}
		        \addplot+[only marks,mark=*,mark options={draw=black,fill=purple},mark size=1.5pt] table[x index=1,y index=2] {fig/mse_huge.csv};
		        \addlegendentry{Enlarged}
			\draw[mybluegray1, very thick, dash pattern=on 2pt] (axis cs:0.00019,\pgfkeysvalueof{/pgfplots/ymax}) -- (axis cs:0.00019,\pgfkeysvalueof{/pgfplots/ymin});
			\draw[mybluegray1, very thick, dash pattern=on 2pt] (axis cs:\pgfkeysvalueof{/pgfplots/xmax},0.00005) -- (axis cs:\pgfkeysvalueof{/pgfplots/xmin},0.00005);
			\fill[mybluegray3, opacity=0.5] (axis cs:\pgfkeysvalueof{/pgfplots/xmin},\pgfkeysvalueof{/pgfplots/ymin}) rectangle (axis cs:0.00019,0.00005);
		\end{axis}	
	\end{tikzpicture}
	\caption{Qualitative method: using the selection criterion $(e_f^*, e_s^*) = (5\times 10^{-5}, 1.9\times 10^{-4})$ from previous analysis, $110$ bad predictions out of 120 outliers are detected.}
	\label{fig:outliers_accept}
\end{subfigure} \quad
\begin{subfigure}[t]{.45\linewidth}
	\centering	
	\begin{tikzpicture}[trim axis left, trim axis right]
		\begin{axis}[	scale=0.8,transform shape, 
					label style={font=\scriptsize}, tick label style={font=\scriptsize}, legend style={nodes={scale=0.7, transform shape}, font=\scriptsize},
					ymin=0, ymax=0.0002, xmin=0, xmax=0.0008,
					xlabel=$e_s$,
					scaled y ticks=base 10:4,
					legend pos=south east,
					legend cell align={left},
					clip=true
					]
		        \addplot+[only marks,mark=*,mark options={draw=black,fill=myorange1},mark size=1.5pt] table[x index=1,y index=2] {fig/mse_test.csv};
		        \addplot+[only marks,mark=*,mark options={draw=black,fill=black},mark size=1.5pt] table[x index=1,y index=2] {fig/mse_polygons.csv};
		        \addplot+[only marks,mark=*,mark options={draw=black,fill=green},mark size=1.5pt] table[x index=1,y index=2] {fig/mse_lower_right.csv};
		        \addplot+[only marks,mark=*,mark options={draw=black,fill=yellow},mark size=1.5pt] table[x index=1,y index=2] {fig/mse_upper_right.csv};
		        \addplot+[only marks,mark=*,mark options={draw=black,fill=purple},mark size=1.5pt] table[x index=1,y index=2] {fig/mse_huge.csv};
		        \addplot+[mybluegray1, very thick,name path=avg,mark=none] coordinates {(0,0) (0.0008,0.0002079738)};
			
			\addplot[mybluegray3, very thick,name path=std21] coordinates {(0,0) (0.0008,0.0003773194)};
			\addplot[mybluegray3, very thick,name path=std22] coordinates {(0,0) (0.0008,0.0002903984)};
			\addplot[mybluegray3, very thick,name path=std23] coordinates {(0,0) (0.0008,0.0001210528)};
			\addplot[mybluegray3, very thick,name path=std24] coordinates {(0,0) (0.0008,0.0000386282)};
			\addplot[mybluegray4,opacity=0.75,forget plot] fill between[of=std22 and std23];
			\addplot[mybluegray4,opacity=0.5,forget plot] fill between[of=std21 and std24];
		        
		\end{axis}	
	\end{tikzpicture}
	\caption{Quantitative method: using the regression line from equation \ref{eq:estimate}, $110$ bad predictions out of 120 outliers fall into the $2\sigma$ confidence interval.}
	\label{fig:outliers_interval}
\end{subfigure}

\caption{\textbf{Prediction and reconstruction error on the outliers.} The qualitative and quantitative methods are illustrated in (a) and (b).}
\end{figure}
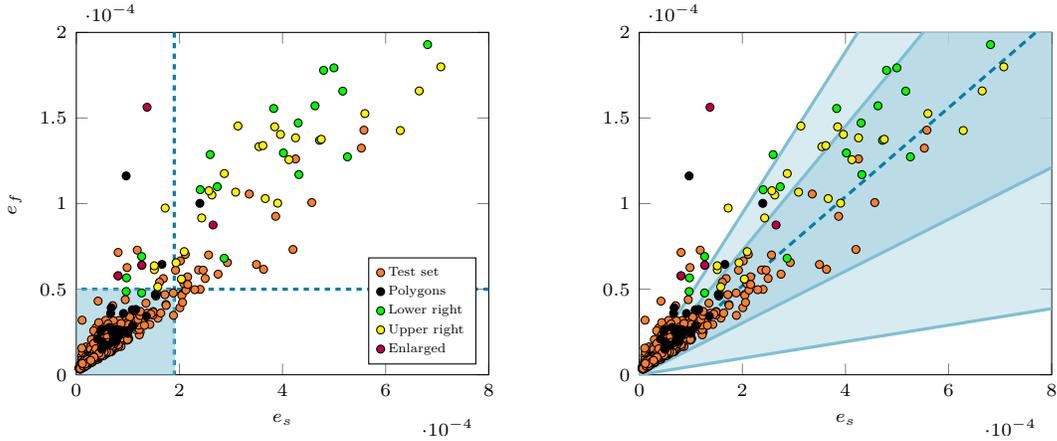

In figure \ref{fig:outliers_interval}, the $(e_s, e_f)$ scatter plot of the test set is shown, along with the position of the 120 outliers, both for the qualitative and the quantitative methods. As can be seen, most polygons are located in the accepted region of the qualitative method, indicating that the higher average relative error shown in table \ref{tab:outliers} is caused by a few polygon outliers with large relative errors. When inspecting the polygons set, it was found out that those with largest $e_f$ errors were presenting edges features that were considerably sharper than the others. Almost all the misplaced and enlarged shapes fall into the rejected area of the qualitative method, while also matching well the $\pm 2\sigma$ interval of the quantitative method ($27$ enlarged shapes out of $30$ are not shown due to their out-of-range MSE.). Still, a handful of outliers presenting low $e_s$ with large $e_f$ are noticed, indicating that the proposed methods are still missing a small amount of outliers. Overall, the qualitative method efficiently detects most of the outliers, with only $10$ out of $120$ bad predictions not detected (\ie 91.6\% of outliers detected). The $\pm 1\sigma$ interval of the quantitative method covers 85 out of 120 (70.8\%) outliers, while $\pm 2\sigma$ interval covers 106 out of 120 (88.3\%) outlier predictions. The results of figure \ref{fig:outliers_interval} also indicate that the uncertainty levels of misplaced and enlarged inputs are partly under-estimated.

Overall, the qualitative method efficiently diminishes the risk of mis-use of the trained model, by catching more than 90\% of the bad inputs. Still, it remains a limited, binary method, and by construction approximately discards 1\% of the dataset points. Conversely, the quantitative method also provides adequate confidence range for almost all the elements from the test set, and for nearly 90\% of the outliers. However, the provided error intervals for inputs leading to very large $e_f$ errors are underestimated. Finally, similarly to the qualitative one, the quantitative method cannot account for a handful of points presenting large $e_f$ values in conjunction with small $e_s$ values.

\begin{table}
\footnotesize
\caption{\textbf{Model performance on test set and polygons}. The comparison is based on the average pixel-level relative error in the $30\times 45$ rectangular zone around the obstacles.}
\label{tab:outliers}
\centering
\medskip
\begin{tabular}{ccccc}
																									\toprule
					&\textbf{Shape rel. error} 	&\textbf{$u$ rel. error} 	&\textbf{$v$ rel. error} 	&\textbf{$p$ rel. error}	\\\midrule
\textbf{Bezier Test Set} 	&$\textbf{3.43\%}$			&$\textbf{3.92\%}$			& $\textbf{3.57\%}$			& $\textbf{3.55\%}$			\\\midrule                   
\textbf{Polygons }        	&$4.16\%$			&$6.11\%$			& $4.74\%$			& $4.70\%$			\\\midrule
\textbf{Lower right}      	&$16.4\%$			&$13.5\%$			& $17.1\%$			& $16.2\%$			\\\midrule
\textbf{Upper right}      	&$16.9\%$			&$11.4\%$			& $18.2\%$			& $14.3\%$			\\\midrule
\textbf{Enlarged}         	&$214\%$				&$244\%$				& $153\%$			& $146\%$			\\\bottomrule
\end{tabular}
\end{table}

\section{Conclusion}

In the present contribution, a twin-AE architecture for 2D incompressible laminar flow prediction with embedded uncertainty estimation was presented. The underlying motivation was to propose a method to naturally incorporate outlier detection and uncertainty estimation in the training procedure, in order to provide a decisional tool to the potential end-user. The embedded uncertainty estimation relies on the coupling of the autoencoder for flow prediction with a second autoencoder for input reconstruction, using well-chosen skip connections. Doing so naturally enforces a quasi-linear relation between the flow prediction error and the input reconstruction error. Building on this particular trait, simple yet effective qualitative and quantitative techniques were proposed to detect outliers and provide uncertainty prediction on any input provided to the trained network.

The proposed architecture was trained on a dataset of 12000 laminar flows around random 2D shapes, generated using Bezier curves. After hyper-parameter calibration, the correlation coefficient between the reconstruction error and flow prediction error reached 0.95 on the test set. The two methods were then tested on true outliers presenting different flaws (polygonal shapes that did not belong to the dataset, shapes from the dataset misplaced in the input domain, shapes significantly larger than those of the dataset), and proved efficient to either reject shapes with high flow prediction errors, or provide adequate uncertainty range. Still, a handful of outliers remained undetected, or their associated uncertainty range was under-estimated. Possible improvements could be brought to these methods, either by improving the network architecture to improve the flow error/reconstruction error correlation level, or by gaining more control on the dataset generation, in order to avoid the inclusion of possible outliers.

These results underline the potential of the proposed approach. Indeed, the implementation of such methods in prediction tasks can significantly lower the risk of the end-user taking decisions based on network predictions using inadequate inputs. Efforts shall be pursued for more accurate input reconstruction. From the experiments on U-Dual-AE, low-level features from its encoder bring great benefits to flow prediction. If the shape decoder of twin-AE provides equivalently beneficial features, we expect an improved flow prediction while keeping the strong correlation between its twin decoders.

\section*{Acknowledgements} 
This work is supported by the Carnot M.I.N.E.S. Institute through the M.I.N.D.S. project.

\appendix

\section{Open source code}
\label{section:open_source}

The code of this project is available on the following github repository: \url{https://github.com/jviquerat/twin_autoencoder}\footnote{The code will be released upon publication of the present manuscript}.

\bibliographystyle{unsrt}
\bibliography{bib}

\end{document}